\documentclass[12pt]{article}

\usepackage{epsf,epsfig}
\usepackage{graphics}
\usepackage{bbold}
\usepackage{dsfont}
\usepackage{xcolor} 
\usepackage{cite}
\usepackage{slashed}
\usepackage{verbatim}
\usepackage{amsmath,amssymb,booktabs}
\usepackage{float}
\def\be{\begin{equation}}
\def\ee{\end{equation}}
\def\bea{\begin{eqnarray}}
\def\eea{\end{eqnarray}}

\def\mbr{$\mathcal{M}$-brane}
\def\mbu{$\mathcal{M}$-bulk}
\def\cbr{$\mathcal{C}$-brane}
\def\cbu{$\mathcal{C}$-bulk}
\def\nn{\nonumber}

\newcounter{MBQ}

\begin{document}

\allowdisplaybreaks
\thispagestyle{empty}  
  
\begin{flushright}  
{\small  
TUM-HEP-1007/15\\
OUTP-15-13P\\
arXiv:1508.01705 [hep-ph] \\[0.1cm] 
August 05, 2015
}  
\end{flushright}  
\vskip1.2cm  

\begin{center}
\Large\bf\boldmath
Lepton flavour violation in RS models with a brane- or 
nearly brane-localized Higgs
\unboldmath
\end{center}

\vspace{0.9cm}
\begin{center}
{\sc M.~Beneke$^{a}$}, {\sc P.~Moch$^{a}$} and 
{\sc J.~Rohrwild$^{b}$}\\[5mm]
{\sl ${}^a$Physik Department T31, James Franck-Stra\ss e 1\\
Technische Universit\"at M\"unchen,\\
D--85748 Garching, Germany}\\
\vspace{0.3cm}
{\sl ${}^b$Rudolf Peierls Centre for Theoretical Physics,\\ 
University of Oxford,
1 Keble Road,\\ Oxford OX1 3NP, United Kingdom}\\[1.5cm]
\end{center}

\date{\today}

\begin{abstract}
\noindent
We perform a comprehensive study of charged lepton flavour violation 
in Randall-Sundrum (RS) models in a fully 5D quantum-field-theoretical 
framework. We consider the RS model with minimal 
field content and a ``custodially protected'' extension as well as 
three implementations of the IR-brane localized Higgs field,  
including the non-decoupling effect of the KK excitations of a 
narrow bulk Higgs. Our calculation provides the first complete result for 
the flavour-violating electromagnetic dipole operator in 
Randall-Sundrum models. 
It contains three contributions with different dependence on the 
magnitude of the anarchic 5D Yukawa matrix, which can all be 
important in certain parameter regions. We 
study the typical range for the branching fractions 
of $\mu\to e\gamma$, $\mu \to 3e$, $\mu N  \to e N$ as well 
as $\tau\to \mu\gamma$, $\tau \to 3\mu$ 
and the electron electric dipole moment by a numerical scan 
in both the minimal and the custodial RS model. The 
combination of $\mu\to e\gamma$ and $\mu N  \to e N$ 
currently provides the most stringent constraint on the parameter 
space of the model.  A typical lower limit on the KK scale $T$ is 
around $2 \,{\rm TeV}$ in the minimal model (up to 4~TeV in the 
bulk Higgs case with large Yukawa couplings), and around $4 \,{\rm TeV}$ 
in the custodially protected model, 
which corresponds to a mass of about 10~TeV for the first 
KK excitations, far beyond the lower limit from the non-observation 
of direct production at the LHC. 
\end{abstract}

\newpage  
\setcounter{page}{1} 


\section{Introduction}
\label{sec:intro}

Rare lepton decays are among the promising indirect probes for physics 
beyond the Standard Model (SM). This is especially true for processes 
featuring lepton-flavour violation (LFV). For such decays the SM contribution 
vanishes for all practical purposes due to the small neutrino masses, 
and the experimental signature is typically very clean. The absence of a 
SM background makes the study of the decays $\mu \to e \gamma$ and 
$\mu \to 3 e$ as well as muon-to-electron conversion in nuclei in 
extensions of the SM also theoretically cleaner than the study of lepton 
observables unrelated to flavour violation. The reason for this is that 
in the latter case the search for new physics involves the search for 
tiny deviations in observables such as $(g-2)_\mu$, which must then be 
measured and predicted to very high precision. 

The warped extra-dimensional Randall-Sundrum (RS) models, originally 
introduced to address the relative weakness of gravity \cite{Randall:1999vf} 
and the gauge-gravity hierarchy problem~\cite{Randall:1999ee}, generically 
also have a rich flavour structure. Further, they provide a geometric 
interpretation for the flavour hierarchies observed in the SM: the wave 
functions of SM fermions in the fifth dimension when localized to varying 
degree naturally generate a hierarchical flavour 
sector~\cite{Grossman:1999ra,Gherghetta:2000qt,Huber:2000ie,Huber:2003tu}. 
Despite the fact 
that the RS model is to some extent protected from large 
flavour-changing neutral currents (FCNCs)~\cite{Agashe:2004cp}, the  
strongest constraint on the scale of the extra dimension comes from 
Kaluza-Klein (KK) gluon mediated $\Delta F=2$ processes---notably in the 
kaon system \cite{Csaki:2008zd}. However, these bounds can, to some degree, 
be avoided by imposing some structure on the quark Yukawa matrices in the 
five-dimensional (5D) theory \cite{Blanke:2008zb} or 
by extending the strong gauge 
group~\cite{Bauer:2011ah}. Nonetheless, direct searches at the LHC 
for KK states are no longer very promising as the scales that can be 
probed in direct production are comparably low. This is precisely 
the situation were low-energy precision observables can help considerably
constraining the model.    

In this work we calculate the lepton-flavour violating four-fermion and 
electromagnetic dipole operators induced by the RS bulk fields at the 
tree- and one-loop level, respectively. We then study the $\mu\to 3e$ and  
$\mu \to e \gamma$ transitions, and muon-to-electron conversion in 
nuclei in the minimal and custodially protected RS model with 
three different implementations of the Higgs field localized on or 
near the TeV brane of the model. Lepton-flavour violating processes have been 
studied in the context of the RS model in the past, beginning 
with~\cite{Huber:2003tu,Kitano:2000wr,Moreau:2006np}. Particularly 
relevant are \cite{Agashe:2006iy}, which gives the first comprehensive 
analysis of charged LFV, and \cite{Csaki:2010aj} with the first 
fully five-dimensional treatment of loop effects, which dominate the 
$\mu \to e \gamma$ decay. However, neither of the two can be said to 
provide a complete description of the loop-induced dipole operator 
coefficient, which leads us to reconsider LFV phenomenology building 
on our recent work \cite{Beneke:2012ie,Beneke:2014sta,Moch:2014ofa} on 
penguin transitions in the RS model. In this approach we integrate 
out the five-dimensional (5D) bulk and match the RS model on the 
SM extended by gauge-invariant dimension-six 
operators~\cite{Buchmuller:1985jz,Grzadkowski:2010es}, which is then 
used to study the processes of interest. This two-step procedure 
is justified, since the scale of the fifth dimension, related to the 
mass of the lightest KK excitation, is already constrained to be 
much larger than the electroweak scale. This approach allows for a 
transparent and complete calculation of gauge- and Higgs-boson exchange 
induced dipole transitions, which has already been applied to the 
flavour-diagonal case of the anomalous magnetic moment of the 
muon \cite{Beneke:2012ie,Beneke:2014sta,Moch:2014ofa}. Here we focus 
on the flavour-changing leptonic transitions. We also add a 5D 
treatment of the bulk Higgs and the non-decoupling effect in the 
localization limit recently pointed out in \cite{Agashe:2014jca}. 
Depending on the exact realization of the RS model loop-effects can be 
sizeable also for $\mu\to 3e$ and muon conversion, which at first glance 
are dominated by tree-level physics, and may introduce 
new correlations between $\mu \to e \gamma$ and those observables. 
In our study we do not impose additional flavour symmetries on the Yukawa 
matrices of the model and stick to the so-called minimal and custodially 
protected models. This implies that we do not attempt here 
a description of neutrino masses and mixing in the RS framework -- 
in the minimal RS model the neutrinos remain massless -- and only consider 
charged lepton flavour violation, which arises in the RS models independent 
from neutrino masses. In order to simultaneously generate hierarchical 
charged lepton masses and large neutrino mixing, RS models with additional 
flavour structure such as minimal flavour violation or discrete 
symmetries were considered~\cite{Chen:2008qg,Perez:2008ee,Csaki:2008qq}, 
as well as mechanisms to suppress lepton flavour 
violation~\cite{Agashe:2008fe,Atkins:2010cc}. These have implications 
for LFV at tree-level, even eliminating tree-level LFV completely, but 
do not modify the structure of loop-induced LFV processes, whose complete 
computation is the main result of this work.

The outline of this paper is as follows. The starting point of 
our analysis in Section~\ref{sec:Lagrangian} is the effective SM Lagrangian 
including dimension-six operators.
We restrict ourselves to operators that can be generated in the 
Randall-Sundrum model at tree-level and to loop-induced dipole operators 
relevant to LFV processes. After expressing the Lagrangian in terms of 
fields in the broken phase, we match onto an effective low-energy theory. 
This allows us to use results from the literature to determine the LFV 
observables as functions of the dimension-six operator short-distance 
coefficients. In Section~\ref{sec:WilsonCoeffs} we discuss three variants 
of localizing the Higgs field on or near the TeV brane and give the results 
for the corresponding short-distance coefficients in the minimal and 
custodially protected RS models using and extending 
results from \cite{Beneke:2012ie,Moch:2014ofa}. LFV phenomenology of 
$\mu\to e\gamma$, $\mu\to 3 e$ and $\mu-e$ conversion and the correlations 
among these observables are presented 
in Section~\ref{sec:pheno}. We further correlate these observables with the 
electron electric dipole moment (EDM) and compare the constraining power of 
present limits on LFV observables and the electron EDM.
We conclude in Section~\ref{sec:conclusion}.


\section{From the 5D theory to low-energy observables}
\label{sec:Lagrangian}

In conformal coordinates the metric for the five-dimensional space-time
of the Randall-Sundrum model is given by
\begin{equation}
ds^2 = \left(\frac{1}{k z}\right)^{\!2}\left(\eta_{\mu\nu} dx^\mu dx^\nu-
dz^2\right), 
\label{metric}
\end{equation}
where $k = 2.44 \cdot 10^{18}\,$GeV is of order of the Planck 
scale $M_{\rm Pl}$. The fifth coordinate $z$ can take values from $z=1/k$ 
(Planck brane) to  $z=1/T$ (IR brane). The scale $T$ determines the mass 
of the KK excitations in the effective 4D spectrum and is assumed to be 
of the order of a few TeV. We denote by $\epsilon =  T/k$ the small ratio of 
the two scales.

The specific RS model is characterised by the particle content and the 
associated boundary conditions on the two branes. The two most common models 
are the minimal and the custodially protected 
model \cite{Agashe:2003zs,Agashe:2006at}. The latter has an extended
particle spectrum and gauge group to enforce a protection mechanism for 
the electroweak $\rho$ parameter and the $Zb\bar{b}$ vertex. In 
both models the Higgs is localized on or close to the IR brane. The 
prescription how the Higgs is localized is part of the model setup itself, 
see also~\cite{Carena:2012fk, Malm:2013jia}. Here we will closely follow 
the notation and conventions of \cite{Beneke:2012ie,Moch:2014ofa}
for the minimal RS model and the model with custodial protection. We 
therefore refer to these references for the exact definition of the 
5D Lagrangians. 

However, it is useful to introduce the parameters of the 5D Lagrangian
relevant for the study of lepton-flavour violation.
The 5D Lagrangian contains two parameter sets related to flavour.
On the one hand there is the 5D mass $M_{\psi_i}$ for each 5D fermion 
field $\psi_i$. The corresponding dimensionless 
5D mass parameter is $c_{\psi_i}= M_{\psi_i}/k$.
To obtain a phenomenologically viable low-energy theory that reproduces the 
measured lepton masses the $c_{\psi_i}$ have to take values not too far 
from $1/2$ for SU(2) doublets and from $-1/2$ for singlets. On the other 
hand, the 5D Higgs Lagrangian incorporates 5D Yukawa couplings $y^{\rm (5D)}$. 
It is customary to work with the dimensionless Yukawa matrices    
$Y=y^{\rm (5D)} \,k$ in the minimal model,\footnote{For the exactly 
brane-localized Higgs field. For the bulk Higgs case, the 
relation between the 5D Yukawa coupling $Y^\beta$ and $Y$ 
is given by (\ref{eq:ybeta}).} and $Y,Y_u$ in the custodial model. We  
generally assume that each is roughly $\mathcal{O}(1)$ and anarchic. 
Furthermore, the Higgs field is localized on or near the IR brane. 
There are several ways to do this.
A genuine brane Higgs is exactly localized on the IR brane and 
its potential 5D structure cannot be resolved by construction. 
An IR localized bulk Higgs can be resolved, and one also needs to 
consider possible effects of 
Kaluza-Klein states of the Higgs as was pointed out in 
\cite{Agashe:2014jca}. If the Higgs is exactly brane-localized
one can in principle introduce a separate Yukawa coupling for the coupling 
of right-handed doublets and left-handed singlet to the Higgs. However, for 
simplicity we assume that these so-called wrong-chirality Higgs 
couplings \cite{Agashe:2006iy, Azatov:2009na} are equal to their ``normal'' 
counterparts. In the following, we only discuss the different Higgs 
localizations separately if they give rise to distinct results.

Ultimately, we match the full 5D Lagrangian onto an effective 4D 
dimension-six Lagrangian following \cite{Beneke:2012ie} and determine the 
physically interesting observables as functions of the corresponding 
Wilson coefficients. The matching calculation is performed in 
Sec.~\ref{sec:WilsonCoeffs}. In the remainder of this section we 
take the SM effective theory including dimension-six operators as the 
starting point to compute the low-energy observables. We will  
consider only operators of dimension six that can be generated by the 
RS model at tree- or one-loop level.\footnote{For recent model-independent  
analyses of LFV using the effective Lagrangian, 
see \cite{Crivellin:2013hpa, Pruna:2014asa}. 
Note that these references use different conventions for 
operator normalisation, covariant derivatives and 
momentum flow compared to 
the present work.}

\subsection{The dimension-six Lagrangian}

Let us start by considering the effective SU(2)$\times$U(1)-symmetric 
Lagrangian up to dimension six, 
\begin{align}
\mathcal{L}= \mathcal{L}_{\rm SM} + 
\frac{1}{T^2}\,\mathcal{L}^{\rm dim-6} +
\text{higher-dimensional operators}\,,
\end{align}
where we did not include the single dimension-five operator as it cannot 
contribute to charged LFV. We also extracted the scale, where the effective 
description breaks down, given by $T$ in the RS model. 
At dimension six the operators relevant to the following analysis of 
LFV observables can be  taken directly 
from \cite{Grzadkowski:2010es, Buchmuller:1985jz} 
\begin{eqnarray}
\label{EffectiveSMOperatorsDim6}
\mathcal{L}^{\rm dim6}_{\rm LFV} &=& 
a^B_{ij} (\bar L_i \sigma^{\mu\nu} E_j) \Phi B_{\mu\nu}
+a^W_{ij} (\bar L_i \tau^A \sigma^{\mu\nu} E_j)  
\Phi W^A_{\mu\nu}+{\rm h.c.}
\nn \\[0.1cm]
&& + \,b^{LL}_{ijkl} (\bar L_i \gamma^\mu L_j) (\bar L_k \gamma_\mu L_l)
+b^{LE}_{ij} (\bar L_i \gamma^\mu L_i) (\bar E_{j} \gamma_\mu E_j)  
+b^{EE}_{ij} (\bar E_i \gamma^\mu E_i) (\bar E_j \gamma_\mu E_j)
\nn \\[0.1cm]
&& + \,c^1_{ij}\Phi^\dagger i \overleftrightarrow{D}^{\!\mu}\Phi 
(\bar E_i \gamma_\mu E_j)
+ c^2_{ij}\Phi^\dagger i \overleftrightarrow{D}^{\!\mu} \Phi 
(\bar L_i \gamma_\mu L_j)
+ c^3_{ij}\Phi^\dagger i \overleftrightarrow{ \tau^A D}^{\mu} \Phi 
(\bar L_i \tau^A \gamma_\mu L_j) 
\nn \\[0.2cm]
&& + \,h_{ij}(\Phi^\dagger\Phi)\bar L_i \Phi  E_j + \mbox{ h.c.}
\nn \\[0.1cm]
&& + \,\sum_{\ell=E,L} \,\sum_{q=Q,U,D}b^{\ell q}_{ij} 
(\bar \ell_i \gamma^\mu \ell_i) (\bar q_j \gamma_\mu q_j)
+ b^{L\tau Q}_{ij} (\bar L_i \tau^A \gamma^\mu L_i) 
(\bar Q_j\tau^A \gamma_\mu Q_j)\,,
\end{eqnarray}
where $L$ ($Q$) denotes doublet and $E$ ($U,D$) singlet lepton (quark) fields. 
Furthermore, $\Phi^\dagger \overleftrightarrow{i \tau^A D}_\mu  \Phi= 
\frac12(\Phi^\dagger i \tau^A D_\mu \Phi - 
\Phi^\dagger i \overleftarrow{D}_\mu\tau^A\Phi)$ and $\Phi^\dagger  
\overleftrightarrow{D}_\mu  \Phi = \frac12(\Phi^\dagger D_\mu \Phi - 
({D}_\mu\Phi)^\dagger \Phi)$ with the covariant derivative defined as
\begin{align}
D_\mu=\partial_\mu - i g^\prime \frac{Y}{2} B_\mu -   i g {T^A} {W^A_{\mu}}\,,
\end{align}
and $T^A$ the SU(2) generators in the appropriate representation 
(e.g.~$T^A=\tau^A/2$ with $\tau^A$ the Pauli matrices for the 
doublet), and $Y$ the hypercharge. The hermitian conjugate 
in~\eqref{EffectiveSMOperatorsDim6} only applies to terms in the same line. 
Of all the operators given in \eqref{EffectiveSMOperatorsDim6} only the 
dipole operators in the first line cannot be generated at tree-level in 
the RS model. The restricted flavour structure of some of the four-fermion 
operators anticipates properties of the RS model at the tree-level. Even 
though we aim to study lepton flavour violation, the operators with two 
quark and two lepton fields have to be included in 
$\mathcal{L}^{\rm dim6}_{LFV}$, since they contribute to muon conversion 
in nuclei.\footnote{It is worth noting that since the 5D 
RS models are non-renormalizable and themselves only effective descriptions 
of some yet more fundamental dynamics, the 5D generalizations of the 
operators appearing in (\ref{EffectiveSMOperatorsDim6}) could already 
be present as higher-dimensional operators in the RS Lagrangian. These 
effects would be suppressed by a factor $(T/\Lambda_{\rm UV})^2\ll 1$, where 
$\Lambda_{\rm UV}$ is the UV cut-off of the RS model, relative to 
the effects we consider here. Since the coefficient of the dipole operator 
is loop-induced in the RS model, the underlying assumption of our treatment 
is that there is no tree-level dipole effect at order 
$1/\Lambda_{\rm UV}^2$ in the UV completion of the RS model, 
which could compete with the loop-induced transition at order $1/T^2$.}

The transition from the theory with unbroken to the one with broken 
electroweak gauge symmetry proceeds via
the standard substitution rules, 
\begin{align}
 \label{SubstitutionForBrokenPhase}
   \Phi \to \begin{pmatrix}
             \phi^+\\
             \frac{1}{\sqrt{2}}(v+h+iG)
            \end{pmatrix} &&
   L_i\to U_{ij}P_L\begin{pmatrix}
             \nu_j\\[0.2cm]
             \ell_j
            \end{pmatrix} &&
   E_i\to V_{ij}P_R \ell_j\\
   Q_i\to   P_L\begin{pmatrix}
             U^u_{ij}u_j\\
             U^d_{ij}d_j
            \end{pmatrix} &&
   U_i\to V^u_{ij}P_R u_j &&  
   D_i\to V^d_{ij}P_R d_j
\end{align}
and 
\begin{align}
\label{CovDerivativeBroken}
 &D_\mu \to \partial_\mu - ie Q A_\mu -\frac{ig}{c_W}(T^3-s^2_W Q) Z_\mu
-\frac{ig}{\sqrt{2}}(T^1+iT^2)W^+_\mu 
-\frac{ig}{\sqrt{2}}(T^1-iT^2)W^-_\mu,
\end{align}
with $Z_\mu = c_W W^3_\mu - s_W B_\mu$, and $A_\mu = c_W B_\mu +s_W W^3_\mu$.
Inserting \eqref{SubstitutionForBrokenPhase} and \eqref{CovDerivativeBroken} 
into \eqref{EffectiveSMOperatorsDim6} generates many operators, most 
of which cannot contribute to the processes we are interested in, 
see \cite{Beneke:2012ie} for details.  We are only concerned with operators 
that can contribute at tree-level (if they are generated at loop- or 
tree-level in the full 5D theory) or at loop-level (if they are generated 
already at tree-level). This leaves us with the Lagrangian
\begin{align}
\label{LFVLagrangianInBrokenPhase}
\mathcal{L}^{\rm dim6}_{LFV}\to  \mathcal{L}^{\rm broken}_{LFV} = 
& \;\frac{\alpha^A_{ij}+{\alpha^A_{ji}}^\star}{2} 
\frac{v}{\sqrt{2}} \,(\bar\ell_i   \sigma^{\mu\nu} \ell_j) F_{\mu\nu} 
+ \; \frac{\alpha^A_{ij}-{\alpha^A_{ji}}^\star}{2} \frac{v}{\sqrt{2}} \,
(\bar\ell_i   \sigma^{\mu\nu} \gamma_5\ell_j) F_{\mu\nu}\nn \\
& + \beta^{EE}_{ijkl} (\bar \ell_i \gamma^\mu P_R \ell_j)  
(\bar \ell_k  \gamma_\mu P_R \ell_l)
+ \beta^{LE}_{ijkl} (\bar \ell_i  \gamma^\mu P_L \ell_j) 
(\bar\ell_k    \gamma_\mu P_R \ell_l)  \nn \\
& + \beta^{LL}_{ijkl} (\bar \ell_i \gamma^\mu P_L \ell_j)  
(\bar \ell_k  \gamma_\mu P_L \ell_l)   \nn \\
& -  {\gamma}^1_{ij} \frac{g v^2}{4c_W}  Z^\mu  
(\bar \ell_i \gamma_\mu P_R \ell_j)
- [{\gamma}^2_{ij}+\gamma^{3}_{ij}] \frac{g v^2}{4c_W}  Z^\mu  
(\bar \ell_i  \gamma_\mu P_L\ell_j) \nn \\
&  +  {\gamma}^3_{ij} \frac{g v^2}{2 \sqrt{2}}  W^{+,\mu}  
(\bar \nu_i \gamma_\mu P_L \ell_j) + {\text{h.c.}} \nn \\
& +\eta_{ij} \frac{3 v^2}{2} \frac{h}{\sqrt{2}} (\bar \ell_i P_R   \ell_j)
 +\eta_{ij} \frac{v^3}{2\sqrt{2}}(\bar \ell_i P_R   \ell_j) 
 + {\text{h.c.}} \nn \\
& + \beta^{Eu}_{ijkl} (\bar \ell_i \gamma^\mu P_R \ell_j) 
 (\bar u_k\gamma_\mu  P_R u_l)
 + \beta^{Ed}_{ijkl} (\bar \ell_i \gamma^\mu P_R \ell_j) 
 (\bar d_k \gamma_\mu P_R d_l)\nn \\
& + \beta^{EQ}_{ijkl} (\bar \ell_i \gamma^\mu P_R \ell_j) 
  (\bar u_k \gamma_\mu P_L u_l)
  + \beta^{EQ}_{ijkl} (\bar \ell_i\gamma^\mu  P_R \ell_j) 
  (\bar d_k\gamma_\mu  P_L d_l)\nn \\
& + \beta^{Lu}_{ijkl} (\bar \ell_i\gamma^\mu  P_L \ell_j) 
  (\bar u_k \gamma_\mu P_R u_l) 
  + \beta^{Ld}_{ijkl} (\bar \ell_i \gamma^\mu P_L \ell_j) 
  (\bar d_k \gamma_\mu P_R d_l)\nn \\
& + (\beta^{LQ}_{ijkl}-\beta_{ijkl}^{L\tau Q}) 
  (\bar \ell_i \gamma^\mu P_L \ell_j) (\bar u_k \gamma_\mu P_L u_l)\nn \\      
& +(\beta^{LQ}_{ijkl}+\beta_{ijkl}^{L\tau Q}) 
  (\bar \ell_i \gamma^\mu P_L \ell_j) (\bar d_k \gamma_\mu P_L d_l)\,,
\end{align}
with $P_{L/R}=\frac12 (1\mp\gamma_5)$, and 
\begin{align}
\alpha^A_{ij} &= [U^\dagger a^A V]_{ij}\,,                
& \gamma^{1}_{ij} &= \sum_{m,n} \,[V^\dagger]_{im} V_{nj} \,c^{1}_{mn} \,,
\nonumber\\
\eta_{ij} &= \sum_{n,m} [U^\dagger]_{im} h_{mn} V_{nj}\,,  
& \gamma^{x}_{ij} &= \sum_{m,n} \,[U^\dagger]_{im} U_{nj} \,c^{x}_{mn} \quad
(x=2,3)\,,
\nonumber\\[0.1cm]
\beta^{LL}_{ijkl} &= \sum_{m,n,o,p} \,
[U^\dagger]_{im} U_{nj} [U^\dagger]_{ko} U_{pl} \,b^{LL}_{mnop} \,,
& \beta^{FF'}_{ijkl}&= \sum_{m,n} \,
[M^\dagger]_{im} M_{mj} [M'^\dagger]_{kn} M'_{nl} \,b^{FF'}_{mn}\,.
\label{couplingdefs}
\end{align}
Here $a^A_{ij}=c_W a^B_{ij} - s_W a^W_{ij}$, $M^{(\prime)}\in 
\{U,V,U^u,V^u,U^d,V^d\}$ are the appropriate 
flavour rotation matrices for the fermion $F^{(\prime)}$, and 
a similar definition applies to $\beta_{ijkl}^{L\tau Q}$.

\subsection{From the effective Lagrangian to LFV observables}
\label{sec:toLFVobservables}

The main observables for charged lepton flavour violation are radiative 
transitions of the type $\ell_1 \to \ell_2 \gamma$, lepton conversion in 
a nucleus, and tri-lepton decays $\ell_1 \to \ell_2\ell\ell$.
These processes are usually studied in high intensity, low energy set-ups. 
The typical energy release of the process is the mass of the initial 
(charged) lepton, a muon or a tau. Starting from the effective 
Lagrangian at the electroweak scale discussed in the previous section, 
we construct an effective low-energy Lagrangian by integrating out the 
heavy gauge bosons and quarks and the fluctuations associated with 
scales above the charged lepton mass. For $\mu\to e\gamma$, muon conversion
and $\mu \to 3e$ this low-energy theory has been discussed in great detail 
in the literature, see e.g.~\cite{Chang:2005ag, Kitano:2002mt}, and 
especially \cite{Kuno:1999jp}. 

We follow \cite{Kuno:1999jp} and consider first the radiative decay 
$\mu \to e \gamma$. The Lagrangian takes the form\footnote{Our 
convention for $A_{L,R}$ and $g_i$ below differs 
from \cite{Kuno:1999jp}, Eq.~(54) 
by the factor $-4 G_F/\sqrt{2}$ and complex conjugation.} 
\begin{align}
\label{eq:MEGlagrangian}
\mathcal{L}_{\mu \to e \gamma} =&
A_R m_\mu \bar \ell_e \sigma^{\sigma\rho}F_{\sigma\rho} P_R \ell_\mu  + 
A_L m_\mu \bar \ell_e \sigma^{\sigma\rho}F_{\sigma\rho} P_L \ell_\mu + 
\mbox{h.c.}\,,
\end{align} 
where the label on the field $\ell$ denotes the lepton flavour. 
This Lagrangian is supposed to be valid at scales below the electron 
mass---all quantum fluctuations involving leptons have been integrated out and 
have been absorbed into the two coefficients of the dipole operators. 
In fact instead of this Lagrangian we might just as 
well consider the general U(1)$_{em}$ invariant vertex function for 
on-shell fermions (the photon momentum $q$ is ingoing)
\begin{eqnarray}
\label{photonvertex}
\Gamma^\mu(p,p^\prime) &=& i e Q_\ell \bar u_\mu (p^\prime,s^\prime)
\bigg[ \gamma^\mu F_1(q^2) + \frac{i \sigma^{\mu\nu}q_\nu }{2 m_\mu} F_2(q^2)
+ \frac{\sigma^{\mu\nu}q_\nu }{2 m_\mu} \gamma_5 F_3(q^2)
\nn \\
&& + \,
\left(q^2 \gamma^\mu - \slashed{q}q^\mu \right)\gamma_5 F_4(q^2)\bigg] 
u_e(p,s)\,.
\end{eqnarray}
The on-shell dipole form factors of the electromagnetic 
muon-electron vertex are related to the coefficients $A_L$ and $A_R$ by 
\begin{align}
A_R= \frac{ Q_\ell e (F_2(0)-i F_3(0))}{4 m_\mu^2} &&
A_L= \frac{ Q_\ell e (F_2(0)+ i F_3(0))}{4 m_\mu^2}\,,
\end{align}
where $Q_\ell=-1$ is the electron charge in units of the positron charge $e$. 
Up to terms suppressed by powers of the electron mass the branching fraction 
can be written as
\begin{equation}
\label{eq:BrMEG}
{\rm{Br}}(\mu \to e\gamma)=  
\frac{m_\mu^5}{4\pi \Gamma_\mu}(|A_L|^2+|A_R|^2)\,. 
\end{equation}
Here $\Gamma_\mu$ is the total decay width of the muon. The generalization 
to $\ell_i\to\ell_j\gamma$ is obvious.

The process $\mu \to 3e$ is described by the 
extended  Lagrangian~\cite{Kuno:1999jp}
\begin{eqnarray}
\label{LowEngergyMuon3e}
\mathcal{L}_{\mu\to 3e} &=&
\,A_R m_\mu \bar \ell_e \sigma^{\sigma\rho}F_{\sigma\rho} P_R \ell_\mu  + 
A_L m_\mu \bar \ell_e \sigma^{\sigma\rho}F_{\sigma\rho} P_L \ell_\mu  
\nn \\
&& + \,g_1 \,\bar \ell_e P_R \ell_\mu \;\bar \ell_e  P_R \ell_e 
+ g_2 \,\bar \ell_e  P_L \ell_\mu\; \bar\ell_e P_L \ell_e  
\nn \\ 
&&+\,g_3 \, \bar \ell_e \gamma^\nu P_R \ell_\mu\;\bar \ell_e 
\gamma_\nu P_R \ell_e 
+  g_4 \,\bar \ell_e \gamma^\nu P_L\ell_\mu \;\bar \ell_e \gamma_\nu 
P_L \ell_e 
\nn \\
&&   +\,g_5 \, \bar \ell_e \gamma^\nu P_R \ell_\mu\;\bar \ell_e 
\gamma_\nu P_L \ell_e
+  g_6 \,\bar \ell_e \gamma^\nu P_L \ell_\mu\;\bar \ell_e 
\gamma_\nu P_R \ell_e +\,\text{h.c.} \,.
\end{eqnarray}
Note that the coefficients 
$A_{L,R}$ and $g_i$ have mass dimension $-2$. The appearance of the 
same coefficients $A_{L,R}$ as in \eqref{eq:MEGlagrangian} indicates 
that all quantum fluctuations are again integrated out. In practice, absorbing e.g.~electron loop diagrams involving a four-fermion operator into 
$A_{L,R}$ and into a loop correction to the $g_i$ is convenient as we do not 
have to treat the different lepton flavours separately. In particular, 
in writing (\ref{LowEngergyMuon3e}) the effect of the 
off-shell ($q^2\neq 0$) form factors in  \eqref{photonvertex} is absorbed 
into the $g_i$ coefficients (see the \cite{Kuno:1999jp} 
for details). In any case, since this represents a loop correction to 
the Wilson coefficients which are already generated at 
the tree-level, we neglect these effects in our calculation.

The branching fraction of $\mu \to 3e$ can easily be expressed through
the coefficients $g_i$ and $A_{L,R}$ \cite{Kuno:1999jp}:\footnote{
The sign of the interference term (second line in \eqref{eq:Brmu3e})
depends on the convention for the covariant derivative. In the convention 
of \cite{Kuno:1999jp} the sign is `+'. This is 
compensated by the Wilson coefficients $A_{L,R}$, the sign of 
which is also convention dependent.}
\begin{eqnarray}
\label{eq:Brmu3e}
{\rm Br}(\mu \to 3 e) &=& \frac{m_\mu^5}{1536\pi^3 \Gamma_\mu}
\bigg[\frac{|g_1|^2+|g_2|^2}{8}+2 (|g_3|^2+|g_4|^2)+|g_5|^2+|g_6|^2 
\nn \\[0.1cm]
&&  - \,8 e\, {\rm{Re}}\left[A_R (2 g_4^*+g_6^*)+A_L(2 g_3^*+g_5^*) \right]
\nn \\[0.1cm]
&& +\,64 e^2 (\ln\frac{m_\mu}{m_e}-\frac{11}{8})(|A_L|^2+|A_R|^2) \bigg]
\end{eqnarray}
with $\Gamma_\mu$ the muon decay width. 
The first line arises from tree-level KK exchange in the RS model, 
while the second and third involve the loop-induced dipole operator 
coefficients. The reason for keeping these formally suppressed terms 
is not only the logarithmic enhancement paired with a large numerical 
coefficient. One-loop corrections to the $g_i$ may have  
a similar logarithmic enhancement and the large factors of 8 and 64 might be  
misleading, since it can be absorbed into the definition of the 
coefficients $A_{L,R}$, which come with a $1/(4\pi)^2$ loop suppression. 
The important point is that one-loop corrections to the $g_i$ introduce 
only a small shifts in the coefficients without altering the general 
properties. On the other hand, the dependence on $A_{L,R}$ implies 
sensitivity to different aspects of the underlying model. In RS models
$A_{L,R}$ have a specific dependence on the Yukawa couplings, which can 
provide an important contribution to the branching fraction in sizeable 
parts of the model parameter space. In these regions
the effect of $A_{L,R}$ should not be neglected, as it will significantly 
alter the signatures of the RS model in flavour observables.

Muon conversion in nuclei is mediated by both, operators containing 
quark fields and electromagnetic dipole operators. The effective 
Lagrangian is \cite{Kuno:1999jp,Crivellin:2013ipa}
\begin{eqnarray}
\label{LowEngergyMuonN}
\mathcal{L}_{\mu N\to e N} &=&
A_R m_\mu \bar \ell_e \sigma^{\sigma\rho}F_{\sigma\rho} P_R \ell_\mu  + 
A_L m_\mu \bar \ell_e \sigma^{\sigma\rho}F_{\sigma\rho} P_L \ell_\mu  
\nn \\
&& + \,\sum_{q=u,d} c^q_{VR}   \bar \ell_e  
\gamma^\nu P_R \ell_\mu \bar q  \gamma_\nu q
+ \sum_{q=u,d} c^q_{VL}   \bar \ell_e  \gamma^\nu P_L \ell_\mu \bar q  
\gamma_\nu q 
\nn \\
&&  + \,\sum_{q=u,d,s} \frac{m_q m_\mu}{M_H^2}  c^q_{SL}   
\bar \ell_e   P_R \ell_\mu \bar q q
+ \sum_{q=u,d,s} \frac{m_q m_\mu}{M_H^2}  c^q_{SR}  \bar \ell_e   
P_L \ell_\mu \bar q q 
\nn  \\                 
&&  + \,\frac{\alpha_s m_\mu}{M_H^2}  c^L_{gg}\bar \ell_e   
P_R \ell_\mu G^{A,\sigma\rho}G^A_{\sigma\rho}
+ \frac{\alpha_s m_\mu}{M_H^2} c_{gg}^R \bar \ell_e   
P_L \ell_\mu G^{A,\sigma\rho}G^A_{\sigma\rho}
+\,{\text{h.c.}}\,. 
\end{eqnarray}
Here $M_H$ denotes the Higgs mass, and $G_{\mu\nu}^A$ is the gluon field 
strength tensor. We do not include operators with pseudo-scalar, 
axial vector or tensor quark currents. Their contributions are suppressed 
by the nucleon number of the target nuclei and can be neglected. 
We also neglect the strange quark in the vector operators, since the 
coefficient is not enhanced by the strange-quark mass. The 
conversion branching fraction depends on properties of the nucleus that 
participates in the reaction. The expression, taken from 
\cite{Crivellin:2013ipa,Kitano:2002mt} and adjusted to match our conventions, 
is
\begin{eqnarray}
\label{eq:BRmToe}
{\rm Br}(\mu N  \to e N)&=&
\frac{m_\mu^5}{4 \Gamma_{\text{capture}}}\bigg| \,A_R \mathcal{D} 
+ 4\bigg[ \frac{m_\mu m_p}{M_H^2}  \!\left(\tilde C^p_{SL}\!-\!12 \pi 
\tilde C^p_{L,gg} \right)  \mathcal{S}^p
+ \tilde C^p_{VL}  \mathcal{V}^p  \nn \\ 
&& + \,\{p\rightarrow n\} \bigg] \bigg|^{\,2} 
+ \left\lbrace L\leftrightarrow R \right \rbrace\,,
\end{eqnarray}
where  $\Gamma_{\text{capture}}$ is the total muon capture rate for nucleus 
$N$. The coefficients $\mathcal{D},\mathcal{V}^{p/n},\mathcal{S}^{p/n}$  
(the superscript refers to the proton and neutron) encode properties of 
the target nucleus, see \cite{Kitano:2002mt}. The tilded coefficients are 
defined as 
\begin{eqnarray}
 \tilde C^p_{SL} &=& \sum_{q=u,d,s} c^q_{SL} f^p_{q},\\
 \tilde C^p_{L,gg} &=&   c_{gg}^L f^p_Q,\\[0.1cm]
 \tilde C^p_{VL} &=&  \sum_{q=u,d} c^q_{VL} f^p_{V_q}\,,
\end{eqnarray}
and analogously for the $p\to n$ and $L\to R$ cases. The form factors 
$f_q^{p,n}$ and $f_{V_q}^{p,n}$ parametrize the coupling strengths of the 
quark scalar and vector currents of flavour $q$ to nucleons, respectively. 
$f_Q^{p,n}$ represent the scalar couplings of heavy quarks ($c,b$ or $t$).

\begin{figure}
\centering
\includegraphics[width=10cm]{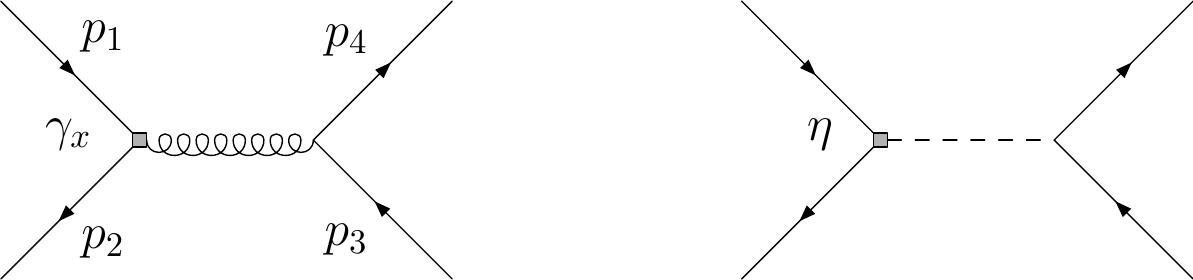}
\vspace*{0.2cm}
\caption{{\it Left:} Matching of LFV couplings of the $Z$-boson onto 
four-fermion operators in the low-energy theory.
{\it Right:} Higgs exchange diagram that contributes to $g_{1,2}$ 
and~$c^q_{SR/L}$.}
\label{fig:Zdiagramm}
\end{figure}

We now determine the coefficients of the low-energy effective Lagrangians 
in terms of the Wilson coefficients of the dimension-six Lagrangian. The 
tree-level matching of the four-fermion operators 
in \eqref{LFVLagrangianInBrokenPhase} to those in \eqref{LowEngergyMuon3e}, 
\eqref{LowEngergyMuonN} is straightforward. Further contributions 
arise from the lepton-flavour violating $Z$-boson interactions 
$\ell_i \gamma_\mu P_{L,R} \ell_j Z^\mu$ together with a SM coupling  
once the intermediate $Z$ is integrated out, see Figure~\ref{fig:Zdiagramm} 
(left). For example, in case of $\mu\to 3e$, the insertion of 
$\ell_e \gamma^\mu P_R \ell_\mu Z_\mu $ evaluates to
\begin{align}
i \mathcal{M}= i \gamma^1_{12}&
\bigg[ \frac{2 s_W^2-1}{2} \bar u_{e}(p_2) \gamma^\mu P_R u_\mu(p_1) \;  
\bar u_{e}(p_4) \gamma_\mu P_L v_e(p_3) 
\nn \\[-0.0cm]
&\;\; + s_W^2\bar u_{e}(p_2) \gamma^\mu P_R u_\mu(p_1) \;  
\bar u_{e}(p_4) \gamma_\mu P_R v_e(p_3) + \mbox{ Fierzed diagram}\bigg]\,, 
\end{align}
which gives a contribution to $g_3$ and $g_5$. Similarly, insertions of 
$\gamma^{2,3}$ will lead to contributions to $g_4$ and $g_6$. Contributions 
to $c^q_{VR}$ follow analogously. Thus we find the relations:
\begin{align}
\label{eq:gCoeffs}
g_1&=g_2=0 \\
g_3&=\frac{1}{T^2}\left( s_W^2 \gamma^1_{12} + \beta^{EE}_{1211}+
\beta^{EE}_{1112}\right) \\
g_4&=\frac{1}{T^2}\left( \frac{2 s_W^2-1}{2} (\gamma^2_{12}+\gamma^3_{12}) 
+ \beta^{LL}_{1211}+\beta^{LL}_{1112}\right)\\
g_5&=\frac{1}{T^2}\left( \frac{2 s_W^2-1}{2} \gamma^1_{12} + 
\beta^{LE}_{1112}\right)\\
g_6&=\frac{1}{T^2}\left( s_W^2 (\gamma^2_{12}+\gamma^3_{12}) + 
\beta^{LE}_{1211}\right) \,,
\end{align}
\begin{eqnarray}
\label{eq:cCoeffs}
 c^u_{VR}&=& \frac{1}{2T^2} \left[\beta^{Eu}_{1211}+\beta^{EQ}_{1211}
+\frac12 \gamma^1_{12}\left( 1-\frac{8}{3}s_W^2 \right)\right] \\
 c^u_{VL}&=& \frac{1}{2T^2} \left[\beta^{LQ}_{1211}- \beta^{L\tau Q}_{1211}
+\beta^{Lu}_{1211}+ \frac12 \left(\gamma^2_{12}+\gamma^3_{12}\right)
\left( 1-\frac{8}{3}s_W^2 \right)\right] \\
 c^d_{VR}&=& \frac{1}{2T^2} \left[\beta^{Ed}_{1211}+\beta^{EQ}_{1211}
+\frac12 \gamma^1_{12}\left(-1+\frac{4}{3}s_W^2 \right)\right]  \\
 c^d_{VL}&=& \frac{1}{2T^2} \left[\beta^{LQ}_{1211}+\beta^{L\tau Q}_{1211}
+\beta^{Ld}_{1211}+ \frac12 \left(\gamma^2_{12}+\gamma^3_{12}\right)
\left(-1+\frac{4}{3}s_W^2 \right)\right] \\
 c^q_{SL}&=& - \frac{v}{\sqrt{2}m_\mu T^2} \,\eta_{12}  \\
 c^q_{SR}&=& - \frac{v}{\sqrt{2}m_\mu T^2} \,[\eta^\dagger]_{12}\,,
\end{eqnarray}
as well as \cite{Shifman:1978zn,Crivellin:2014cta}
\begin{equation}
 c^L_{gg}= - \frac{1}{12 \pi } \sum_{q=c,b,t} c^q_{SL},\quad\qquad
 c^R_{gg}= - \frac{1}{12 \pi } \sum_{q=c,b,t} c^q_{SR} \,.
\end{equation}

It should be noted that $g_1$ and $g_2$ receive contributions from the 
tree-level Higgs exchange diagram, Figure~\ref{fig:Zdiagramm} (right diagram),
with an insertion of one flavour-changing Higgs operator $h \bar\ell_i P_R 
\ell_j + \mbox{ h.c.}$, but these are suppressed by powers of the 
electron mass (light lepton mass in the general case) and we neglect them. 
The same diagram (with the two fermion lines on the right being quarks) 
also generates $c^q_{SL}$. Here we should comment on a (well-known) 
subtlety. Naively, the operator $\bar L_i\Phi E_j \Phi^\dagger \Phi$
in (\ref{EffectiveSMOperatorsDim6}) modifies the Yukawa interaction 
according to 
\begin{align}
\frac{y_{ij}}{\sqrt{2}}  \,h \bar \ell_i P_R \ell_j \to 
\frac{y_{ij}}{\sqrt{2}}  \, h \bar \ell_i P_R \ell_j  -  
h_{ij} \frac{3v^2}{2\sqrt{2}T^2} \,h \bar \ell_i P_R \ell_j
\end{align}
after electroweak symmetry breaking but before flavour rotations. 
Here $y_{ij}$ is not the SM Yukawa coupling but the coefficient of the 
operator $\bar L_i\Phi E_j$ in the dimension-four Lagrangian.
However, the fermion mass matrix is also modified by dimension-six operator, 
\begin{align}
\label{eq:changesmass}
\frac{ y_{ij} v}{\sqrt{2}}\to m_{ij}=\frac{ y_{ij}v}{\sqrt{2}} 
-  h_{ij} \frac{v^3}{2\sqrt{2}T^2}\,.
\end{align}
Since the flavour rotation matrices $U$ and $V$ by construction diagonalize 
the modified mass term $m_{ij}$, we have to rewrite the shift of the Yukawa couplings as (see also \cite{Azatov:2009na})
\begin{align}
\left( \frac{1}{\sqrt{2}} y_{ij}  - h_{ij} \frac{3v^2}{2\sqrt{2}T^2}\right) 
h \bar \ell_i P_R \ell_j
\to \left( \frac{m_{ij}}{v} - h_{ij}\frac{v^2}{\sqrt{2}T^2}\right)  
h \bar \ell_i P_R \ell_j\;.
\end{align}
As a consequence the factor $3/2$ in the flavour-violating Higgs interaction 
$h \bar\ell_i P_R \ell_j + \mbox{ h.c.}$ of 
(\ref{LFVLagrangianInBrokenPhase}) must be replaced by 1 for the 
computation of $c^q_{SL}$, $c^q_{SR}$ above.

We did not include the effect of the RS bulk on the effective gluon-Higgs 
interaction. The operator $\alpha_s h G^{\mu\nu}G_{\mu\nu}$ 
receives sizeable contributions from KK fermions in the 
loop \cite{Azatov:2010pf,Carena:2012fk,Malm:2013jia,Archer:2014jca}.
These corresponding contributions to $c^{L,R}_{gg}$ after integrating out 
the Higgs boson are formally of the order $v^2/T^4$, but are enhanced by  
the traces of products of 5D Yukawa matrices. We only work in leading order 
in the $1/T^2$ expansion, so we drop these terms based on power counting 
arguments. In practice, it turns out that $c^{L,R}_{gg}$ (with or without 
this additional correction) gives a much smaller contribution to the 
LFV branching fractions than, e.g., $c^q_{VR}$, and could be ignored 
altogether.

\begin{figure}
 \centering
 \includegraphics[width=0.25\textwidth]{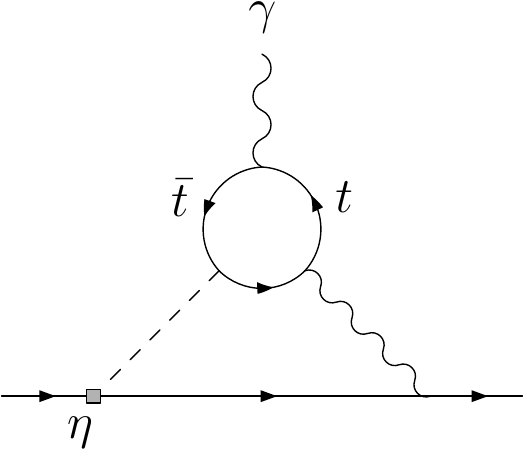}
\caption{Example of a Barr-Zee type diagram. 
The box denotes the insertion of the lepton-flavour violating Higgs 
interaction. The internal gauge boson can be a $Z$ or a photon.}
 \label{fig_BZxmp}
\end{figure}

The determination of the coefficients $A_{L,R}$ is more complicated. 
One can identify three contributions: (1) from tree or one-loop diagrams 
involving the operators in the dimension-six 
Lagrangian \cite{Beneke:2012ie,Moch:2014ofa}. (2) from dimension-eight 
operators, which may become relevant if the dimension-six contributions 
are suppressed. We will discuss them later specifically in the 
context of the RS model. (This contribution can effectively be included 
via a modification of the $a^{ij}_{B,W}$ Wilson 
coefficients.) (3)~from enhanced two-loop ``Barr-Zee type diagrams'' 
with a flavour-changing Higgs coupling \cite{Barr:1990vd}, 
see \cite{Chang:1993kw} 
for a discussion in the context of $\mu \to e \gamma$.  An example 
diagram, which avoids the coupling of the Higgs boson to a light 
lepton through the coupling to a top or gauge-boson loop, is shown in 
Figure~\ref{fig_BZxmp}. These terms 
are known to give sizeable contributions in models where the Higgs 
interactions are the dominant sources of new flavour violation. In the RS 
model this is generally not the case. Nonetheless we include these 
terms as they may become relevant in specific scenarios.
We obtain\footnote{In practice, we can drop the terms proportional to the 
lighter lepton mass, here $m_e$.}
\begin{align}\label{ARtotal}
m_\mu A_R =  &\;\alpha_{12}^A\frac{v}{\sqrt{2}T^2}
 -\sum_{k=1,2,3}\frac{Q_\ell e}{16\pi^2 T^2} \,m_{\ell_k} \beta^{EL}_{1kk2}
\nn \\
&\;-\frac{Q_\ell e}{3(4 \pi)^2 T^2}\left( s_W^2 
\left[m_\mu (\gamma_{12}^{2}+\gamma_{12}^{3}) +m_e \gamma_{12}^1\right] 
 + m_\mu \gamma_{12}^2 -\frac32 m_\mu \gamma_{12}^3 
-\frac{3}{2} m_e \gamma_{12}^1\right)
\nn \\
  &\; + A_{BZ}\left[\eta_{12}\frac{v^2}{\sqrt{2}T^2}\right]\\
\label{ALtotal}
 m_\mu A_L =  &\;[\alpha^{A\dagger}]_{12}\frac{v}{\sqrt{2}T^2}
 -\sum_{k=1,2,3}\frac{Q_\ell e}{16\pi^2 T^2} \,m_{\ell_k} 
\beta^{EL}_{k21k}
\nn \\
 &\;-\frac{Q_\ell e}{3(4 \pi)^2 T^2}\left( s_W^2 \left[m_e (
\gamma^2_{12}+\gamma_{12}^{3})+m_\mu \gamma_{12}^1\right] 
+ m_e  \gamma^2_{12} -\frac32 m_e \gamma^3_{12} 
-\frac{3}{2} m_\mu \gamma_{12}^1\right)
\nn \\
  &\; + A_{BZ}\left[\eta^\dagger_{12}\frac{v^2}{\sqrt{2}T^2}\right] 
\end{align}
where~\cite{Chang:1993kw}
\begin{eqnarray}
 A_{BZ} &=& \frac{Q_\ell e \alpha_{\rm em}\sqrt{2} G_{\!F} v}{ 32 \pi^3 } 
\Bigg[
2 N_c Q_t^2 f(r_t)-3 f(r_W)-\frac{23}{4} g(r_W) - \frac{3}{4} h(r_W)
 \nn \\ 
&& - \,\frac{f(r_W)-g(r_W)}{2r_W} +\frac{1-4 s_W^2}{4s_W^2} 
\,\bigg\{\frac{1-4Q_t s_W^2}{4 c_W^2}\,2 N_c Q_t \tilde{f}(r_t,r_{t_Z}) 
\nn \\ 
&& - \,\frac{1}{2}\,(5-s_W^2/c_W^2) \tilde{f}(r_W,r_{WZ}) 
-\frac{1}{2}\,(7-3s_W^2/c_W^2 )\tilde{g}(r_W,r_{WZ})-\frac{3}{4} g(r_W) 
\nn \\
&&-\,\frac{3}{4} h(r_W)  
-\frac{1-s_W^2/c_W^2}{4 r_W}\,\Big(\tilde{f}(r_W,r_{WZ})-
\tilde{g}(r_W,r_{WZ})\Big)\bigg\}          
\nn \\
&&-\,\frac{1}{4 s_W^2} \Big(D_e^{(3a)}(r_W) +D^{(3b)}_{e}(r_W)+
D^{(3c)}_{e}(r_W)+D^{(3d)}_{e}(r_W)+D^{(3e)}_{e}(r_W)
\nn \\
&&+ \,D_e^{(4a)}(r_Z) +D^{(4b)}_{e}(r_Z)+D^{(4c)}_{e}(r_Z)  \Big) 
\Bigg]
\end{eqnarray}
with $r_t=m_{\rm top}^2/M_{\rm Higgs}^2$, $r_{t_Z}=m_{\rm top}^2/m_{Z}^2$, 
$r_{WZ}=m_{W}^2/m_{Z}^2$ , $r_W=m_{W}^2/M_{\rm Higgs}^2$, 
$r_Z=m_{Z}^2/M_{\rm Higgs}^2$, and $N_c=3, Q_t=2/3$. The functions $f,g,h$ 
and $\tilde f, \tilde g$ can be found in \cite{Chang:1993kw} and the 
functions $D_{a}^{X}$ in \cite{Leigh:1990kf}.
Note again that relative signs in \eqref{ARtotal}, \eqref{ALtotal} depend 
on the convention for the covariant derivative.

All expressions for $\mu\to3e$ and $\mu \to e \gamma$  in this section 
can be trivially extended to $\tau\to e \gamma$
or $\tau \to 3 \mu$ by exchanging the appropriate flavour 
indices, masses and widths. We note that we do 
not take into account the running of the Wilson coefficients
between the high and the low scale, see 
e.g.~\cite{Jenkins:2013zja,Jenkins:2013wua,Alonso:2013hga}
for the anomalous dimensions of the dimension-six operator basis.

\section{Wilson coefficients for lepton flavour violation in the RS model}
\label{sec:WilsonCoeffs}

Several of the  dimension-six operator Wilson coefficients have already 
been computed in \cite{Beneke:2012ie} for the minimal RS model and 
in \cite{Moch:2014ofa} for the custodially protected model. These 
works focus on the anomalous magnetic moment of the muon 
and therefore only included a subset of the operators that are needed for 
studies of lepton flavour violation.  We extend the computation to the 
flavour-off-diagonal transitions here.

\subsection{Treatment of the 5D Higgs field}

Before going into the details of the determination of the missing Wilson 
coefficients we need to address the treatment of the Higgs field. It is 
now well established that the physics of the RS model with an IR-brane 
localized Higgs depends on how the localization is implemented, see 
e.g.~\cite{Azatov:2010pf,Carena:2012fk,Beneke:2012ie,Malm:2013jia,Beneke:2014sta}.
In effect one needs to specify whether the 5D structure of the Higgs field, 
that is, its 5D wave function, can be resolved within the model or not. 
If one regularizes the delta function in the fifth coordinate $z$, which 
localizes the Higgs close to the IR brane, by a narrow box-shaped profile, 
\begin{align}
\label{eq:HiggsRegulator}
\delta(z-1/T)=\lim_{\delta\to 0} \,\frac{T}{\delta} 
\Theta(z-\frac{1-\delta}{T})\,, 
\end{align} 
this is equivalent to specifying the order of limits for the regulator 
$\delta$ and the regulator of the 4D loop integrals, for example a 
dimensional regulator ($\epsilon\to 0$) or a cut-off ($\Lambda\to \infty$). 
Removing the regulator $\epsilon$ first while keeping $\delta$ finite
corresponds to the case where the Higgs localization width 
$\delta/T$ can be resolved by the modes propagating in 
the loop.\footnote{This is more readily seen with a momentum 
cut-off $\Lambda$. For $\Lambda\to \infty$ with fixed small but finite 
$\delta$ the particles propagating in the loop can resolve the Higgs 
localization as the loop momentum can be larger than $T/\delta$.} 
If $\delta\to 0$ first, the Higgs width remains unresolved. The latter case 
corresponds to a truly brane localized Higgs, whereas the 
first scenario assumes a ``narrow bulk Higgs'' (in the terminology of 
\cite{Malm:2013jia}). Due to these non-commuting limits the RS model is 
only fully defined with a prescription on how the order of limits should 
be taken. 

When the Higgs field permeates the bulk, even if only close to the IR 
brane, the question arises what is the effect of the KK Higgs states. 
Since their masses are of order $T/\delta$ these effects have previously 
been assumed to be small and have not received much attention. However, 
recent work \cite{Agashe:2014jca} has shown that the sum of all 
KK Higgs contributions does not decouple in the localization limit. 
In the following we include 
a computation of Higgs KK effects in the 5D formalism. In order to 
have a consistent description of KK Higgs modes, we abandon the 
ad-hoc regularization (\ref{eq:HiggsRegulator}) and implement the Higgs 
field as a full 5D scalar doublet. Since we require both a vacuum 
expectation value and a zero-mode profile that is strongly localized near 
the IR brane we need to introduce additional brane potentials on both the 
IR and UV brane. We will follow the construction of \cite{Cacciapaglia:2006mz} 
(see also \cite{Davoudiasl:2005uu}) and thus use the same Higgs profile as 
in \cite{Agashe:2014jca}. The details of this realization together with useful 
formulae are collected in Appendix~\ref{sec:Higgs}. The 5D profile of the 
vacuum expectation value takes the form
\begin{align}
  v(z)=\sqrt{\frac{2(1+\beta)}{1-\epsilon^{2+2\beta}} } \,
  k^{3/2} T^{\beta+1} v_{\rm SM} \,z^{\beta+2}, 
\label{eq:betaprofile}
\end{align}
where $v_{\rm SM}\equiv v\simeq 246\,\mbox{GeV}$ denotes the SM Higgs 
vacuum expectation value (vev),
and the zero mode profile is, up to small corrections of order $v^2/T^2$, 
proportional to $v(z)$. The parameter $\beta$ is related to the 5D mass of 
the Higgs field and determines the degree of IR localization; the larger 
$\beta$ the stronger the localization. Since we start with a genuine 
bulk field it is always implied that $\beta$ is finite until all other 
regulators have been removed.

In order to obtain the correct SM parameters in the low-energy limit
the Yukawa matrices and the Higgs self-coupling must themselves depend on 
$\beta$. For the Yukawa matrices we indicate this dependence 
by a superscript $\beta$, while no superscript refers to the 
$\beta$-independent, dimensionless matrix. The relation is (see 
Appendix~\ref{app:yukawascaling}) 
\begin{align}
Y^{\beta} = \frac{Y}{\sqrt{k}} \frac{2-c_{L_i}+c_{E_j}+\beta}
{\sqrt{2(1+\beta)}}
\label{eq:ybeta}
\end{align}
with $c_{L_i}$, $c_{E_j}$ the 5D mass parameters of the lepton fields 
in the Higgs-Yukawa interaction.  

Ultimately, we are interested in large values of $\beta$. Whenever 
we give a result for the bulk Higgs case that does not show an 
explicit dependence on $\beta$, we tacitly assume that the $\beta\to \infty$ 
limit has been taken, and the result should be valid up to corrections 
of ${\cal O}(1/\beta)$. We will consider three 
different implementations of the Higgs field:
\begin{itemize}
\item an exactly brane localized Higgs, that is we use 
(\ref{eq:HiggsRegulator}) (necessary to avoid ambiguities in the 
calculation), but take $\delta\to 0$ first.
\item a delta-function localized narrow bulk Higgs, that is 
we use (\ref{eq:HiggsRegulator}), but keep $\delta$ finite until 
all other regulators are removed, then $\delta\to 0$. No 
Higgs KK modes are considered.
\item a true bulk Higgs with the $\beta$-profile (\ref{eq:betaprofile}) 
and KK modes.
\end{itemize}
The second scenario is somewhat inconsistent as a Higgs field 
with a resolvable width should be accompanied by resolvable KK 
excitations. We will still consider it, as it turns out that this 
precisely captures the effect of the bulk Higgs zero-mode in the
IR-localized $\beta\to \infty$ limit of the third scenario. Below we will 
discuss explicitly the different localization prescriptions only 
if they lead to a difference in the Wilson  coefficients.

\subsection{Tree-level dimension-six operators}

\subsubsection{Four-fermion operators}
\label{sec:fourfermion}

\begin{figure}
\begin{center}
\includegraphics[width=0.22\textwidth]{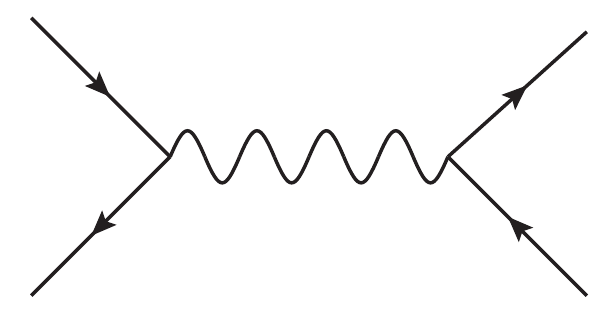}
\vskip0.3cm
\caption{  \label{fig_generic4fermion}
Generic topology of 5D diagrams 
that give rise to the four-fermion operators upon integrating out 
the exchanged particle. External states can be doublets or singlets. 
Consequently the intermediate boson can be a B or Z$_X$ or the 
SU(2) $W$ gauge boson, if all external states are doublets.
Due to the chirality of the external states the 
fifth component of the boson cannot propagate.  }
\end{center}
\end{figure}

The tree-level diagram contributing to the matching of the 
Wilson coefficients of four-fermion operators is shown in generic form 
in Figure~\ref{fig_generic4fermion}. The exchanged 
particle could be an off-shell KK gauge boson or a KK Higgs 
excitation. The latter vanishes for $\beta \to \infty$ and 
can safely be ignored, which can be verified by explicit analytic 
calculation, see Appendix~\ref{sec:example}.  
The contribution from the remaining gauge-boson exchange diagram can be 
inferred from known results \cite{Beneke:2012ie, Moch:2014ofa} by 
adjusting hypercharge and weak isospin factors. In case of $b_{LL}^{ijkl}$ 
and $b_{EE}^{ij}$, there are two contractions giving rise to an 
additional ``t-channel'' diagram. In the following we summarize the 
results for the four-fermion operators appearing in 
(\ref{EffectiveSMOperatorsDim6}).

The Wilson coefficient of the four-lepton operator 
$\left(\bar L_i \gamma_\mu L_i\right)  
\left(\bar E_j \gamma^\mu E_j\right)$ is given by 
\begin{equation}     
b^{LE}_{ij} =  \frac{Y_L Y_E}{4} 
\left[{g^\prime}^2 \left(b_0 + b_1(c_{L_i}) 
+ b_1(-c_{E_j}) + b_2(c_{L_i},c_{E_j}) \right) 
+ (g^2-{g^\prime}^2) \,b_2(c_{L_i},c_{E_j}) \right]
\label{b2coeff}
\end{equation}
with
\begin{align}
&b_0 = - \frac{1}{4}\,\frac{1}{\ln(1/\epsilon)} \,,
\\
& b_1(c) = - \frac{1}{4}\,\frac{(5-2 c)(1-2 c)}{(3-2
  c)^2} \,\frac{\epsilon^{2 c-1}}{1-\epsilon^{2 c-1}}\,,
\\
 & b_2(c_L,c_E) = 
 - \frac{1}{2}\,
\frac{(1-2 c_L) (1+2 c_E)(3-c_L+c_E)}
{(3-2 c_L) (3+2 c_E)(2-c_L+c_E)}
 \,\ln\frac{1}{\epsilon}\,
 \frac{\epsilon^{2 c_L-1}}{1-\epsilon^{2 c_L-1}}
 \,\frac{\epsilon^{-2 c_E-1}}{1-\epsilon^{-2 c_E-1}}\,.
\end{align}
As in \cite{Beneke:2012ie} we drop terms suppressed by the 
tiny ratio  $\epsilon=T/k$. The result for the minimal RS model can be 
obtained from this expression by setting the coupling to the $Z_X$ to 
zero, which corresponds to removing the $({g^2-{g^\prime}^2})$ 
term in (\ref{b2coeff}). Using these expressions the Wilson coefficient 
of the operator 
$\left(\bar E_i \gamma_\mu E_i\right) \left(\bar E_j \gamma^\mu E_j\right)$ 
takes the form
\begin{equation}
\label{b3coeff}
 b^{EE}_{ij} = \frac{Y_E}{2 Y_L} b^{LE}_{ij}(c_{L_i}\to -c_{E_i})\,. 
\end{equation}
Here and above $Y_E$ and $Y_L$ are the hypercharges of singlet and doublet 
lepton field, respectively.
For the operator $(\bar L_i \gamma^\mu L_j) (\bar L_k \gamma_\mu L_l)$ 
there are contributions from abelian $Z_X$ or $B$ bosons as above, and 
additionally from the exchange of a $W$ boson. The abelian contribution 
due to $Z_X$, $B$ exchange is given by
\begin{align}
\label{b1coeff}
b^{LL}_{ijkl,{\text{B+Z}}}& = 
\frac{Y_L}{2 Y_E}\,\delta_{ij} \delta_{kl}  \,
b^{LE}_{ik}(c_{E_k}\to -c_{L_k})\,
\qquad\mbox{(no sum over $i$, $k$)}
\end{align}
The non-abelian bosons generate the operator
$\left(\bar L_i \tau^A \gamma^\mu L_i\right)  
\left(\bar L_j \tau^A \gamma_\mu L_j\right)$, 
which is not part of our basis, and has to be rewritten using 
the SU(2) Fierz identity
\begin{align}
\left(\bar L_i \tau^A \gamma^\mu L_i \right) 
\left(\bar L_j \tau^A \gamma_\mu L_j\right) =
2 \left(\bar L_i  \gamma^\mu L_j\right) 
\left(\bar L_j  \gamma_\mu L_i\right)
-  \left(\bar L_i  \gamma^\mu L_i\right) 
\left(\bar L_j  \gamma_\mu L_j\right)\,.
\end{align}
We then find the Wilson coefficient of $(\bar L_i \gamma^\mu L_j) 
(\bar L_k \gamma_\mu L_l)$ to be
\begin{eqnarray}
b^{LL}_{ijkl} &= &b^{LL}_{ijkl,{\text{B+Z}}} + 
\frac{g^2}{4} \left(b_0 + b_1(c_{L_i}) + b_1(c_{L_j}) + 
b_2(c_{L_i},-c_{L_j})\right)\delta^{il} \delta^{kj} \nn \\
&&- \,\frac{g^2}{8} \left(b_0 + b_1(c_{L_i}) + b_1(c_{L_k}) + 
b_2(c_{L_i},-c_{L_k}) \right)\delta^{ij} \delta^{kl}\,.
\end{eqnarray} 

The Wilson coefficients of the seven quark-lepton four-fermion operators 
are even simpler to compute as there are never two identical fields
and all operators but one, $\left(\bar L_i \gamma^\mu \tau^A L_j\right) 
\left(\bar Q_k \tau^A \gamma_\mu Q_l\right)$,
are generated via the exchange of an abelian gauge boson. The result 
is
\begin{eqnarray}
b^{\ell q}_{ij}&= & \frac{Y_\ell Y_q}{4}{g^\prime}^2 
 \Big[b_0 + b_1(s_\ell c_{\ell_i}) + b_1(s_q c_{q_j}) + 
b_2(s_\ell c_{\ell_i},-s_q c_{q_j})\Big] \nn\\
&& + \frac{Y_\ell Y^X_{q}}{4}\,
(g^2-{g^\prime}^2) b_2(s_\ell c_{\ell_i},- s_q c_{q_j} ) 
\label{bllqqcoeff}\\
 b^{L\tau Q}_{ij}&=&\frac{g^2}{4} 
(b_0 + b_1(c_{Q_j}) + b_1(c_{L_i}) + b_2(c_{L_i},-c_{Q_j}) )
\label{bLtauQcoeff}
\end{eqnarray}
with $\ell \in (L,E)$ and $q \in (Q,U,D)$. $s_f$ is $-1$ for a singlet 
fermion $f$ and $+1$ for a doublet, $Y_f$ is the hypercharge of fermion 
$f$, and $Y^X_q=T_R^3 - 4 \tan^2\Theta_W/(3(1-\tan^2\Theta_W))$
with $T_R^3=\{-1,-2,0\}$ for $q={Q,D,U}$. 
The second line in \eqref{bllqqcoeff} is only present in the custodially 
protected model. The dependence on the 5D mass parameters of the quarks 
shows that muon conversion depends not only on the model parameters of 
the lepton sector. However, ultimately we only need operators which are 
built of light quarks fields after EWSB, and of these only the quark-flavour 
diagonal part. Since both the up- and the down-quark sector masses are 
hierarchical, the RS Froggatt-Nielsen mechanism generates hierarchical 
flavour rotation matrices in the quark sector 
(see e.g.~\cite{Casagrande:2008hr}). Consequently, the $b_2(c_x,c_y)$ 
terms---the only terms that are simultaneously sensitive 
to 5D quark parameters and contribute to the flavour-non-diagonal lepton 
couplings---are suppressed for light quarks, and we neglect them. 
The only unsuppressed sources 
of LFV are then the terms $b_1(c_{L_i})$ or  $b_1(-c_{E_i})$.

\subsubsection{Higgs-Fermion operators}

\begin{figure}[t]
\begin{center}
\includegraphics[width=0.55\textwidth]{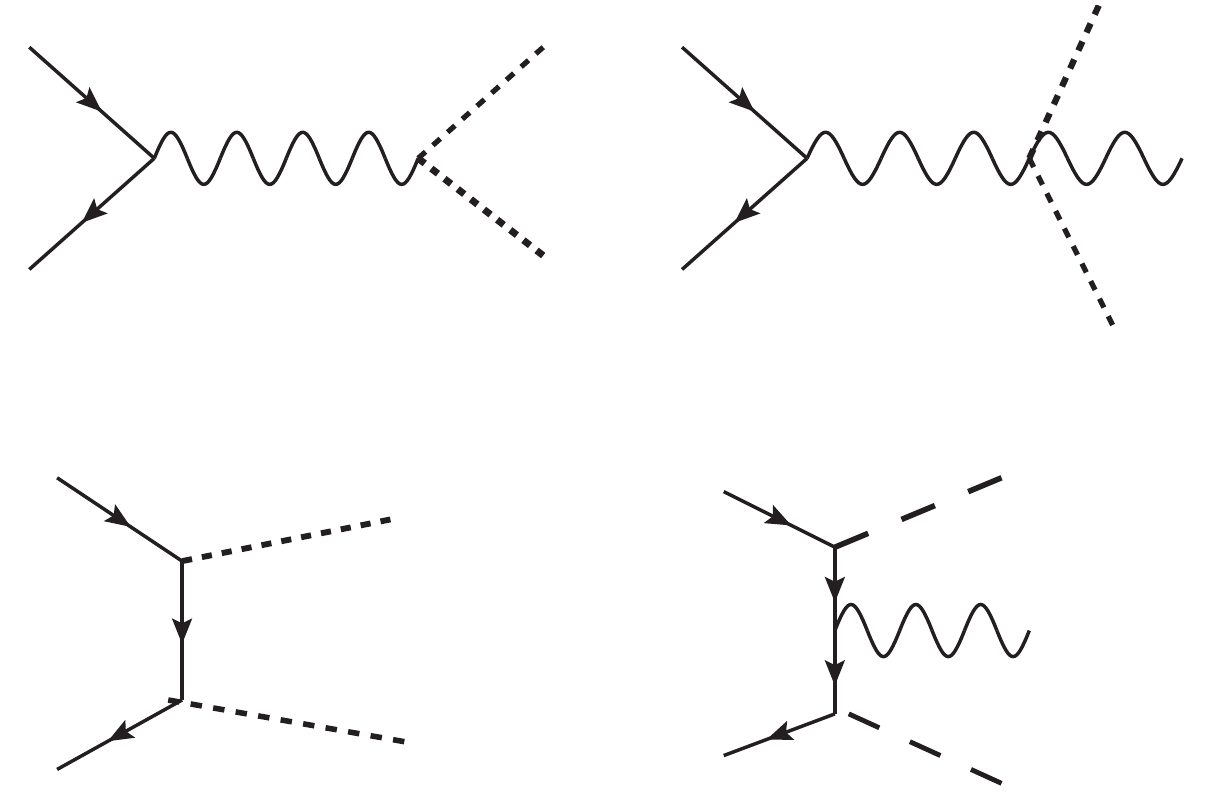}
\vskip0.2cm
\caption{\label{fig_GeneralPhi2psi2} 
Generic topologies that contribute to operators of the type
$\Phi^\dagger i {D}_\mu\Phi \,
(\bar \psi_i \gamma_\mu \psi_j)$. External fermion 
states can be either $E$ or $L$. Intermediate and external gauge bosons 
can be abelian or non-abelian, the external Higgses are 
indicated by dashed lines.}
\end{center}
\end{figure}

The tree-level matching coefficients of the Higgs-fermion operators 
$\Phi^\dagger i \!\overleftrightarrow{D}^{\!\mu} \Phi \,
(\bar \psi_i \gamma_\mu \psi_j)$ follow from the diagrams in 
Figure~\ref{fig_GeneralPhi2psi2}, where the ones with an external gauge 
field are related to those without by gauge invariance.

The diagrams in the first row of Figure \ref{fig_GeneralPhi2psi2} have 
already been computed in \cite{Beneke:2012ie,Moch:2014ofa} for the minimal 
and custodial RS model. Their contribution to the Wilson coefficients
$c^a_{ij}=c^{a}_{i}\delta_{ij},\; a=1,2,3$ is given by
\begin{eqnarray}
c^{1}_{i}&= &\frac{{g^\prime}^2 Y_E}{8} \,\Bigg(1-\frac{1}{\ln1/\epsilon}  
    - \left[ \frac{(1+2c_{E_i})(5+2c_{E_i})}{(3+2c_{E_i})^2} 
    -\frac{2(1+2c_{E_i})\ln1/\epsilon}{(3+2c_{E_i})} \right]
    \frac{\epsilon^{-2c_{E_i}-1}}{1-\epsilon^{-2c_{E_i}-1}}\Bigg) \nn \\
&&+ \,\frac{(g^2-{g^\prime}^2)Y_E }{4}  \left[ \frac{(1+2c_{E_i})
\ln1/\epsilon}{(3+2c_{E_i})}\right]
\frac{\epsilon^{-2c_{E_i}-1}}{1-\epsilon^{-2c_{E_i}-1}}\,,
\label{c1coeff}
\\
c^{2}_{i}&=&\frac{{g^\prime}^2 Y_L}{8} \,\Bigg(1-\frac{1}{\ln1/\epsilon}  
    - \left[ \frac{(1-2c_{L_i})(5-2c_{L_i})}{(3-2c_{L_i})^2} 
    -\frac{2(1-2c_{L_i})\ln1/\epsilon}{(3-2c_{L_i})} \right]
    \frac{\epsilon^{2c_{L_i}-1}}{1-\epsilon^{2c_{L_i}-1}}\Bigg) \nn \\
&& + \,\frac{(g^2-{g^\prime}^2)Y_L}{4}  
\left[ \frac{(1-2c_{L_i})\ln1/\epsilon}{ 3-2c_{L_i} }\right]
        \frac{\epsilon^{2c_{L_i}-1}}{1-\epsilon^{2c_{L_i}-1}}\,,
\label{c2coeff}\\
c^{3}_{i}&=&\frac{g^2}{8} \,\Bigg(1-\frac{1}{\ln1/\epsilon}  
    - \left[ \frac{(1-2c_{L_i})(5-2c_{L_i})}{(3-2c_{L_i})^2} 
    -\frac{2(1-2c_{L_i})\ln1/\epsilon }{ 3-2c_{L_i} } \right]
    \frac{\epsilon^{2c_{L_i}-1} }{1-\epsilon^{2c_{L_i}-1} }
\Bigg)\,.
\label{c3coeff}
\end{eqnarray}
As in the case of the four-fermion operators the minimal RS model 
results can  be obtained by removing the terms proportional 
$(g^2-{g^\prime}^2)$. The Wilson coefficients are independent of 
the Higgs localization provided the 
limit $\beta \to \infty$ is taken in the bulk Higgs case. 

The diagrams in the second row  of Figure~\ref{fig_GeneralPhi2psi2} also 
exist, but it turns out that they are numerically small compared to the 
previous contribution. Hence, we only give the explicit expression 
for the minimal RS model:
\begin{align}
\delta c^1_{ij}   = &\;  - \frac{T^8}{k^8} \,g_{E_i}(1/T) g_{E_j}(1/T) 
F(c_{{L}_k}) \,Y^\dagger_{ik} Y_{kj} \\
\delta c^2_{ij}   = &\; \delta c^3_{ij} =   
\frac{1}{2}\frac{T^8}{k^8} \,f_{L_i}(1/T) f_{L_j}(1/T) F(-c_{{E}_k})  
\,Y_{ik}Y^\dagger_{kj}  
\end{align}
with   
\begin{equation}
F(c)= -\frac{k^4}{T^5}\, 
\frac{(1+2c) + (3-2c)\epsilon^{2-4c} -(1+2 c)(3-2 c)\epsilon^{1-2c}}
{(1+2 c)(3-2 c)(1-\epsilon^{1-2c})^2 }\,.
\label{eq:F}
\end{equation}
A similar expression is found in the custodially protected model. 
The smallness of this 
contribution arises from the zero-mode profiles of the light external 
leptons. We ignore the Yukawa contributions $\delta c^a_{ij}$
in the subsequent analysis.

\subsubsection{Yukawa-type operators}

\begin{figure}
\begin{center}
 \includegraphics[width=0.40\textwidth]{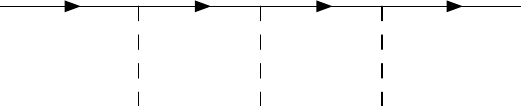}
\vskip0.2cm
\caption{\label{fig_GeneralHiggs} Diagram 
topology that gives the dominant contribution to the operator
$(\Phi^\dagger\Phi)\bar L_i \Phi  E_j$.}  
\end{center}
\end{figure}

The dominant contribution to the Wilson coefficient of the dimension-six 
Yukawa-like operators $(\Phi^\dagger\Phi)\bar L_i \Phi  E_j$ is generated 
by diagrams of the type shown in Figure~\ref{fig_GeneralHiggs}.
In the minimal RS model there is only one diagram as the two intermediate 
fermions must be a doublet and a singlet lepton. In the custodially protected 
model both triplet fermions, $T_3$ and $T_4$, can substitute the singlet.
The contribution to the Wilson coefficient is then given by
\begin{align}
\label{eq:YukawaLikeWC}
h_{ij} &=  \frac{N_{cs}}{3}\times \frac{T^3}{k^4}f^{(0)}_{L_{i}}(1/T)
[YY^\dagger Y]_{ij} g^{(0)}_{E_{j}}(1/T)       
\end{align}
where  $N_{cs}$ equals one in the minimal and two in the 
custodially protected model. 

For completeness we remark that the diagrams in the second line 
of figure \ref{fig_GeneralPhi2psi2} also contribute to the Wilson coefficient 
of $(\Phi^\dagger\Phi)\bar L_i \Phi  E_j$ through derivative terms that 
can be eliminated by the fermion equation of motion, such as 
$\slashed{D} L_i = y_{ij}\Phi E_j$. In the minimal model we find
\begin{align}
\delta h_{ij} = -&  \frac{1}{2}\,\frac{T^8}{k^8} g_{E_l}(1/T) g_{E_j}(1/T) 
F(c_{{L}_k}) y_{il}Y^\dagger_{lk} Y_{kj}  \nn\\  
                -&  \frac{1}{2}\,\frac{T^8}{k^8} f_{L_i}(1/T) f_{L_l}(1/T) 
F(-c_{{E}_k})  Y_{ik} Y^\dagger_{kl} y_{lj} \,.
\end{align}
Due to the appearance of the small SM lepton Yukawa matrix $y$ this 
contribution is tiny. This also holds true in the custodially protected 
model, and hence in the numerical analysis we neglect this term. 
However, in studies of flavour violation involving third generation 
quarks (notably top quarks) the contribution can be sizeable and must 
be included.

\subsection{Loop-induced dipole operators}
\label{sec:genDipole}

The dipole operators are generated by genuine 5D one-loop penguin diagrams. 
We distinguish between two classes of diagrams---those with internal 
gauge-boson exchange proportional to one Yukawa coupling $Y$ and those with 
Higgs exchange, which involve three Yukawa couplings. A diagram such as 
shown in Figure~\ref{Higgssamplediagram} below counts as gauge-boson 
exchange, since it involves only a single $Y$. As we only need the 
electromagnetic dipole operator for the LFV processes under consideration, 
we reduce the number diagrams needed for the one-loop coefficient 
$a_{ij}$ by imposing that the external gauge boson is a photon.
In addition, we set the Higgs doublet in the operators 
$\bar L_i \Phi \sigma^{\mu \nu} E_j  B_{\mu\nu}$, $ \bar L_i \tau^A \Phi 
\sigma^{\mu\nu} E_j   W^A_{\mu\nu}$ to its vacuum expectation value. 
The complete set of non-vanishing diagrams can be found in 
\cite{Beneke:2012ie} for the minimal RS model and in \cite{Moch:2014ofa} 
for the custodial RS model.

\subsubsection{Internal gauge boson exchange}

We start the discussion with the gauge-boson contribution. 
There are three different regions of 4D loop momentum that have 
to be distinguished. In the first region all propagators 
are zero--mode propagators. This is only possible for the diagrams that are 
already present in the minimal RS model as the additional fields in the 
custodially protected model do not have massless modes \cite{Agashe:2004cp}.
This region corresponds to a SM contribution and must be removed. 
As explained in \cite{Beneke:2012ie} this can be done by subtracting 
the zero-mode from a single 5D gauge-boson propagator.

When the 4D loop momentum is much smaller than the KK scale $T$ and the loop 
contains at least one KK mode propagator, this propagator can be contracted 
to a point. The resulting diagram corresponds to a four-dimensional 
diagram with an insertion of a higher-dimensional tree-level operator. 
The leading dimension-six contributions are in one-to-one correspondence 
with the one-loop matrix elements of the operators considered in the previous 
subsection. 

In the last region the 4D loop momentum is 
of the order of the KK scale $T$. Only this region  
contributes to the dipole matching coefficient $a^A_{ij}$. We extract 
this contribution by expanding the propagators in the external momenta. 
This automatically removes the low-momentum region 
and prevents double-counting of the insertions of tree 
operators \cite{Beneke:2012ie}. As the amplitude of the dipole operator is 
proportional to $\sigma^{\mu \nu}q_\nu$ we only have to expand to 
linear order in the external momenta. The remaining calculation 
requires the numerical evaluation of integrals over the modulus of the 
4D loop momentum and several bulk coordinate integrals in the interval 
$[1/k,1/T]$. The complete numerical calculation was performed for the 
minimal RS model in \cite{Beneke:2012ie}  and for the custodially protected 
model in \cite{Moch:2014ofa}. Here we use the results of \cite{Moch:2014ofa} 
as the routines used there for the numerical evaluation have been 
considerably improved compared to~\cite{Beneke:2012ie}. 
We perform the calculation in 5D $R_\xi$ gauge and 
use the spurious dependence of the numerical result on the gauge-fixing 
parameter as an additional estimate of the numerical uncertainty. From this 
we conclude that the numerical accuracy is below $0.1\%$, which 
translates to an error of about 5\% for the flavour-non-diagonal LFV terms 
after rotation to the standard mass basis. 
This is sufficient for our purposes. 

For completeness we remark that some diagrams induce a scheme-dependence 
of the dipole coefficient $a^A_{ij}$ via finite but IR-sensitive 
$\varepsilon  \times 1/\varepsilon$ terms. This scheme dependence 
is cancelled by the scheme dependence of the 4D one-loop 
penguin diagrams with insertions of $\beta^{LE}_{ij}$, $\gamma^{a}_{ij}$ 
\cite{Beneke:2012ie}.

\begin{figure}
 \centering
 \includegraphics[width=0.3\textwidth]{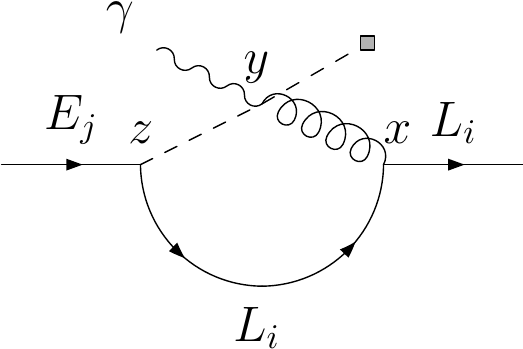}
 \caption{Example diagram with internal KK Higgs modes.
\label{Higgssamplediagram}}
 \end{figure}

In principle there could be an additional momentum region. 
The width of the Higgs localization introduces the new scales $T/\delta$ and
$\beta T$ for a delta-localized and a bulk Higgs, respectively.
In \cite{Beneke:2014sta} it was shown that the effect of 
this additional momentum scale does not contribute to the 
gauge-boson exchange diagrams for $\delta \to 0$
(or, equivalently, $\beta \to \infty$) when only the Higgs zero-mode 
is considered. For the bulk Higgs case it still needs to be shown that
the contribution of the infinite tower of Higgs KK modes also vanishes 
for $\beta\to \infty$.
To this end let us examine the diagram shown in 
Figure~\ref{Higgssamplediagram}. Up to a constant prefactor it is given 
by
\begin{align}
\int\frac{d^dl\;l^2}{(2\pi)^d} \int_{\frac{1}{k}}^{\frac{1}{T}} \!\!
\frac{dz dx dy}{k^{13} z^5 x^5 y^3} \,Y^\beta_{ij} 
f_{L_i}^{(0)}(z) f_{E_j}^{(0)}(x) \Phi^{(0)}(y) \Delta_\Phi^{\rm ZMS}(l,z,y)
F^+_{L_i}(l,x,z)
\frac{\partial}{\partial l^2}\Delta_{B}^{\rm ZMS}(l,y,x) \,.
\end{align}  
For the explicit expressions for the zero-mode subtracted gauge boson 
propagator $\Delta_{B}^{\rm ZMS}$ and  the fermion propagator $F^+_{L_i}$ 
we refer to \cite{Beneke:2012ie}.
The Higgs propagator $\Delta_\Phi(l,z,y)$, its zero-mode subtracted version  
$\Delta^{\rm ZMS}_\Phi(l,z,y)$, the Higgs zero-mode $\Phi^{(0)}(y)$, 
and the Yukawa coupling $Y^\beta$ are discussed in Appendix~\ref{sec:Higgs}.

We now show that the KK Higgs contribution is ${\cal O}(1/\beta)$ 
and therefore can be neglected for large $\beta$.
The Yukawa matrix $Y^\beta$ and zero-mode profile $\Phi^{(0)}(y)$ both 
scale as $\sqrt{\beta}$. Since the zero-mode profile is localized 
near the IR brane, the associated 5D coordinate integral over $y$ is 
effectively restricted to the interval $[(1-1/\beta) 1/T,1/T]$ 
of length $1/(\beta T)$. Hence the $y$ integration introduces a factor of 
$1/\beta$. The integration over $y$ then compensates the factor
$\beta$ from the product $Y^\beta \Phi^{(0)}(y)$ independent of
the magnitude of the 4D loop momentum $l$.
For  $l\ll T$ and $l\sim T$, the Higgs propagator scales 
as $1/\beta$ and, after a change of integration variables from
$\{x,y,z\}$ to $\{y,y-z,z-x\}$, one finds that the integrand is 
dominated by the region where the distance $z-y$ is 
of the order $1/\beta$ (see also 
Appendix~\ref{sec:example}). Putting all factors together, we conclude 
that the integrand scales as  $1/\beta^2$ for small loop momenta, and hence 
the integral over these momentum regions also 
vanishes for $\beta \to \infty$. 
For loop-momenta $l$ of order $\beta T$, we can expand the fermion 
and boson propagator for large momenta, in which case they become simple 
and their dependence on the loop momentum can 
readily be extracted. The Higgs propagator is more complicated, but it 
can only depend on the scale $\beta T$ and therefore scales 
as $1/(\beta T)$. 
We find that the product of all three propagators together with the 
derivative $\partial/\partial l^2$, 
which counts as $1/(\beta T)^2$, 
compensates the factor $l^5\sim (\beta T)^5$ from 
$d^4 l \,l^2\sim dl l^5$.
We are left with the two integrals over $y-z$ and $z-x$. For $l \sim \beta T$
the integrand is exponentially suppressed for $|z-x|>1/l$ and $|y-z|>1/l$, 
and hence each of the coordinate difference integration regions is 
effectively restricted to size $1/(\beta T)$. We then  
find that the total scaling of the integrand in this momentum region 
is $\propto 1/\beta^2$.
The integral over $dl$ can only compensate one inverse power of $\beta$
and we conclude that the integral over the region $l\sim \beta T$
vanishes as well for $\beta \to \infty$. 
For very large loop momentum $l\gg \beta T$  we can expand 
all propagators. Now all bulk coordinate differences are 
constrained to be within about $1/l$ ($l$ is now the largest scale) 
and the 5D Higgs propagator scales as $1/l$. This ensures the 
convergence of the integral as the integrand vanishes as $1/l^2$ for 
$l\to \infty$. The parameter $\beta$ only enters through the integral 
over $y$, which is cancelled by Higgs profile and Yukawa coupling, hence 
the integrand is independent of $\beta$.
This universal behaviour allows for a straightforward determination 
of the contribution of the region $l\gg \beta T$:
\begin{equation}
 \int_{\beta T}^\infty \frac{dl}{l^2} = \frac{1}{\beta T}\;.
\end{equation}
Hence the integral over this region vanishes in the large $\beta$ limit. 
Since this holds in all regions, we conclude that KK Higgs contribution 
vanishes as $1/\beta$. 

\begin{figure}
\centering
\includegraphics[width=0.49\textwidth]{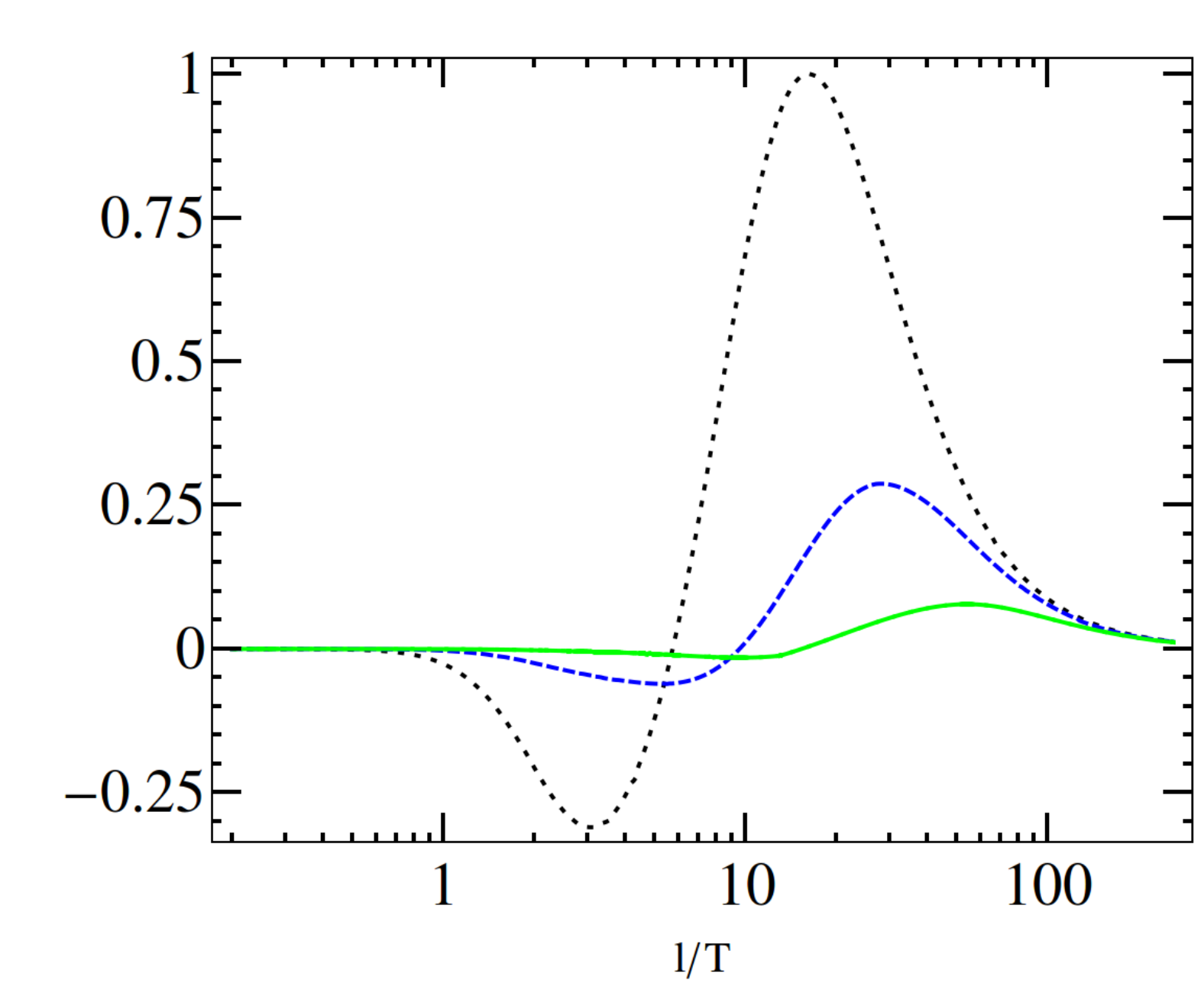}
\caption{Integrand as a function of the loop-momentum $l$ for 
$\beta=10$ (black, dotted), $\beta=20$ (blue, dashed) and $\beta=40$ 
(green, solid). For clarity all curves have been rescaled relative to 
the maximum of the integrand for $\beta = 10$. For loop momenta in 
excess of $\beta T$ the integrands show a universal $1/l^2$ behaviour. 
\label{fig:W8IntegrandKKHiggs}}
\end{figure}

This can be verified numerically as shown 
in Figure~\ref{fig:W8IntegrandKKHiggs}. The three curves correspond to 
different values of $\beta$ (10, 20 and 40, respectively). For better 
visibility all curves are normalized to the maximum of the $\beta =10$ 
curve.  The maximum of the integrand is close to $l \sim \beta T$
and exemplifies the $1/\beta^2$ scaling of the integrand in that region.
For large modulus of the (euclidean) loop momentum the three curves 
lie on top of each other consistent with the $\beta$ independent 
asymptotic expression. Consequently, the
integral  over $l$ as well as the contribution to the dipole 
operator coefficient vanishes for $\beta \to \infty$.

\begin{figure}
\begin{center}
\includegraphics[width=0.6\textwidth]{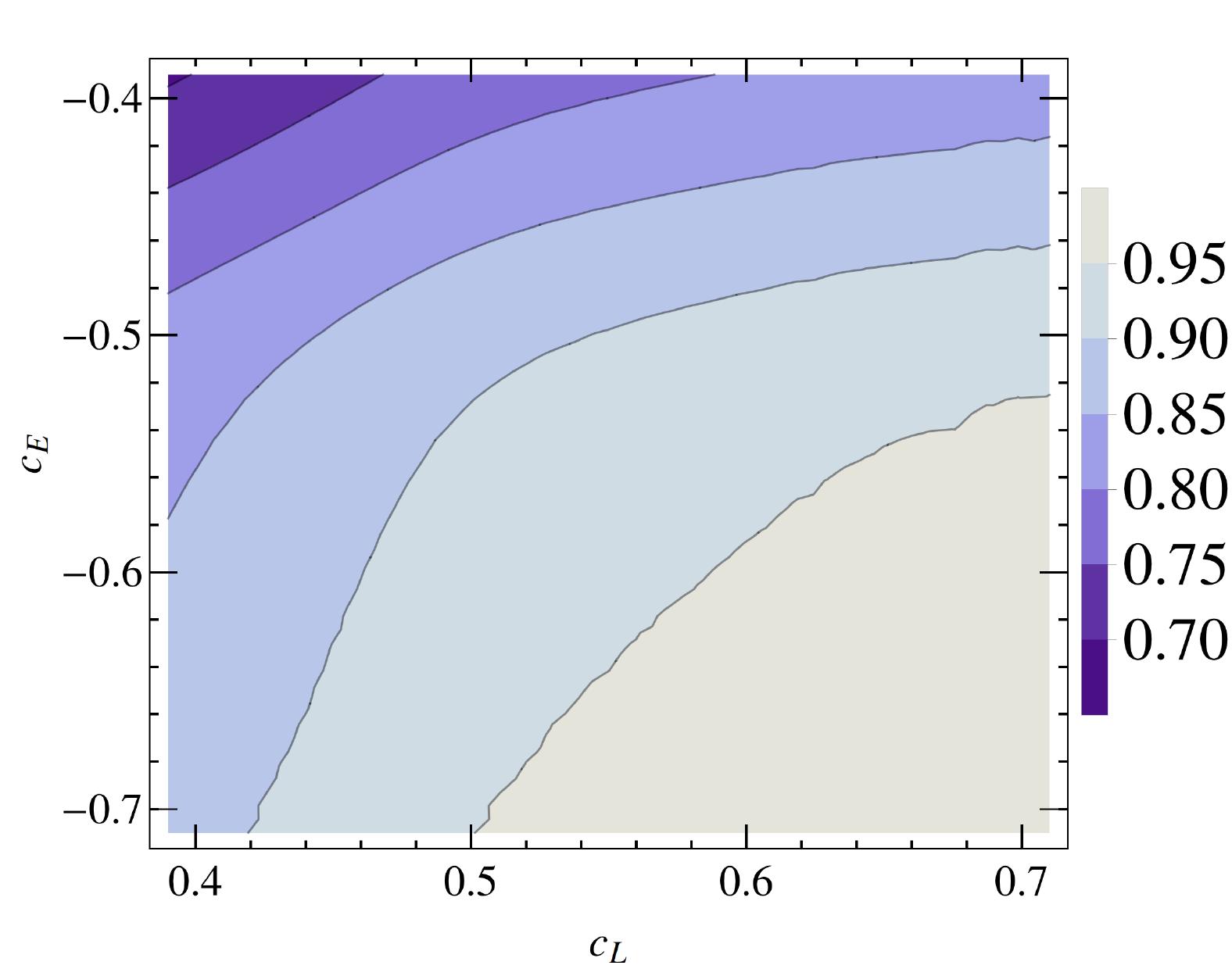}
\end{center}
\vskip-0.3cm
\caption{\label{fig:MassGaugeDep}
Contour plot of $\mathcal{A}_{ij}$  in the custodially protected RS model 
normalized to its value for $c_{L}=|c_{E}|=0.7$ as a function of the 5D 
mass parameters $c_{L}$ and $c_{E}$ for $T=1\,\rm TeV$, 
$k=2.44\cdot 10^{18}\,$GeV.}
\end{figure}

To understand the numerical size of the Wilson coefficient $a^{g}_{ij}$ 
generated by the gauge-boson contribution it is convenient to factorize 
all terms that combine to the 4D Yukawa matrix before rotation to the 
mass basis:
\begin{align}
\label{gaugedipole}
  a^{g}_{ij} = Y_{ij} \frac{T^3}{k^4} f^{(0)}_{L_i}(1/T) g^{(0)}_{E_j}(1/T) 
  \mathcal{A}_{ij} = y_{ij}\mathcal{A}_{ij}\;.
\end{align}
The remaining short-distance function $\mathcal{A}_{ij}$ depends only on 
the 5D bulk masses of the external fermion fields with flavours $i$, $j$ 
and the RS scales $k$ and $T$. $\mathcal{A}_{ij}$ can be interpreted as a 
measure of the misalignment between the mass matrix of 
the lepton sector and the dipole coefficient $ a^{g}_{ij}$ before  
rotation to the mass basis. If the $\mathcal{A}_{ij}$ were all equal, 
no LFV would be generated by the gauge-boson 
exchange diagrams. Figure \ref{fig:MassGaugeDep} shows the result of 
the numerical computation of $\mathcal{A}_{ij}$ for the custodially 
protected model at the KK scale $T=1\, \rm TeV$. There is a small asymmetry 
in the dependence of $\mathcal{A}_{ij}$ on the bulk mass parameters of the 
external lepton fields, which arises from 5D diagrams with 
non-abelian gauge bosons as the $W$ bosons do not couple equally to 
singlet and doublet fields. 
To reproduce the 4D lepton mass matrix the bulk mass parameter $c_L$ 
of the doublet muon (electron) has to be around
$0.57$ ($0.66$) and the masses of the corresponding singlets 
around $-0.57$ ($-0.66$), if
the SM mass hierarchy is carried by both singlet and doublets.
As illustrated in the figure the variation of $\mathcal{A}_{ij}$ in
this region is around $\pm (2\text{--}3)\,\%$. In an extreme case where 
e.g.~all singlets are ``delocalized'' with bulk mass parameter 
$c_E=-0.5$, the bulk mass of the doublet muon (electron) has to 
be around $0.64$ ($0.8$), and the variation is less pronounced.
For the minimal RS model the dependence of $\mathcal{A}_{ij}$ on the 
bulk mass parameters is slightly smaller in the region of mass parameters 
relevant to muons and electrons than in the custodially protected 
model~\cite{Beneke:2012ie,Beneke:2014sta}.  It follows that 
the gauge-boson exchange contribution $\alpha^{g}_{ij}$ to the 
dipole coefficient has smaller off-diagonal elements by a factor 30 to 50 
compared to the flavour-conserving diagonal entries\footnote{This factor is 
responsible for the larger numerical error for the flavour-violating 
transitions mentioned above.}---the 
RS model has a built-in protection from large gauge-boson induced 
LFV transitions. It is interesting to 
note that the variation of $\mathcal{A}_{ij}$  increases for decreasing 
absolute value of both bulk masses. Since typically the absolute
values of the 5D bulk masses decrease with decreasing 
magnitude of the 5D Yukawa couplings, a smaller absolute value of the 5D 
Yukawa couplings leads to more pronounced LFV transitions from internal 
gauge-boson exchange.

\subsubsection{Internal Higgs exchange \label{sec:Higgsexch}}
 
Unlike gauge-boson exchange the internal Higgs-exchange contribution depends 
strongly on the Higgs localization. 

\subsubsection*{\boldmath Delta-function localized Higgs}

We first consider the delta-function 
localized Higgs (\ref{eq:HiggsRegulator}).   
For the {\it minimal} RS model the result for a delta-localized bulk 
Higgs without KK modes was determined in \cite{Beneke:2012ie} to be 
\begin{align}\label{eq:Higgsresults}
a^H_{ij}=& \phantom{+} \frac{ Q_\mu e}{192\pi^2} \frac{T^3}{k^4}
\cdot \frac{T^8}{2 k^8} \left(  F_L -    F_E \right)
\nn \\
& +  \frac{Q_\mu e}{192 \pi^2}  \frac{T^3}{k^4} \,  
f^{(0)}_{L_i}(1/T)
[ YY^{\dagger} Y]_{ij} \, g^{(0)}_{E_j}(1/T) ,
\end{align}
where 
\begin{align}
F_E =&\hphantom{+} f^{(0)}_{L_i}(1/T) Y_{ik} F(-c_{{E}_k})
Y^{\dagger}_{kh}f^{(0)}_{L_h}(1/T)^2 Y_{hj} g^{(0)}_{E_j}(1/T), \nn \\
F_L = &      \hphantom{+} f^{(0)}_{L_i}(1/T) Y_{ik} g^{(0)}_{E_k}(1/T)^2
Y^{\dagger}_{kh}F(c_{{L}_h}) Y_{hj} g^{(0)}_{E_j}(1/T)\, .
 \end{align}
Here the three times repeated indices $k$ and $h$ are summed over only once and the
function $F(c)$ is defined in \eqref{eq:F}.
The expression in the first line arises via ``off-shell contributions''
(see~\cite{Beneke:2012ie,Beneke:2014sta}) and is suppressed by fermion 
zero-mode factors.
The last, numerically dominant line (\ref{eq:Higgsresults}) 
is not present for the exactly brane-localized setup. As a consequence  
the Higgs-exchange contribution, which involves three Yukawa matrices, 
is suppressed in the minimal model with an exactly brane-localized Higgs 
boson.

\begin{figure}
\centering
\includegraphics[width=0.7\textwidth]{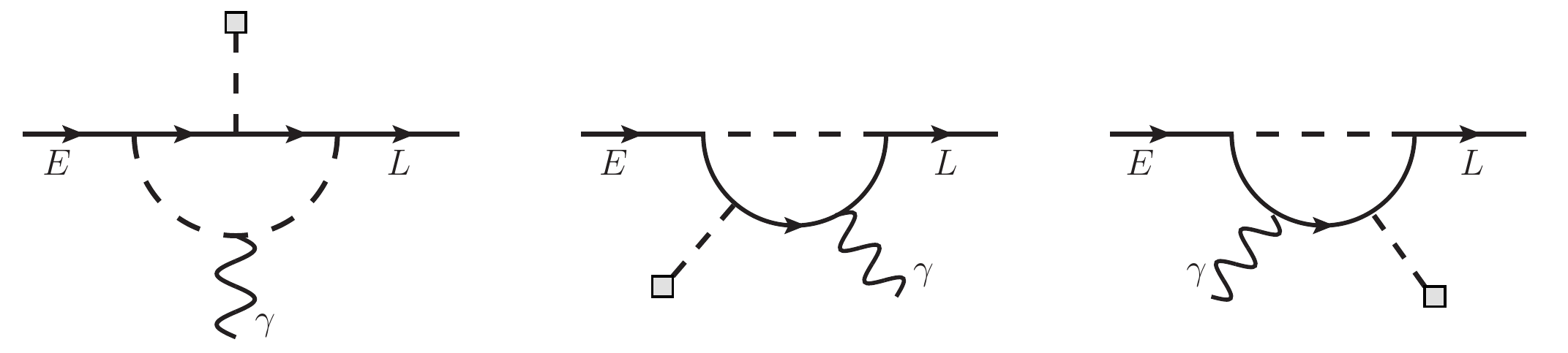}
\caption{Additional Higgs-exchange diagrams in the custodially protected 
model.}
\label{fig_AdditionalDiagrams}
\end{figure}

The corresponding result for the custodially protected model was partially 
determined in \cite{Moch:2014ofa}. However, due to a misplaced SU(2) 
index the two diagram topologies (absent in the minimal 
model) shown in Figure~\ref{fig_AdditionalDiagrams} were missed.\footnote{
Note that for external quarks, i.e. for the calculation 
of the electromagnetic dipole coefficient of quarks, the diagrams
are present even in the minimal RS model.} These diagrams have a 
non-trivial dependence on the Higgs localization. In the following 
we compute these missing diagrams. Note that the exchanged Higgs 
refers to the zero mode, since for now we have adopted the theta-function 
regularized delta-function Higgs profile. 

To illustrate the computation we consider explicitly the sum 
of the two right-most diagrams in Figure~\ref{fig_AdditionalDiagrams}
where the photon couples to the fermion line. Up to prefactors 
that depend on the U$(1)$ and SU$(2)$ charges of the fermions,  
and after some simplifications the contribution to the 
dipole operator structure is given by the integral 
\begin{eqnarray}
\label{eq_extradiagram}
\mathcal{I}_{new}&= & \frac{T^3}{\delta^3}\frac{T^9}{k^{12}}
\int_0^{\Lambda^2} \!\!\frac{d l^2}{16\pi^2} 
\int^{1/T}_{(1-\delta)/T}\!\!\!\!dx\,dy\,dz\,
f^{(0)}_{L_i}(x) Y_{ik}Y^\dagger_{kh} Y_{hj}g^{(0)}_{E_j}(z)
\nn\\
&&\hspace*{-1.4cm} \times \,\frac12 \,\bigg[\,\partial_{l^2}\Big( l^4 
\partial_{l^2} \!\left(F_{X_k}^-(l,x,y)  F_{Y_h}^+(l,y,z)\right)\!\Big)  
+ \,l^2 \partial_{l^2} \partial_{l^2} \!
\left(d^-F_{X_k}^+(l,x,y)  d^-F_{Y_h}^+(l,y,z)\right) \bigg]\,,
\qquad
\end{eqnarray}
where we used the notation of \cite{Beneke:2012ie} for the different 
fermion propagator functions and applied a momentum cut-off $\Lambda$ to 
the loop integral. The fermion $X$ can be part of the SU(2)$_L$ singlets 
$\xi_2, T_3, T_4$  while $Y$ is part of the custodial bi-doublet $\xi_1$. 
Both propagator functions $d^-F^+$ in the second line obey Dirichlet boundary 
conditions at $y=1/T$, while the propagator functions $F^+$ and $F^-$ have 
Neumann boundary conditions on the IR brane. That is, the second term 
in square brackets contains wrong-chirality Higgs 
couplings \cite{Agashe:2006iy}, whereas the first features 
right-chirality Higgs couplings 
(correct-chirality in the language of \cite{Agashe:2014jca}).

Consider first the first term the square brackets 
in \eqref{eq_extradiagram}. The $l^2$ integral 
can be carried out trivially, since it is a total derivative. Only the 
upper limit gives a non-zero contribution. Then we use that for 
$l^2\sim \Lambda^2 \gg T^2$ and $x,y$ close to $1/T$, the fermion propagators 
can be simplified,
\begin{align}
 F^-_X(l,x,y)\approx  F^+_Y(l,x,y) \propto \frac{ i}{l} 
\cosh(l(1/T-x))e^{l(y-1/T)}\Theta(x-y) +\{x \leftrightarrow y \}\,,
\end{align}
which allows all coordinate integrals to be evaluated analytically. 
The result is quite involved and depends crucially on the expression 
$e^{-(\Lambda \delta)/T}$, such that it vanishes for $\Lambda \to 
\infty$ at fixed, finite $\delta$, but is equal 
to $1$ for $\delta\to 0$ with fixed $\Lambda$.\footnote{In dimensional 
regularisation the calculation is more tedious. The second line of 
\eqref{eq_extradiagram} can be written as a total derivative plus an 
evanescent term $\propto (d-4)$.  The non-commuting limits manifest 
themselves in form of factors of $\delta^\epsilon$ or in combinations 
of incomplete Gamma functions like $\Gamma(-1+\epsilon, \delta)$.
The final result coincides with the one obtained above with the cut-off 
regulator.}
Hence, for a narrow bulk Higgs (small but finite $\delta$, 
$\Lambda\to\infty$ first) the first term in square brackets  
in \eqref{eq_extradiagram}  does not contribute to 
the dipole Wilson coefficient. This reproduces known 
results~\cite{Agashe:2014jca, Delaunay:2012cz} for the right-chirality Higgs
couplings. On the other hand, for the exactly brane-localized Higgs,  
the dipole coefficient receives a finite unsuppressed contribution. 
This effect is only present in the custodially protected model, since in 
the minimal model the diagrams do not exist.
The contribution of the second term in the square brackets 
in \eqref{eq_extradiagram} 
can be computed following the approach of \cite{Beneke:2014sta}. 
Here situation is exactly opposite: the narrow bulk Higgs leads to a finite 
contribution, whereas the integrals vanish when the Higgs width $\delta$ 
is taken to zero before the regulator of the loop integral is removed, 
i.e.~for the exactly brane-localized Higgs. The total result is 
\begin{align}
\mathcal{I}_{new}= \left\lbrace 
\begin{array}{rl}
\displaystyle
\phantom{-} \frac{1}{32 \pi^2 T^2}\cdot 
\frac{T^3}{k^4} f^{(0)}_{L_i}(1/T) [YY^\dagger Y]_{ij}g^{(0)}_{E_j}(1/T)
\qquad
& \text{``exactly localized''}
\\[0.5cm]
\displaystyle
-\frac{1}{96 \pi^2 T^2}\cdot \frac{T^3}{k^4} f^{(0)}_{L_i}(1/T) 
[YY^\dagger Y]_{ij}g^{(0)}_{E_j}(1/T)
\qquad
&
\text{``narrow bulk''}
\end{array}
\right.
\end{align}
The contribution of the left diagram in Figure~\ref{fig_AdditionalDiagrams} 
can be obtained analogously.

Putting all diagrams together and including the SU(2) and hypercharge factors
we find in the {\it custodially protected} RS model for the narrow bulk Higgs 
\begin{align}
\label{eq:HiggsresultsCustodialbulk}
a^H_{ij}=& \frac{Q_\mu e}{192 \pi^2}  \frac{T^3}{k^4} \cdot 2\cdot
f^{(0)}_{L_i}(1/T) [YY^{\dagger} Y]_{ij} g^{(0)}_{E_j}(1/T)\,,    
\end{align}
and for the exactly brane-localized Higgs
\begin{align}
\label{eq:HiggsresultsCustodialbrane}
a^H_{ij}= & \frac{Q_\mu e}{192 \pi^2}  \frac{T^3}{k^4}\cdot (-3) \cdot
f^{(0)}_{L_i}(1/T) [YY^{\dagger} Y  + Y_u Y_u^{\dagger} Y]_{ij} 
g^{(0)}_{E_j}(1/T)\,.    
\end{align}
In both equations we neglected terms similar to the first line 
in~\eqref{eq:Higgsresults}, which are suppressed by lepton masses 
and/or lepton zero-mode profiles. These terms are always subleading in the 
custodially protected model irrespective of the Higgs localization, and 
we ignore them in the further analysis. Note that $a^H_{ij}$ has opposite 
signs for the narrow bulk and exactly brane-localized Higgs.

\subsubsection*{\boldmath Bulk Higgs with a $\beta$ profile}

\begin{figure}[t]
\centering
\includegraphics[width=0.3\textwidth]{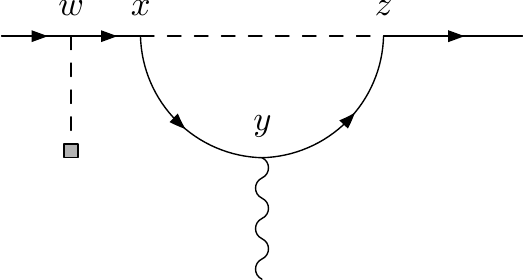}
\vskip0.2cm
\caption{Example diagram with an internal Higgs exchange and 
three Yukawa couplings. 
The internal Higgs can represent the zero-mode or the tower of 
KK modes. 
\label{fig_HiggsDiagramEXP}}
\end{figure}

Next we consider the case of a bulk Higgs with the 
$\beta$ profile \eqref{eq:betaprofile}. The dominant contributions were 
studied in some detail in \cite{Agashe:2014jca} and numerical estimates 
were obtained by summing a large number of KK modes.\footnote{
At this point it is worth recalling that the dipole transitions 
are not sensitive to the UV cut-off $\Lambda_{\rm UV}$ of the RS model. 
That is, whether the summation of KK modes is extended to infinity or 
truncated when the KK mass reaches $\Lambda_{\rm UV}$, results in 
differences suppressed by inverse powers of $\Lambda_{\rm UV}$, which can 
be absorbed into the coefficient functions of higher-dimensional 
operators. Equivalently, in the 5D treatment the integrations can be done 
over the entire loop momentum and bulk space without imposing an 
explicit cut-off. This 
is completely analogous to the standard practice of performing the loop 
integral over all four-momenta for UV-finite observables in the Standard 
Model, even though the Standard Model is most likely only valid up to 
a certain scale. When this logic is applied to summing KK modes of 
a narrow bulk Higgs in the RS model, since the non-decoupling effects is 
related to the new scale $\beta T$, it is, of course,  
implied that $T\ll \beta T\ll \Lambda_{\rm UV}$. Otherwise the 
issue of Higgs localization could not be separated from the UV completion 
of the RS model.} 
Using 5D propagators 
the effect of the Higgs zero mode can be computed analytically for  
large $\beta$. To see this let us focus on the simplest diagram, shown in 
Figure~\ref{fig_HiggsDiagramEXP}. Other contributions can be obtained 
analogously, but may require appropriate 
expansions of the fermion propagators for a fully analytic result.
For light external fermions the dominant contribution can be 
written as 
\begin{eqnarray}
\label{eq:HiggsDiagram}
\mathcal{I}_H &=& Q_\mu e \,Y^\beta_{ih} \left[Y^{\beta}\right]^\dagger_{hk} 
Y^\beta_{kj} \int \frac{d^4l}{(2\pi)^4}
\int_{\frac{1}{k}}^{\frac{1}{T}}\frac{ dz\;dy\;dx\;dw}{k^{19}x^5 y^4 z^5 w^5} 
\,f_{L_i}^{(0)}(z) g_{E_j}^{(0)}(w) d^{-}F^+_{L_k}(p,x,w) 
\nn \\
&& \times \,\Delta_\Phi(l,x,z) \Phi^{(0)}(w) \left[d^-F^+_{E_h}(p'-l,z,y) 
F^+_{E_h}(p-l,y,x)\gamma^\mu (\slashed{p}-\slashed{l}) \right. 
\nn \\[0.2cm]
&& \left.+\, F^-_{E_h}(p'-l,z,y) d^-F^+_{E_h}(p-l,y,x)
(\slashed{p}'-\slashed{l})\gamma^\mu \right]
\end{eqnarray} 
where  we chose $p, p'$ for the incoming and outgoing fermion momentum, 
respectively. The integral over the $w$ coordinate can be taken right 
away as we can set $p$ to zero in the external fermion propagator:
\begin{eqnarray}
&&\mathcal{E}(x,\beta,c_{L},{c_{E}})\equiv
\int_{\frac{1}{k}}^{\frac{1}{T}}\frac{dw}{k^{5} w^5} \,
g_{E}^{(0)}(w)\, d^{-}F^+_{L}(p=0,x,w) \,\Phi^{(0)}(w) 
\\
&& = i\,\sqrt{\frac {1 + 2 c_ {E}} {1 - \epsilon^{1 + 2 c_ {E}}}}
\sqrt{\frac {2(1+\beta)}{1 - \epsilon^{2 +2\beta}}}
\frac {1}{2 - c_ {L} + c_ {E} + \beta}
\frac {(T x)^{2 + c_L}\epsilon^{-5/2}} {1- \epsilon^{2 c_ {L}-1}} \nn \\
&&\phantom{=} 
\times\Big[(Tx)^{2 - c_L + c_E + \beta} 
(1 - \epsilon^{2 c_L-1}) +
 (Tx)^{1-2 c_L} (\epsilon^{2 c_L-1} - \epsilon^{1 + c_L + c_E + \beta}) 
- (1 - \epsilon^{1 + c_L +  c_E + \beta}) \Big].
\nn \end{eqnarray}
After expanding the remaining integrand for small $p,p'$ 
we perform the integral over the photon vertex bulk position $y$ using 
the completeness and orthogonality relations. We then find 
\begin{eqnarray}
\label{eq:HiggsDiagram2}
\mathcal{I}_H &=& Q_\mu e Y^\beta_{ih} \left[Y^{\beta}\right]^\dagger_{hk} 
Y^\beta_{kj}
\int \frac{d^4l}{(2\pi)^4} \int_{\frac{1}{k}}^{\frac{1}{T}}
\frac{ dz\;dx}{k^{10}x^5 z^5} \,
f_{L_i}^{(0)}(z) \mathcal{E}(x,\beta,c_{L_k},{c_{E_j}})\Delta_\Phi(l,x,z) 
\nn \\
&&\times\,\left[\frac{i}{2} l^2 \partial_{l^2}^2 d^-F^+_{E_h}(l,z,x) 
\right] (p^\mu+{p'}^\mu)\,.
\end{eqnarray}
This leaves us with only three integrals over $x$, $z$ and the loop 
momentum.
 
Let us first consider the Higgs zero-mode contribution 
by substituting $\Delta_\Phi(l,x,z) \to i/l^2\times 
\Phi^{(0)}(x)\Phi^{(0)}(z)$. Since $\beta$ is large but finite until 
all integrals have been carried out and all regulators removed, we can 
perform the momentum integral directly in $d=4$ dimensions. To this 
end, we switch temporarily to the mode picture for the fermion 
propagator, evaluate the integral
\begin{align}
\int \frac{d^4l}{(2\pi)^4} \frac{1}{(l^2-m_n^2)^3}=-\frac{i}{2(4\pi)^2}
\frac{1}{m_n^2}\,,
\end{align}
and resum the mode expansion back into 5D propagators, which results in 
\begin{eqnarray}
\label{eq:HiggsExchangegAnalytic}
\mathcal{I}_H^{(0)} &=& -i Q_\mu e \,Y^\beta_{ih} 
\left[Y^{\beta}\right]^\dagger_{hk} 
Y^\beta_{kj}\int_{\frac{1}{k}}^{\frac{1}{T}}\frac{ dz\;dx}{k^{10}x^5 z^5} 
\,f_{L_i}^{(0)}(z)
 \mathcal{E}(x,\beta,c_{L_k},{c_{E_j}})\Phi^{(0)}(x)\Phi^{(0)}(z)
 \nn \\ 
&& \times\,\frac{1}{2 (4 \pi)^2}
 d^-F^+_{E_h}(0,z,x) (p^\mu+{p'}^\mu)\,.
\end{eqnarray}
Since the zero-momentum limit of the fermion propagator has a simple form, 
the two remaining integrals are elementary. The final analytic expression is 
lengthy and valid for any positive value of $\beta$.
We refrain from giving the explicit expression. However, the limit 
$\beta \to \infty$ is straightforward. After using \eqref{YukawaRelation} 
to relate the Yukawa matrices for the bulk Higgs to the 
couplings for the delta-regularized Higgs we recover the same answer as 
in \cite{Beneke:2012ie}, 
\begin{align}
\mathcal{I}_H^{(0)}\to  - \frac{i Q_\mu e}{96 \pi^2 T^2} \frac{T^3}{k^4}
f_{L_i}^{(0)}(1/T) Y_{ih} Y^{\dagger}_{hk} 
Y_{kj} g^{(0)}_{E_j}(1/T)\,
 (p^\mu+{p'}^\mu)\,.
\end{align}
This observation is general: the Higgs zero-mode contribution of the 
bulk Higgs in the $\beta \to \infty$ limit is always equal to the one 
of the theta-function regularized brane Higgs, 
see \eqref{eq:Higgsresults}
and \eqref{eq:HiggsresultsCustodialbulk}. In other words, the localization 
limit of the bulk Higgs is independent of the bulk profile at finite 
Higgs localization width.
 
We still have to determine the contribution 
from the tower of KK Higgs excitations in the loop. To illustrate the 
computation in the 5D framework, we consider again the diagram shown in 
Figure~\ref{fig_HiggsDiagramEXP}. The KK contribution is 
obtained by the replacement $\Delta_\Phi(l,x,z) 
\to \Delta^{\rm ZMS}_\Phi(l,x,z)$ in \eqref{eq:HiggsDiagram2}. 
An analytical evaluation seems difficult even for $\beta\gg 1$. Before 
turning to the numerical calculation we shall first show that 
the KK contribution does not go to zero for large $\beta$ despite the 
fact that the lowest KK masses are of order $\beta T$. This confirms 
the non-decoupling effect found in \cite{Agashe:2014jca}, now in the 
5D framework. To this end we look at the different 
loop-momentum regions separately. There are two relevant scales,
the KK scale $T$ and the Higgs localization scale $\beta T$.
This leads to several momentum regions that allow for various expansions 
of the propagators. The expanded forms can then either be integrated 
directly or at least their $\beta$ scaling can be determined.
\begin{itemize}
\item For small loop momenta $l\ll T$ we can expand both the fermion 
and the Higgs propagator around $l=0$. We can then analytically integrate 
the $x$ and $z$ coordinates as in~\eqref{eq:HiggsExchangegAnalytic}.
In this region the scaling with $\beta$ must be the same as the 
scaling of the Wilson coefficient of the four-fermion operator
discussed in Section~\ref{sec:fourfermion} and in Appendix \ref{sec:example}. 
That is, for large $\beta$ the integrand scales as $1/\beta$. Hence, 
the total contribution from this region vanishes for $\beta \to \infty$.
\item The second region is $l \sim T$. For the Higgs propagator we
can use the same expansion for small euclidean momenta as for $l\ll T$ but 
the fermion propagator can no longer be expanded. Nonetheless,  
$d^-F^+_E(l,x,z)$ does not introduce an additional $\beta$ dependence
in this momentum region.  We recover the overall scaling $\propto 1/\beta$ 
for fixed values of $l$ just as for $l\ll T$.  The only difference to the 
region with $l\ll T$ is the scaling of the integrand with the loop 
momentum $l$, which no longer is a simple power law. However, 
for $l\sim T$, the scaling of the integral with $\beta$ is the same 
as the integrand, that is $1/\beta$, and hence the contribution from 
this region also vanishes for $\beta \to \infty$.
\item For loop momentum $l$ of the order $\beta T$ we can make use of 
an expansion of  modified Bessel functions of the form 
$\mathcal{I}_{\beta}(\beta x)$ and   $\mathcal{K}_{\beta}(\beta x)$ for 
large $\beta$, given by 
\begin{equation}
\mathcal{I}_{\beta}(\beta x) \sim
\sqrt{ \frac{1}{2\pi \beta }}\frac{e^{\beta f(x)}}{(1+x^2)^{1/4}} \,g(x)\,,
\qquad
\mathcal{K}_{\beta}(\beta x) \sim
\sqrt{\frac{\pi}{2 \beta }}\frac{e^{-\beta f(x)}}{(1+x^2)^{1/4}}
\,\tilde{g}(x)\,.
\end{equation}
The exact expressions for the functions $f,\,g$ and  $\tilde g$ can be found 
in~\cite{DLMF,OLBC10}. Here we only need that $f$, $g$ and $\tilde g$ 
depend on $\beta$ only via terms that vanish at least as fast as $1/\beta$
for $\beta \to \infty$ and that $f(x)$ is a strictly monotonically increasing
function of $x$. 
Using these expansions one can show that the Higgs propagator retains the 
same $1/\beta$ scaling as in first two regions. Taking into 
account the behaviour of the fermion propagator for $l\gg T$ we find that 
$d^4l\,l^2\,\partial_{l^2}^2 d^-F^+(l,z,x)$ counts as a factor of $dl \,l$ 
or equivalently $dl\cdot(\beta T)$. This cancels the $1/\beta$ from the 
Higgs propagator and leaves us with the coordinate integrals.
Their counting is easier to determine when the integral 
over $w$ has not yet been carried out.
The integral over $w$ then cancels the $\sqrt{\beta}$ factors from 
the Higgs zero-mode profile and one Yukawa coupling. 
Every integral over a coordinate difference counts as $1/\beta$
(compare the discussion of KK effects in the gauge contribution). 
Including the two remaining Yukawa couplings, we find that the 
integrand scales as $1/\beta$ in the region $l\sim \beta T$. 
Hence the integral over the domain $l \sim \beta T$ 
takes a constant value for $\beta \to \infty$. 
\item 
Finally, for  $l\gg \beta T$  we expand the Higgs propagator for large 
momenta, since it is now dominated by the scale $l$ and no 
longer by $\beta T$. Consequently, the Higgs propagator scales as $1/l$, and 
the distance $|x-z|$ is limited to be of order $1/l$. This 
effectively trades two powers of $1/(\beta T)$ for two powers of $1/l$ 
compared to result in the $l \sim \beta T$ region, resulting in 
the scaling $\propto \beta/l^2$ of the integrand. The final integral over 
the modulus of $l$ is therefore convergent and since
\begin{equation}
\int_{\beta T}^\infty \!\!{\rm d}l\; \frac{\beta}{l^2} = \frac{1}{T}\,,
\end{equation}
the high-momentum region also gives a finite $\beta$-independent
contribution to the dipole operator coefficient.          
\end{itemize}
Since in every region the integral over $l$ either vanishes 
($l\ll T$, $l \sim T$) or converges ($l\gg \beta T$ and $l \sim \beta T$)
to a constant, the contribution to the dipole Wilson coefficient
due to the Feynman diagram in Figure~\ref{fig_HiggsDiagramEXP} tends  
to constant for large $\beta$ as announced.  For large values of $\beta$
the integral is further dominated by the high-momentum regions and therefore 
the 5D masses of the fermions enter predominantly via the external 
zero-modes. 

\begin{figure}
\vskip-0.8cm
\centering
\includegraphics[width=0.95\columnwidth]{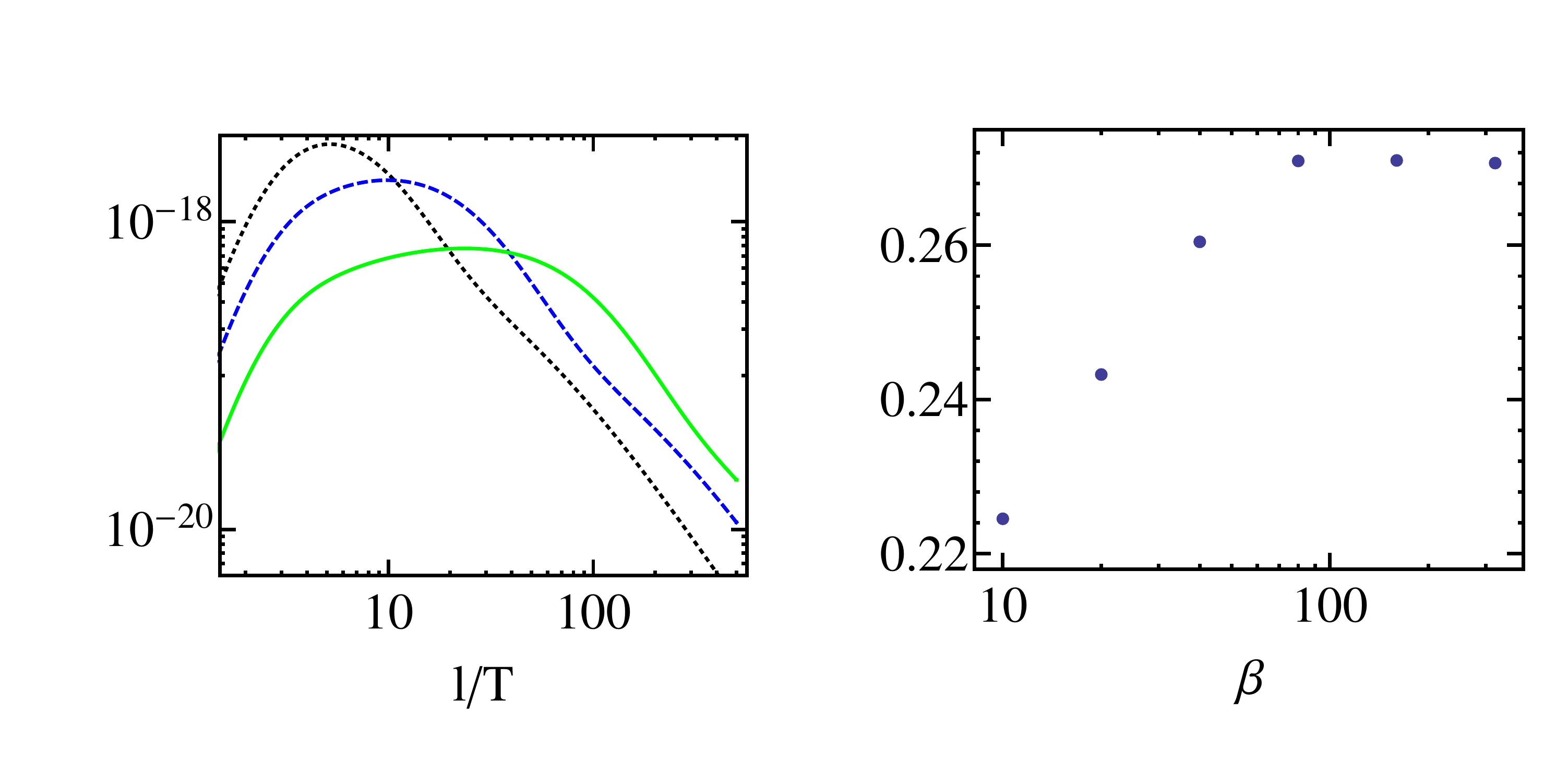} 
\vskip-0.8cm
\caption{\label{fig_HiggsKKcontribution} {\it Left:} 
Absolute value of the integrand for the diagram of 
Figure~\ref{fig_HiggsDiagramEXP} with zero-mode subtracted Higgs propagator 
as a function of the loop momentum $l$. The curves correspond to $\beta =10$ 
black (dotted), $40$ blue (dashed) and $160$ green (solid). The 
KK scale was set to $T=1\;\rm TeV$.
{\it Right: } KK Higgs contribution to the dipole operator as a function of 
$\beta$ normalized to the $\beta\to \infty$ limit of the Higgs 
zero-mode contribution. }
\end{figure}

The left panel of Figure~\ref{fig_HiggsKKcontribution} shows the  
numerical result for the integrand as a function of the loop momentum $l$   
and demonstrates the expected 
inversion of the order of the curves for different $\beta$ values 
from the intermediate to the high-momentum regions.\footnote{
Note that the solid curve for $\beta=160$ does not reach the 
asymptotic region of very large loop momentum $l\gg \beta T$, while 
$\beta=10$ is on the small side for the $\beta\gg 1$ scaling to hold. 
When taking the coordinate integrals analytically
(possible in some of the momentum regions) we encounter 
ratios of $\Gamma$ functions such as~$\Gamma(6+\beta)/\Gamma(7+\beta)$, 
which scale as $1/\beta$
for large $\beta$, but $\beta\sim 10$ is not quite large enough 
to make this manifest.} The right panel shows 
the KK Higgs contribution as a function of $\beta$ 
normalized to the zero-mode contribution
in the $\beta\to \infty$ limit. The plot illustrates the 
approach of KK contribution to a constant.
The relatively fast convergence with increasing $\beta$ is a feature 
of the simple diagram topology under consideration. 
The plot shows that the KK Higgs contribution while somewhat smaller than the 
corresponding zero-mode contribution is of the same order of 
magnitude \cite{Agashe:2014jca}.

A similar scaling analysis can be applied to all other diagrams
involving KK Higgs modes. We will not discuss them in detail, as we 
have to resort to a numerical evaluation. In Appendix~\ref{sec:KKHiggs} 
we give the numerical ratio of the KK tower to the zero-mode contribution 
for each diagram topology.  The final result for dipole operator 
coefficient generated by the exchange of the KK Higgses is 
\begin{align}
\label{eq:HiggsKKmin}
a^{H,\rm KK}_{ij} = 
\frac{Q_\mu e}{192 \pi^2}  \frac{T^3}{k^4} \cdot \mathcal{A}_{\rm KK} \cdot 
f^{(0)}_{L_i}(1/T) 
[YY^{\dagger} Y]_{ij} g^{(0)}_{E_j}(1/T) 
\end{align}
in the minimal model and 
\begin{equation}
\label{eq:HiggsKKcs}
a^{H,\rm KK}_{ij} = \frac{Q_\mu e}{192 \pi^2}  \frac{T^3}{k^4} 
\cdot f^{(0)}_{L_i}(1/T) \,\big[\mathcal{A}^{cs}_{\rm KK} YY^{\dagger} Y + 
\mathcal{B}^{cs}_{\rm KK}Y_u Y_u^{\dagger} Y \big]_{ij} \,g^{(0)}_{E_j}(1/T) 
\end{equation}
in the custodially protected model, which should be compared to the second 
line of (\ref{eq:Higgsresults}) and (\ref{eq:HiggsresultsCustodialbulk}), 
respectively. Here we again dropped the suppressed ``off-shell terms''
similar to those in the first line of \eqref{eq:Higgsresults}. The numerical 
values of the coefficients are 
\begin{align}
\mathcal{A}_{\rm KK}= 0.46(0.04) && \mathcal{A}^{cs}_{\rm KK}=1.4(0.2) 
&&\mathcal{B}^{cs}_{\rm KK}= 0.1(0.05)\,,
\end{align}
where the number in parenthesis shows the estimated error due to the 
extrapolation to $\beta=\infty$. 
The sizeable relative uncertainty in $\mathcal{B}^{cs}_{\rm KK}$ 
comes from large cancellations among the various contributions to the 
coefficient. In the minimal (custodial) model the KK contribution is 
about 50\% (75\%) of the zero-mode contribution.

Irrespective  of the Higgs localization, the dipole coefficient $a^H_{ij}$ 
generated by Higgs exchange is in general misaligned relative 
to the mass matrix. For the bulk Higgs case the numerically dominant 
terms scale as $Y Y^{\dagger} Y$ in both the minimal and custodially 
protected RS model (which includes $Y_{u} Y^{\dagger}_{u}Y$). 
After rotation to the mass basis this potentially 
generates large LFV transitions. For the same 
reason, even after the rotation to the mass basis, unlike the gauge boson 
contribution, the Higgs 
contribution depends strongly on the values of the 5D bulk mass parameters 
and the 5D Yukawa matrices. It usually increases with the magnitude 
of the Yukawa matrix entries. In the minimal RS model with an 
exactly brane-localized Higgs, however, $a^H$ is much smaller than 
for the bulk Higgs case (recall that in this scenario only the first 
line of \eqref{eq:Higgsresults} is present).

\subsection{Dimension-eight operators}
\label{sec:dim8}

The effects of dimension-eight operators are suppressed relative to the 
dimension-six ones by a factor of ${\cal O}(v^2/T^2)$ and therefore 
negligible. However, for LFV observables this counting can be numerically 
upset, as noted in \cite{Csaki:2010aj}, since the leading dimension-6  
contribution to the dipole operator from gauge-boson exchange is suppressed 
by a factor of 30-50 due to the near-alignment discussed 
above and in~\cite{Beneke:2012ie, Moch:2014ofa} . 
Relevant dimension-eight effects can arise directly from 
dimension-eight operators and indirectly from $v^2/T^2$
corrections to the field rotation to the mass basis.

The first class corresponds to the descendant
$(\bar L_i \sigma^{\mu\nu} E_i)  \Phi X_{\mu\nu} \Phi^\dagger \Phi$ 
($X=B,W$) of the dimension-six dipole operator 
$(\bar L_i \sigma^{\mu\nu} E_i)  \Phi X_{\mu\nu}$, which 
after EWSB give rise to the same dipole vertex structure. However, 
the dimension-eight operator has a coefficient function 
proportional to $Y Y^\dagger Y$ even for the internal gauge-boson exchange 
contribution, and 
does not suffer from the alignment suppression of terms proportional 
to $Y$. Depending on the value of $T$, the dimension-eight 
contribution may then be the dominant source of flavour violation. 
This is relevant only for the case of an 
exactly brane localized Higgs in the minimal RS model, 
where the contributions to the dimension-six dipole Wilson 
coefficient cubic in the Yukawa coupling due to Higgs exchange are 
also suppressed (see previous subsection). 

The second class of dimension-eight effects arises from the 
fact that the tree-level relation  
\begin{align}\label{SimpleMassrelation}
\frac{v}{\sqrt{2}}\,U^\dagger_{ij} \,\sqrt{\frac{1-2 c_{L_j}}
{1-\epsilon^{1-2c_{L_j}}}}Y_{jk} \sqrt{\frac{1+2 c_{E_k}}
{1-\epsilon^{1+2c_{E_k}}}}\,V_{kn}=\text{diag}\{m_e,m_\mu,m_\tau \} 
\end{align}
that defines the rotations $U$, $V$ to the mass 
basis \cite{Grossman:1999ra} receives corrections due to multiple 
Higgs vev insertions.\footnote{The square root factors arise from 
the explicit expressions for the lepton zero-mode profiles.} 
The diagonalization condition has the form 
\begin{align}\label{TrueMassrelation}
      \frac{v}{\sqrt{2}} U^\dagger_{ij}  \sqrt{\frac{1-2 c_{L_j}}
      {1-\epsilon^{1-2c_{L_j}}}}\left[Y-\frac{v^2}{6T^2} YY^\dagger 
      Y \right]_{jk} \sqrt{\frac{1+2 c_{E_k}}{1-\epsilon^{1+2c_{E_k}}}} V_{kn} 
  =   \text{diag}\{m_e,m_\mu,m_\tau \} \,,
\end{align}
cf.~\eqref{eq:changesmass}. The modified $U$ and $V$ field rotation 
matrices applied to the Lagrangian \eqref{EffectiveSMOperatorsDim6}
generate an additional source of LFV which formally enters at the same level 
in the $v/T$ counting as dimensions-eight operators, which can be 
taken into account by the substitution 
\begin{align}
\label{rotationcontributions}
a_{ij}^g \to a_{ij}^g + 
\frac{v^2}{6 T^2}\left. a_{ij}^g\right|_{Y\to YY^\dagger Y}
\end{align}

The direct effect of the dimension-eight operators is more difficult to 
estimate. We have to evaluate the contributions to the dipole-like  
operators that appear at the dimension eight level, i.e., 
\begin{align}
 \label{dimeightLagrangian}
\mathcal{L}^{\rm dim-8}\supset \frac{1}{T^4} a^{B,\rm dim-8}_{ij} 
(\bar L_i\Phi \sigma^{\mu\nu} E_j)  B_{\mu\nu} \Phi^\dagger \Phi+
\frac{1}{T^4} a^{W,\rm dim-8}_{ij} (\bar L_i \tau^A \Phi \sigma^{\mu\nu} E_j)
W^A_{\mu\nu} \Phi^\dagger \Phi\,.
\end{align}
The computation of the electromagnetic dipole coefficient 
$a^{\rm dim-8}_{ij}=\cos \Theta_W \, a^{B,\rm dim-8}_{ij} -
\sin \Theta_W\, a^{W,\rm dim-8}_{ij}$ would require the computation of 
roughly 150 different diagrams in the 5D theory for the minimal RS model 
alone.

Fortunately, only some of these diagrams actually contribute. For the 
following we consider only the {\it minimal} RS model with an exactly 
brane-localized Higgs. For the other Higgs localizations the dimension-six 
dipole is always dominant and dimension-eight terms are negligible as 
discussed above. We then have two fundamentally different contributions: from 
the so-called wrong-chirality Higgs couplings (WCHC) and from the ordinary 
Higgs couplings to lepton modes with the same chirality as the SM zero modes. 
It turns out that for the exactly brane-localized Higgs the WCHC 
contribution can be computed analytically and is simply given by
\begin{align}
\label{dim8mapping}
 a^{\rm dim-8,WCHC}_{ij}=  -\frac13 
\left. a_{ij}^g\right|_{Y\to YY^\dagger Y} 
\end{align}
in terms of the dimension-six gauge-boson exchange contribution.

\begin{figure}
\centering
\includegraphics[width=0.88\textwidth]{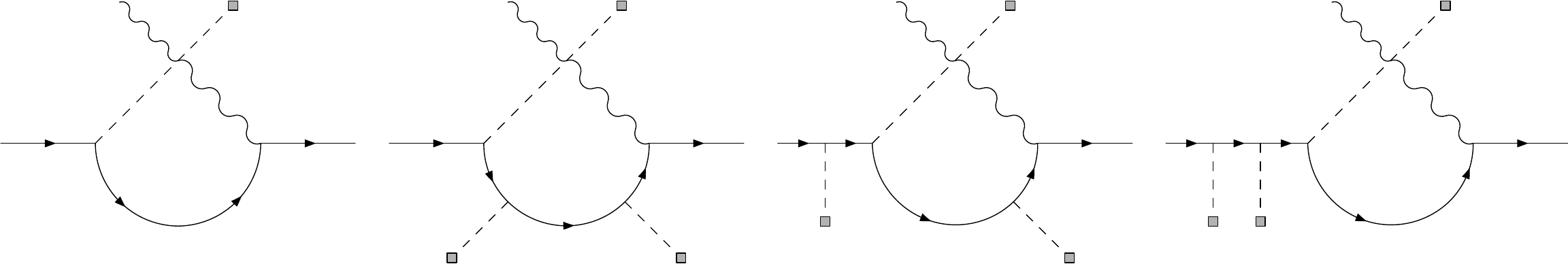} 
\caption{Example of a diagram contributing to the matching onto 
the dimension-six dipole operator and several related diagrams that 
contribute to the dimension-eight operator.}
\label{fig:dim8original}
\end{figure}

To illustrate how this result arises let us consider the left-most 
diagram in Figure~\ref{fig:dim8original} 
(W8 in the notation of \cite{Beneke:2012ie}), which contributes to the 
matching of the $a^W_{ij}$ coefficient. There are 10 ways to add 
two additional external Higgs lines to the fermion line. However, since 
$\delta/T$ ($\delta$ being the Higgs localization regulator) is much smaller 
than the dimensional regulator or, equivalently, than the inverse loop 
momentum cut-off, we find that only the three diagrams shown 
to the right in Figure~\ref{fig:dim8original} 
give a non-vanishing WCHC contribution for $\delta\to 0$. 
In each case the integrals over the Higgs vertices can be taken analytically. 
In the above example the WCHC contributions of the two right-most diagrams 
cancel, and the remaining diagram can be expressed in terms of the 
associated dimension-six diagram as shown in~\eqref{dim8mapping}. 
Similarly the descendants of all other dimension-six diagrams can be 
shown to satisfy~\eqref{dim8mapping}. 

Hence the effect of the WCHC can be included via the redefinition 
\begin{align}
\label{WCHCcontributions}
 a_{ij}^g \to a_{ij}^g -\frac{v^2}{6 T^2}\left. a_{ij}^g
\right|_{Y\to YY^\dagger Y}
\end{align}
where we used that the Higgs fields will assume their vacuum expectation 
value ($\Phi^\dagger\Phi\to v^2/2$). Combining this with 
(\ref{rotationcontributions}), we find that the direct and indirect 
contribution cancel. That is, at the dimension-eight level the WCHCs 
do not generate sizeable flavour-changing transitions by 
lifting the misalignment suppression and can be ignored.  

This leaves us with the dimension-eight contributions that have no WCHCs. 
In the minimal model as defined in \cite{Beneke:2012ie} there are no
such contributions from the diagrams with non-abelian vertices. Then there 
are only seven non-vanishing diagrams that involve an internal $W$ boson, 
but about 50 diagrams with a hypercharge boson. Fortunately, the limited 
particle content of the minimal model allows us to recast the expressions 
of all diagrams in the form of the original dimension-six diagram with 
modified fermion lines. For instance, the second diagram in 
Figure~\ref{fig:dim8original} has terms without WCHCs, but differs from  
the original diagram only by the two additional (zero-momentum) 
Higgs insertions that modify one fermion propagator. This can easily 
be calculated as the Higgs vertices can be treated analytically. 
Since the flavour-dependence of the fermion propagators (excluding zero-modes)
is relatively mild, one can use the single-flavour approximation, where 
the Yukawa matrices are the only flavour-dependent quantities.  It is 
then straightforward to compute the contribution to the dimension-eight 
coefficients. We find 
\begin{align}
a^{\rm dim-8}_{ij} \approx  -0.4 \left. 
a_{ij}^g\right|_{Y\to YY^\dagger Y}\,.
\end{align}
This size is in agreement with the estimate given on the basis of 
a subset of diagrams in \cite{Csaki:2010aj}, 
where the non-abelian contribution was found to be 
 $a^{W,dim-8}_{ij}\approx -0.31  \left. a^{W}_{ij}\right|_{Y\to YY^\dagger Y}$.
The minimal model requires a KK scale $T>4 \;{\rm TeV}$
in order to pass the constraints set by electroweak precision observables
\cite{Casagrande:2008hr}. Hence the dimension eight contribution to 
$\alpha_{ij}$ is suppressed by an additional factor $v^2/T^2$ of at 
least $1/500$. We therefore neglect the contribution of the dimension-eight 
terms to the off-diagonal elements of $\alpha_{ij}$, since it is 
smaller than the effect of the Barr-Zee diagrams which also feature three 
Yukawa couplings without the need to 
take the dimension-eight term into account. 

For the custodially protected model the dimension-eight coefficient would  
be much harder to compute. Not only does the number of non-trivial Feynman 
diagram topologies increase significantly, but the larger particle content 
leads to numerous non-vanishing possibilities to assign the various fermion
species to each topology. However, independent of the Higgs localization 
there always exists an unsuppressed dimension-six contribution proportional 
to $Y Y^\dagger Y$, hence the dimension-eight terms are never 
relevant.

\section{Phenomenology}
\label{sec:pheno}

The Standard Model in its original form \cite{Weinberg:1967tq} does not 
allow for flavour violation in the lepton sector. Even after introducing 
neutrino masses, the charged LFV 
processes are suppressed by the tiny neutrino masses and too small to be 
detected in any foreseeable experiment. Any signal of charged LFV is 
a clear sign of physics beyond the SM.
In our analysis of LFV in RS models, we focus on the three processes with 
the highest current experimental sensitivity, $\mu \to e \gamma$, 
$\mu \to \bar e e e$ and muon conversion in a gold atom. The best 
current limits for the branching fractions 
are \cite{Adam:2013mnn,Bellgardt:1987du,Bertl:2006up}
\begin{align}
 {\rm Br}\left(\mu \to e\gamma \right)<5.7 \times 10^{-13}         
&&&  \text{MEG      }\\
 {\rm Br}\left(\mu \to \bar eee \right)<1 \times 10^{-12}   
 &&&  \text{SINDRUM  }\\
 {\rm Br}^{\rm Au}\left(\mu N \to e N \right)<7 \times 10^{-13}  
&&&  \text{SINDRUM II}\,.
\end{align}
The limit on muon conversion in gold is more stringent than current 
limits extracted using other nuclei, see e.g., 
\cite{Ahmad:1988ur,Dohmen:1993mp} (titanium) or \cite{Honecker:1996zf} (lead).

In Section~\ref{sec:toLFVobservables} these three observables were 
expressed in terms of the Wilson coefficients of the dimension-six 
SM effective theory Lagrangian, which in turn were determined by integrating 
out the fifth dimension of the RS model. We now calculate the branching 
fractions of the observables for a given set of 5D input parameters.  
Before performing a scan through the parameter space of the model, it is 
useful to have some qualitative understanding of the effects of the 
various dimension-six Wilson coefficients. We will generally assume that the 
5D Yukawa matrices are anarchic, without imposing any additional flavour 
symmetries.

\subsection{Estimates}
\label{sec:simpleestimates}

We first consider the effect of the dimension-six dipole operator, where 
we distinguish two different contributions: from the Higgs-exchange diagrams, 
which involve three Yukawa matrices, and from gauge-boson exchange, 
which involves only one. The gauge contribution leads to
naturally suppressed flavour-violating couplings, whereas the 
Higgs contribution does not have a built-in flavour protection. 
For not too small Yukawa couplings the Higgs contribution is 
dominant. We mainly focus on $\mu \to e$ transitions, for which the 
dipole coefficients $\alpha_{12}$  and $\alpha_{21}$ are relevant. 

To obtain an estimate of the Higgs-exchange 
contribution let us start with  Wilson coefficient 
[see (\ref{eq:HiggsresultsCustodialbulk}) and 
(\ref{eq:HiggsKKcs})]
\begin{align}
\label{eq:totalHiggsDipole}
a_{ij}^H= \frac{Q_\mu e }{192 \pi^2}\frac{T^3}{k^4} \cdot 
f_{L_i}^{(0)}(1/T)[(2+\mathcal{A}_{\rm KK}^{cs})YY^\dagger Y + 
\mathcal{B}_{\rm KK}^{cs} Y_u Y_u^{\dagger} Y]_{ij} g^{(0)}_{E_j}(1/T)
\end{align}   
in the custodially protected model with a bulk Higgs. An analogous 
expression holds in the minimal model [(\ref{eq:Higgsresults}) and 
(\ref{eq:HiggsKKmin})]. For an exactly brane-localized Higgs, 
$a_{ij}^H$ is of similar size as above for the custodially protected model, 
cf.~(\ref{eq:HiggsresultsCustodialbrane}), but suppressed in the minimal 
model due to the absence of the second line of (\ref{eq:Higgsresults}).
Now we recall that the relation of fermion zero-mode profiles, the 5D 
Yukawa matrix and SM Yukawa matrix (before rotation into the mass basis) 
is given by
\begin{equation}
y_{ij}=\frac{T^3}{k^4} \, f^{(0)}_{L_i}(1/T) Y_{ij} g^{(0)}_{E_j}(1/T) \;.
\end{equation}
If the fermion mass hierarchy of the diagonalized SM Yukawa matrix is 
carried democratically by left- and right-handed fermion modes, i.e. 
\begin{equation}
\label{yijestimate}
y_{ij} \sim \frac{\sqrt{m_i m_j}}{v/\sqrt{2}} \,,
\end{equation}
we arrive at the estimate
\begin{align}
\label{eq:HiggsestimateDipole}
a_{ij}^H\sim \frac{Q_\mu e \sqrt{2 m_i m_j}}{192 \pi^2 v } 
\left[(2+\mathcal{A}_{\rm KK}^{cs})Y_\star^2 + 
\mathcal{B}_{\rm KK}^{cs} Y_{u,\star}^2 \right]\,,
\end{align}  
where we assume that 
\begin{equation}
 Y^2_\star\equiv \frac{[YY^\dagger Y]_{ij}}{Y_{ij}}\quad\qquad
 Y^2_{u,\star}\equiv \frac{[Y_uY_u^\dagger Y]_{ij}}{Y_{ij}}
\label{ystardef}
\end{equation}
are approximately independent of $ij$ ("anarchy"). For anarchic Yukawa 
matrices we also expect that the rotation matrices $U$ and $V$ follow the same 
hierarchy $|U_{ij}|\sim |V_{ij}| \sim \mbox{min}\,(\sqrt{m_i/m_j}, 
\sqrt{m_j/m_i})$ and hence, barring accidental cancellations, 
it follows from (\ref{couplingdefs}) that $\alpha_{12} 
= [U^\dagger]_{1k} a_{kl} V_{l2} \sim a_{12}$.

Further using that $\mathcal{A}_{\rm KK}^{cs} \approx 1.4 \gg 
\mathcal{B}_{KK}$ we obtain\footnote{Our estimates always yield 
$\alpha_{21}\sim\alpha_{12}$, hence we only give $\alpha_{12}$ explicitly.}
\begin{align}
\alpha_{12}^H\sim \frac{5 Q_\mu e \sqrt{ m_e m_\mu}}
{192 \pi^2 v } \,Y_\star^2\,,
\end{align}
which yields
\begin{align}
\label{eq:estimateM2e}
{\rm Br}\left(\mu \to e\gamma \right)_{|\text{Higgs dipole}} \sim 
5 \cdot 10^{-9} \times\frac{1 \,{\rm TeV}^4}{T^4}\, {Y_\star}^4  \,.
\end{align}
If the dipole also dominates $\mu \to 3 e$ one can combine \eqref{eq:BrMEG} 
and \eqref{eq:Brmu3e} to obtain the relation
\begin{align}\label{eq:dipoleDomRelation}
  \frac{ {\rm Br} \left(\mu \to 3 e       \right) }
       { {\rm Br} \left(\mu \to  e \gamma \right) } = 
\frac{2 \alpha_{\rm em}}{3 \pi}
       \left[\log\frac{m_\mu}{m_e}-\frac{11}{8} \right]\approx 0.006\,,
\end{align}   
which translates into an estimate of 
\begin{align}
\label{eq:estimateM3e}
  {\rm Br}\left(\mu \to 3 e \right)_{|\text{Higgs dipole}} \sim
       3 \cdot 10^{-11}  \times\frac{1 \,{\rm TeV}^4}{T^4} \, {Y_\star}^4   \,.
\end{align}
For muon conversion one finds
\begin{align}
\label{eq:estimateMe}
  {\rm Br^{Au}}\left(\mu N \to e N \right)_{|\text{Higgs dipole}} \sim
       1.5 \times 10^{-11}  \times
\frac{1 \,{\rm TeV}^4}{T^4}\,  {Y_\star}^4   \,.
\end{align}
We emphasize that these are crude estimates. Even in the  
anarchic case the random phases of the different elements can lead to 
cancellations or add coherently. However, they provide useful guidance 
to the results of the numerical scan discussed below.

The Barr-Zee contribution is similar to the Higgs contribution, since the 
dominant contribution to the $\eta_{ij}$ Wilson coefficient 
is also proportional to a product of three Yukawa factors. Comparing the 
prefactors  in \eqref{ARtotal} we find that the Barr-Zee contribution 
to the dipole coefficient is smaller  
by a factor of about $170$ than the contribution from the 5D Higgs loops. 
Thus we expect a $\mu \to e\gamma$ branching fraction of about 
\begin{align}
\label{eq:estimateMEGBZ}
  {\rm Br}\left(\mu \to e   \gamma \right)_{|\text{BZ}}     
  \sim   2 \cdot 10^{-13}  \times \frac{1 \,{\rm TeV}^4}{T^4}\,Y_\star^4\,,
\end{align}
if only the BZ contribution existed. The BZ contribution to the other 
processes is also smaller by a factor of about $170^2$. 
  
\begin{figure}
\begin{center}
\includegraphics[width=0.45\textwidth]{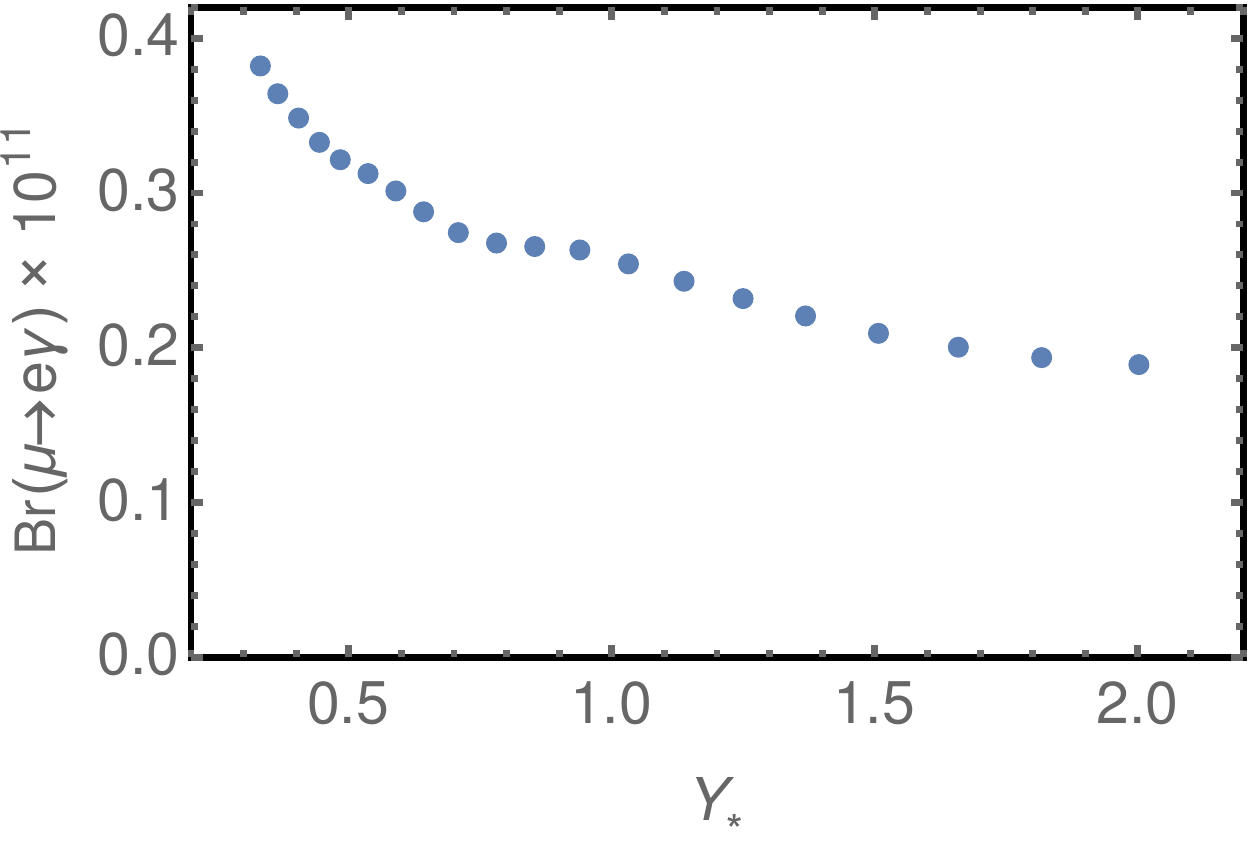} 
\hspace*{0.3cm}
\includegraphics[width=0.45\textwidth]{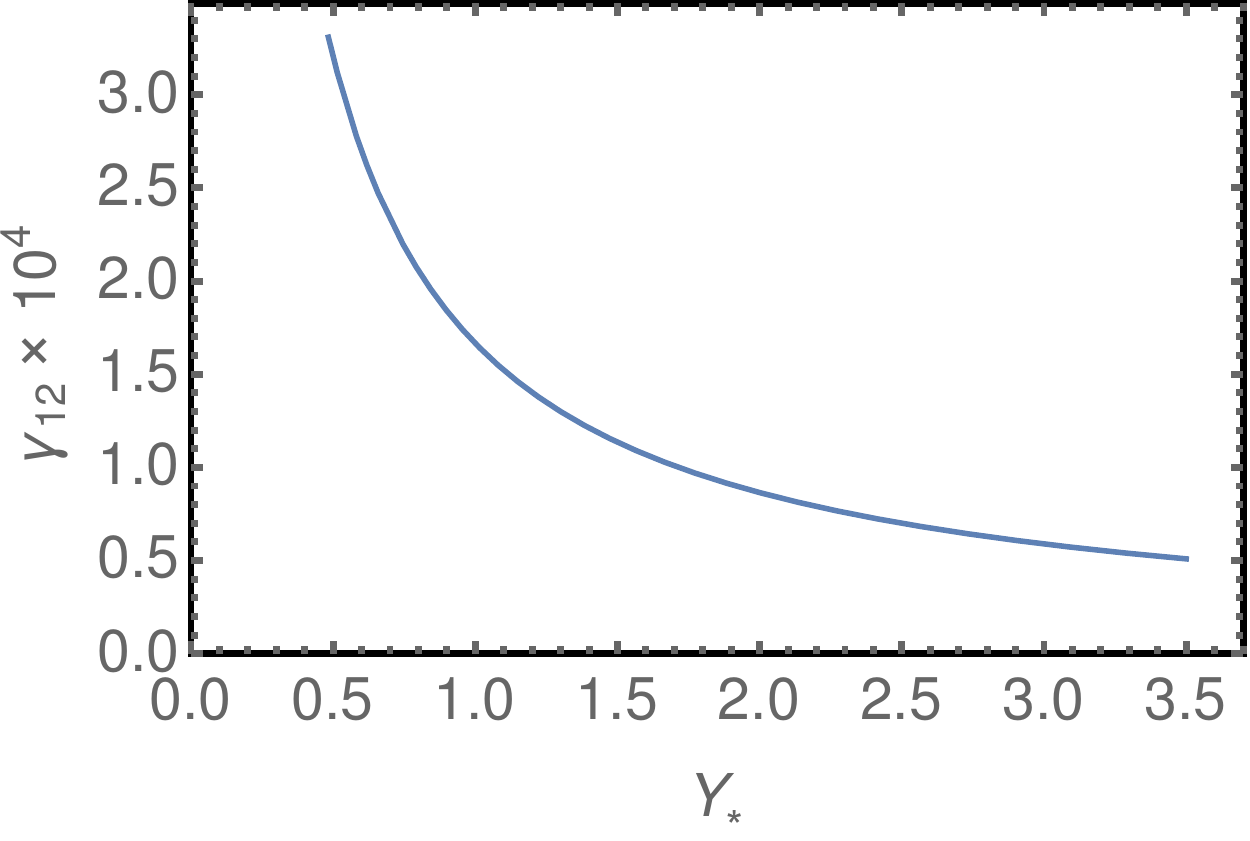} 
\end{center}
\vskip-0.4cm
\caption{\label{fig:GaugeDIpoleEffectSym} {\it Left:} Gauge contribution 
to ${\rm Br}\left(\mu \to e \gamma \right)$ ($T=1\;\rm TeV$) for fixed
Yukawa structure as a function of the absolute Yukawa coupling size. 
{\it Right:} The Wilson coefficient $\gamma_{12}^1$ following the 
approximation \eqref{eq:approxWCc} as a function of Yukawa coupling size for 
$T=1\;\rm TeV$. The $\mathcal{O}(m_\mu/m_\tau)$ term is not included.}
\end{figure}

Due to the $Y_\star^4$ dependence the Higgs-exchange induced dipole operator 
is less important for small Yukawa coupling. In this case, and also for  
the special case of the brane-localized Higgs in the minimal
RS model, the dipole operator generated by gauge-boson exchange becomes 
crucial.  We do not have an analytical expression for the gauge-boson 
contribution, but we know that there would be no flavour violation from it, 
if the function $\mathcal{A}_{ij}$ in (\ref{gaugedipole}) was independent 
of $ij$. The 5D mass parameters must decrease with the absolute values of 
the Yukawa couplings in order to guarantee the correct values for the SM 
masses fermion masses. $\mathcal{A}_{ij}$ varies more strongly for smaller 
absolute values of the 5D mass parameters, 
see Figure~\ref{fig:MassGaugeDep}, and therefore the flavour-changing 
gauge-boson contribution should increase with decreasing Yukawa coupling. 
To verify this we fix the Yukawa matrix structure, that is the ratios of 
all matrix elements, and scale the maximal entry $Y_{\rm max}$ from 
$2$ to $0.3$. For simplicity we assumed symmetric 5D mass parameters 
$c_{L_i}=-c_{E_i}$. The resulting $\mu\to e\gamma$ branching fraction 
from $a_{ij}^{g}$ alone in the minimal model is shown in 
Figure~\ref{fig:GaugeDIpoleEffectSym} (left). 
The precise value of ${\rm Br}\left(\mu \to e \gamma \right)$ 
obviously depends on the arbitrarily chosen Yukawa matrix structure, 
but the variation with the size $Y_\star$ of the Yukawa couplings is 
not very large compared to the fourth-power law of the Higgs-exchange 
contribution. For the Yukawa matrix used in 
Figure~\ref{fig:GaugeDIpoleEffectSym}
we find a $\mu \to e \gamma$ branching fraction of a 
$\text{few}\times 10^{-12}$.

This agrees with the estimate based on the functional form of 
the gauge-boson induced dipole coefficient $a^{g}_{ij}$.
The numerical value of the Wilson coefficient 
is \cite{Beneke:2012ie,Moch:2014ofa}
\begin{align}
a^{g}_{ij} \approx - 6 \,(19) \cdot 10^{-4}  \,y_{ij}.
\end{align} 
The value without (in) parenthesis is valid for the minimal (custodial) 
model and is independent of the details of the Higgs localization.   
$y_{ij}$ is the 4D Yukawa matrix in the flavour eigenbasis. 
The matrix relation $a^g \propto y$ is only violated
by corrections of about (2-3)\% as discussed in Section \ref{sec:genDipole}.
This violation is the source of charged LFV
as it introduces small off-diagonal elements in the dipole   
coefficients $\alpha_{ij}$ in the mass eigenbasis after EWSB. 
Using (\ref{yijestimate}) and applying a factor $2/100$ for the 2\% of 
misalignment between $y_{ij}$ and $a^{g}_{ij}$, we estimate
\begin{align}
\label{eq:alphaGaugeEstimate}
\alpha_{A,12} \sim  2.6\,(8.1) \cdot 10^{-8}\times 
\frac{2}{100}
\end{align}
for the coefficient relevant to $\mu\to e$ transitions.
Again, we regard this as a rough estimate, since there may be cancellations 
when the rotation into the mass basis is performed.  
We then find: 
\begin{eqnarray}
\label{eq:estimategaugeMEG}
{\rm Br}\left(\mu \to \bar e\gamma \right)_{|\text{gauge dipole}} &\sim& 
0.5 \,(5)\cdot 10^{-11}\times\frac{1 {\rm TeV}^4}{T^4} \\
\label{eq:estimategaugeM3E}
{\rm Br}\left(\mu  \to \bar e e e  \right)_{|\text{gauge dipole}}  &\sim& 
0.3\,(3) \cdot 10^{-13}\times\frac{1 {\rm TeV}^4}{T^4}\\
\label{eq:estimategaugeME}
{\rm Br}^{\rm Au}\left(\mu N \to e N  \right)_{|\text{gauge dipole}}  &\sim& 
0.2\,(2.2) \cdot  10^{-13}\times\frac{1 {\rm TeV}^4}{T^4}
\end{eqnarray}
Note that this contribution is independent of the typical size of 
anarchic Yukawa coupling up to the $\mathcal{O}(1)$ variation shown in 
Figure~\ref{fig:GaugeDIpoleEffectSym}. It is typically smaller than the 
Higgs contribution, but provides the ``gauge-boson floor'' to the 
dipole coefficient, since it is less 
sensitive to 5D model parameters than the Higgs contribution 
and always present. In the custodially protected model the rate is a factor 
of 10 larger than in the minimal model.

The previous estimates were based on the assumption that the dipole 
operator dominates the LFV amplitudes. This is not always the case, 
especially for the $\mu\to 3 e$ and muon conversion process. 
Next, we therefore consider the impact of the four-fermion and fermion-Higgs 
operators, which are generated at tree-level. In both cases the dimension-six 
Wilson coefficients are independent of the 5D Yukawa matrices. However, a 
dependence on the Yukawa matrices enters through the rotation to the mass 
basis after EWSB. For illustration we consider the operator 
$(\bar E_i \gamma^\mu E_j) \,\Phi^\dagger i\overleftrightarrow{D_\mu} \Phi$ 
with Wilson coefficient $c^{1}_{ij} = c^{1}_{i}\delta_{ij}$ and restrict 
ourselves to the minimal model.  
For all three muon flavour-violating processes the 
relevant matrix elements are $V^\dagger_{1j}c^{1}_{jk}V_{k2}$. 
Flavour violation arises, because $c^{1}_{i}$ depends on the 
bulk mass parameter $c_{E_i}$, hence $c^{1}_{ij}$ while diagonal 
is not proportional to the unit matrix in flavour space. We can estimate 
$V^\dagger_{1j}c^{1}_{jk}V_{k2}$ by 
making use of hierarchical fermion zero-mode functions. 
Assuming $f_{E_1}(1/T)\ll f_{E_2}(1/T)\ll f_{E_3}(1/T)$ and 
 symmetric mass parameters we can 
employ the rough estimate $|V_{ij}|\sim {\rm min}\,(\sqrt{m_{i}/m_{j}},\sqrt{m_{j}/m_{i}})$ with $m_i$ being the SM lepton masses to obtain
\begin{align}\label{eq:approxWCc}
\gamma_{12}=V^\dagger_{1j}c^1_{jk}V_{k2}
\sim \sqrt{\frac{m_e}{m_\mu}}\,(c^{1}_{2}-c^{1}_{1}
+\mathcal{O}(m_\mu / m_\tau) )\,.
\end{align}
We can use this formula to study the dependence of $\gamma_{ij}^a$ 
on the size of the 5D Yukawa couplings.
Since the product of Yukawa matrix and 5D fermion zero-mode profiles
must reproduce the SM mass matrix to leading order in $v/T$, the 5D 
profiles and therefore the 5D mass parameters are correlated 
with the Yukawa matrix. The simplest estimate (assuming symmetric mass
parameters) yields the correlation $1/\sqrt{Y_\star} \sim 
f^{(0)}_{E_i}(1/T)$, where here $Y_\star$ is the generic size of the anarchic 
Yukawa matrix element. In the following we do not distinguish this 
$Y_\star$ from the one defined in (\ref{ystardef}) for order-of-magnitude 
estimates. Since the $c^a_i$ Wilson coefficients arise from a coordinate
integral over a single fermion-gauge boson vertex they will 
roughly scale as $[f^{(0)}_{E_i}(1/T)]^2$, that is $1/Y_\star$.
This behaviour was already observed and explained in \cite{Agashe:2006iy}.
The right panel of Figure~\ref{fig:GaugeDIpoleEffectSym} shows
$\gamma^1_{12}$ as a function of $Y_\star$ 
(keeping the lepton masses fixed). The curve can be 
fitted by ${Y_\star}^{-0.94}$ confirming the above scaling.
This scaling is quite general, although if
the mass hierarchy is mainly driven by the right-handed modes, 
the mass factors in the estimate
\eqref{eq:approxWCc} must change to account for the 
change in the relation (\ref{yijestimate}).

Similar estimates can be obtained for the four-fermion operator 
coefficients. Here we have terms with different dependencies on flavour. 
The Wilson coefficient $b^{LE}_{ij}$ of $(\bar L_i \gamma_\mu L_i)\,
(\bar E_j \gamma^\mu E_j)$ has three contributions denoted by $b_0$, 
$b_1$ and $b_2$, see \eqref{b2coeff}. $b_0$ does not depend on the 5D 
masses and hence does not contribute to flavour-changing processes. 
$b_1$ depends on a single bulk mass parameter and has the same scaling 
$\propto 1/{Y_\star}$ as the $c^{a}$ Wilson coefficients. The 
$b_2$ function depends on two bulk mass parameters and scales 
roughly as  $1/{Y_\star}^2$. However, as discussed below (\ref{bLtauQcoeff}), 
for light leptons this term is suppressed and not relevant. 
This would be different for processes involving fermions with zero-modes 
that are IR brane localized such as the  (right-handed) top  quark, in 
which case the ratio  of exponentials in $b_2$ no longer compensates the 
logarithmic enhancement factor and $b_2$ becomes the dominant
term in \eqref{b2coeff}.

From \eqref{ARtotal}, (\ref{eq:gCoeffs}ff) and (\ref{eq:cCoeffs}ff) 
we see that the four-fermion coefficients usually appear in combination 
with the coefficients of the Higgs-lepton operators. For a typical RS 
model parameter point, which reproduces the lepton masses, the 
Higgs-lepton operator coefficients are larger by a factor $\log\epsilon$ 
relative to the four-fermion operator coefficients. This allows us to 
use  \eqref{eq:approxWCc} to estimate the effect of the tree-level operators 
on the generically tree-dominated LFV observables. We find
\begin{eqnarray}
\label{eq:estimateTreeM3E}
{\rm Br}\left(\mu \to \bar eee \right) &\sim& 
{\text{few}} \cdot 10^{-12}\times \frac{1 {\rm TeV}^4}{T^4}\,
\frac{1}{{Y_\star}^2}
\\
\label{eq:estimateTreeME}
{\rm Br}^{\rm Au}\left(\mu N \to e N \right) &\sim&  
{\text{few}} \cdot 10^{-9}\times \frac{1 {\rm TeV}^4}{T^4}\,
\frac{1}{{Y_\star}^2}\,,
\end{eqnarray}
where we used the parameters given in Table \ref{tab:InputParameters}.
We stress again that these numbers are rough estimates, which depend 
strongly on the precise structure of the flavour rotation matrices 
$V$ and $U$. 
Thus we have three separate contributions to $\mu\to \bar{e} e e$ and 
muon conversion with different dependence on the size of the 5D anarchic 
Yukawa coupling ($Y_\star^4$, $Y_\star^0$, $Y_\star^{-2}$).

\begin{table}[t]
\begin{center}
\begin{tabular}{|c|cc||c|cc|}\hline
$m_\mu$ & $0.105658$ GeV & $\;\;$ \cite{PDG} $\;\;$&   
$m_e$  & $5.10998 \cdot 10^{-4}$ GeV &  \cite{PDG} \\
$s_W^2$ & $0.231 $             &   \cite{PDG}          &   
$M_H$ &  $125.7 $ GeV   & \cite{PDG} \\
$M_Z$  & $ 91.187 $   GeV           &   \cite{PDG}          &   
$M_W$ &  $80.385$ GeV   &  \cite{PDG} \\
$m_t$  & $ 173 $   GeV           &   \cite{PDG}          &  
$\Gamma_\mu$  & $2.99598 \cdot 10^{-19}$ GeV &  \cite{PDG} \\
$\mathcal{D}[\text{Au}]$ & $0.189$   & \cite{Kitano:2002mt} & 
$\mathcal{S}^p[\text{Au}]$ & $0.0614$ & \cite{Kitano:2002mt} \\
$\mathcal{S}^{n}[\text{Au}]$ & $0.0918$   & \cite{Kitano:2002mt} & 
$\mathcal{V}^p[\text{Au}]$ & $0.0974$ &  \cite{Kitano:2002mt}\\
$\mathcal{V}^{n}[\text{Au}]$ & $0.146$   &\cite{Kitano:2002mt}  &   
$\Gamma^{\text{Au}}_{\text{capture}}$ &  
  $8.71\cdot 10^{-18}$ GeV $^\star$   & \cite{Filippas:1963,Suzuki:1987jf} \\
$\mathcal{D}[\text{Al}]$ & $0.0362$   & \cite{Kitano:2002mt} & 
$\mathcal{S}^p[\text{Al}]$ & $0.0155$ & \cite{Kitano:2002mt} \\
$\mathcal{S}^{n}[\text{Al}]$ & $0.0167$   & \cite{Kitano:2002mt} & 
$\mathcal{V}^p[\text{Al}]$ & $0.0187$ &  \cite{Kitano:2002mt}\\
$\mathcal{V}^{n}[\text{Al}]$ & $0.0173$   &\cite{Kitano:2002mt} &
$\Gamma^{\text{Al}}_{\text{capture}}$ & 
  $4.64\cdot 10^{-19}$ GeV & \cite{Suzuki:1987jf}  \\
$f^u_{V_p}$ & $2$   &    & $f^d_{V_p}$ & $1$ &   \\
$f^u_{V_n}$ & $1$   &    & $f^d_{V_n}$ & $2$ &   \\
$f^u_{p}$ & $0.018$   &  & $f^d_{p}$   & $0.034$ &  \\
$f^u_{n}$ & $0.016$  &   & $f^d_{n}$   & $0.038$ &  \\
$f^s_{n}$ & $0.043$  &   & $f^s_{n}$   & $0.043$ & \\ \hline 
\end{tabular}
\end{center}
\caption{Input parameters for the numerical analysis. 
For the couplings of scalar quark currents to the nucleons we use the 
results of \cite{Crivellin:2013ipa} and fix the value of the 
nucleon-pion $\sigma$-term to $50\;\rm MeV$. The $^\star$ indicates that 
we use the average of the values given in the references. 
\label{tab:InputParameters}}
\end{table}
 
\subsection{Numerical analysis}

In the previous subsection we attempted to give an idea about the size and 
the relative importance of the various contributions to our three main 
LFV observables. While such estimates are useful to understand the rough
dependence of our results on the input parameters, especially
the Yukawa coupling size, they cannot replace a study of the full 
parameter dependence. To this end we next perform a numerical scan 
over the ``generic'' parameter space.  
We analyze four RS models: the minimal RS model (as defined 
in~\cite{Beneke:2012ie}) as well as a custodially protected 
model (as defined in~\cite{Moch:2014ofa}), each with either an exactly 
brane-localized or a bulk Higgs including its KK excitations 
in the $\beta \to \infty$ limit. We refer to these models 
as  \mbu\ (minimal, bulk), \mbr\ (minimal, exactly brane-localized), 
\cbu\ (custodial, bulk) and \cbr\ (custodial, exactly brane-localized).

The 5D input parameters needed for the numerical evaluation of 
the dimension-six Wilson coefficients are the 5D Yukawa matrices $Y$ 
(and $Y_u$ in the custodially protected model),
the 5D bulk mass parameters  $c_{\psi} = M_\psi/k$ of the leptons and 
the KK scale $T$. In case of the exactly brane-localized Higgs the 
wrong-chirality Yukawa couplings can in principle differ from the ``standard''
correct-chirality Yukawa couplings, but for simplicity we assume them 
to be equal. The quark flavour parameters would affect our analysis 
only through suppressed terms, which are omitted 
(see Section \ref{sec:WilsonCoeffs}).

Since we do not want to give up the idea of ``natural'' Yukawa matrices,
we further assume that the moduli of the complex Yukawa 
matrix entries are $\mathcal{O}(1)$ and anarchic.
To illustrate how the size of the 5D Yukawa matrix entries affects the 
different observables, we adopt two scan strategies. In both the 
modulus of the 
matrix elements is larger than $0.1$, but in the first (second) scan the 
maximal modulus $Y_{\rm max}$ is bounded by $0.5$ ($3$ for the second). 
The phases are arbitrary. Further, we require that the 
measured values of the charged lepton masses are reproduced by the 
chosen sets of 5D parameters. Contrary to the minimal RS model the 
custodially protected RS model includes a right-handed neutrino and 
a Dirac mass term for the neutrinos. Here we only require the neutrino 
masses to be below 0.1~eV. However, we do not demand that the PMNS 
matrix is reproduced by the 5D parameter sets, since the precise values 
of the neutrino masses do not affect charged LFV, and since, 
as mentioned in the introduction, the 
explanation of neutrinos masses and mixings is considered to lie outside 
the present model frameworks, as must obviously be the case for the 
minimal models. In practice, we randomly 
generate 5D Yukawa matrices within the above mentioned constraints, 
and then fix the 5D mass parameters $c_{\psi_i}$ such that the correct 
lepton masses are obtained. For fixed value of the KK scale $T$ and 
given scan strategy, we generate 
about $2 \cdot 10^5$ Yukawa matrices. For each of these we calculate the 
Wilson coefficients and then the branching fractions of $\mu\to e \gamma$, 
$\mu \to e$ conversion, $\mu \to 3 e$, $\tau\to \mu \gamma$, and 
$\tau\to 3 \mu $. The required low-energy parameters are shown in 
Table \ref{tab:InputParameters}. We also added the material constants of 
aluminium, which serves as the target for the 
next generation of muon conversion experiments.

\subsubsection{Minimal model}

The results of our numerical scan through the constrained parameter space
are best illustrated in two-dimensional scatter plots, which visualize the 
typical range of values for the branching fractions and correlations between 
the observables. It is important to keep in mind that the point densities 
in these scatter plots should not be used as a measure for the likelihood of 
the corresponding value in a given model. 

\begin{figure}[t]
\begin{minipage}{0.47\textwidth}
   \includegraphics[width=1\textwidth]{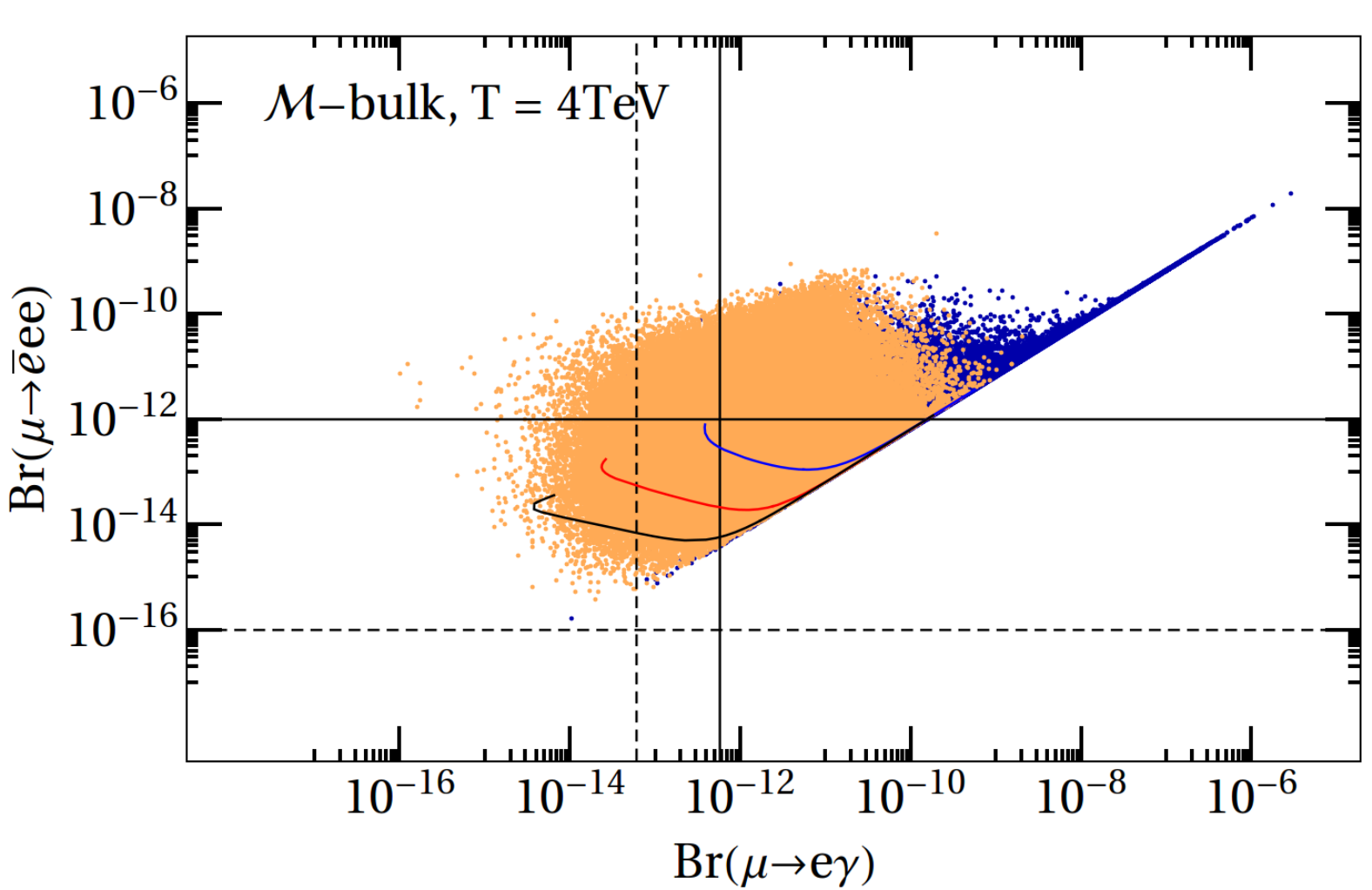}
\end{minipage}
\begin{minipage}{0.47\textwidth}
   \includegraphics[width=1\textwidth]{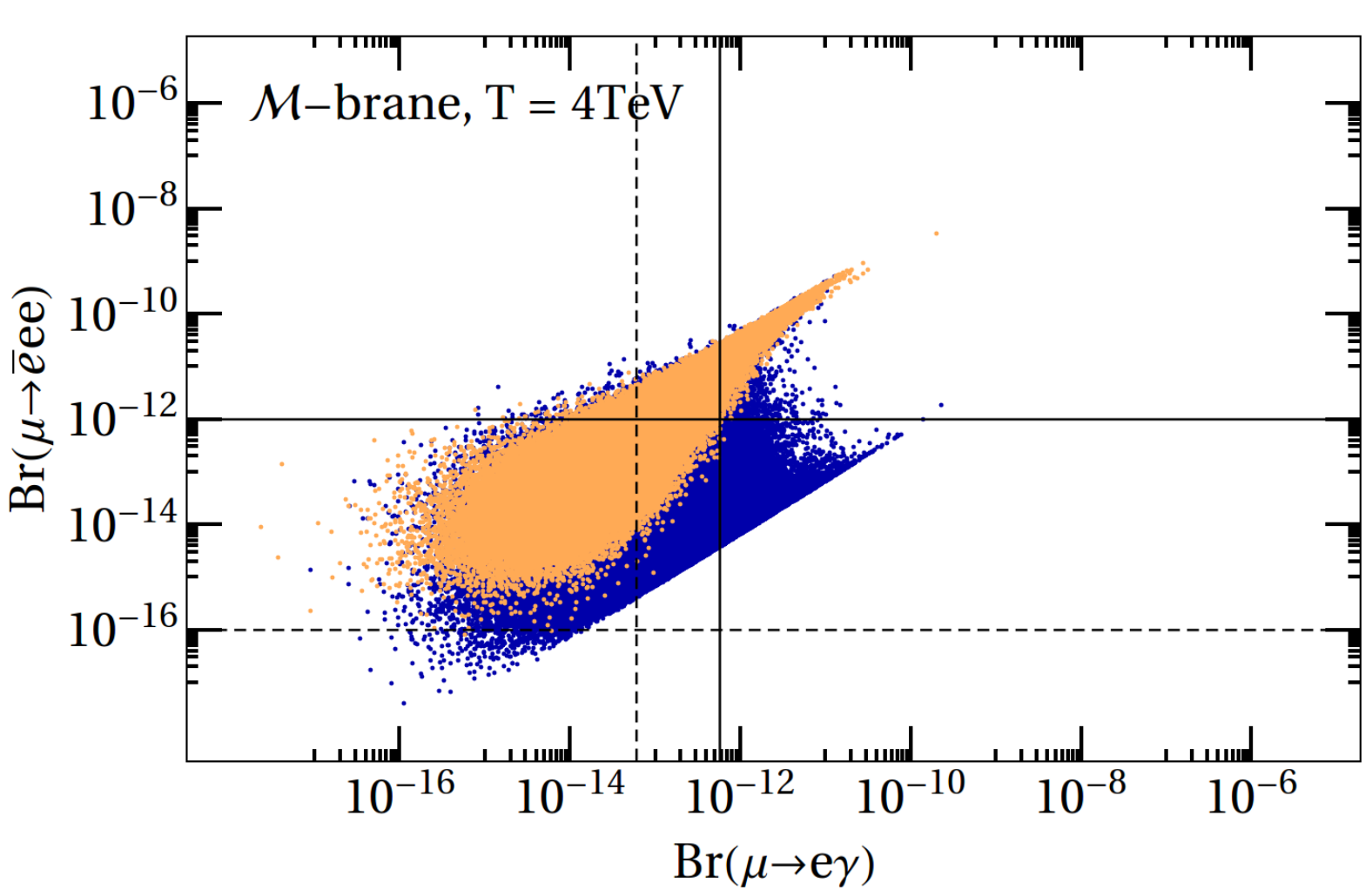}
\end{minipage}
\vskip0.3cm
\begin{minipage}{0.47\textwidth}
   \includegraphics[width=1\textwidth]{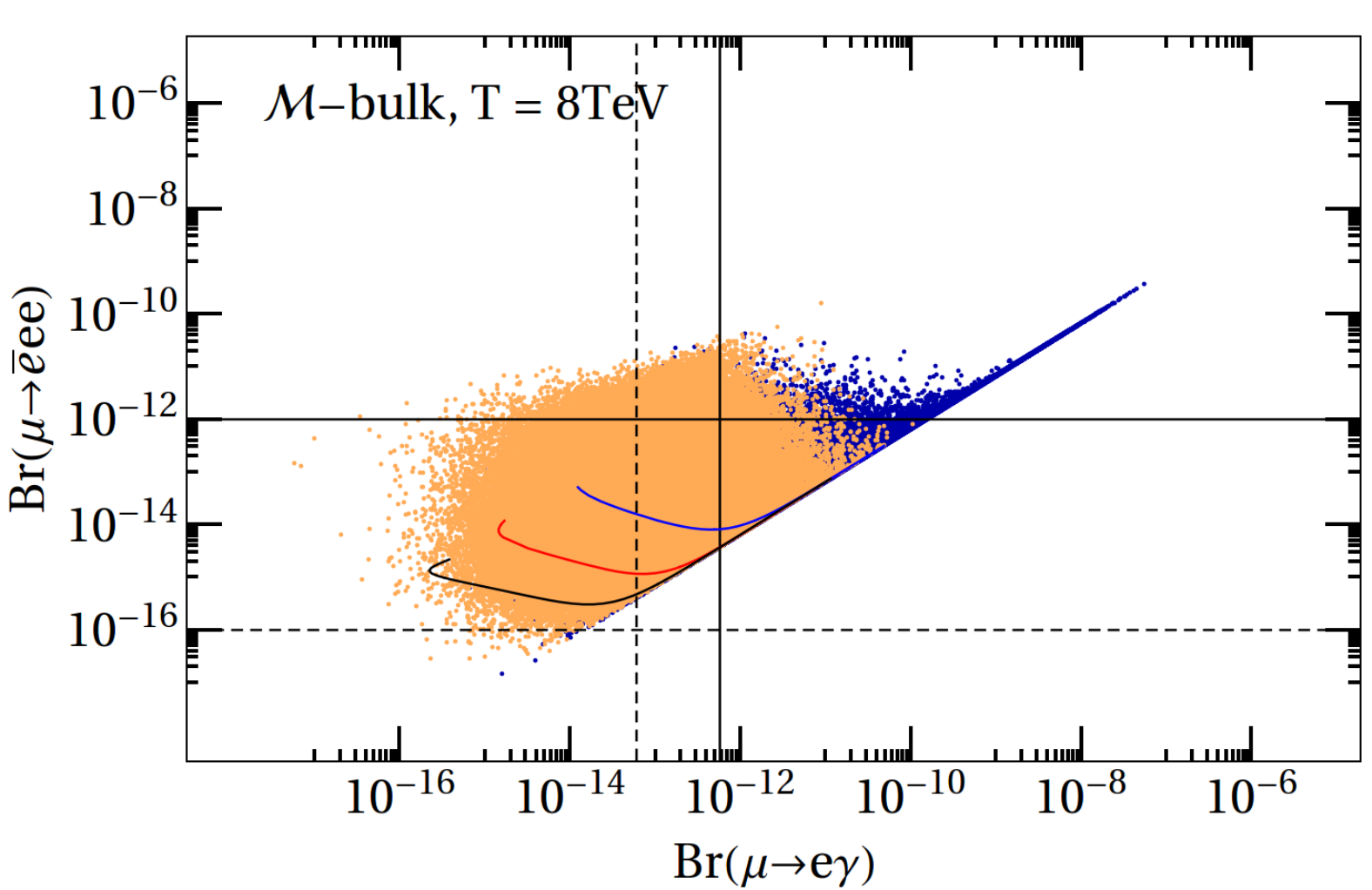}
\end{minipage}
\begin{minipage}{0.47\textwidth}
   \includegraphics[width=1\textwidth]{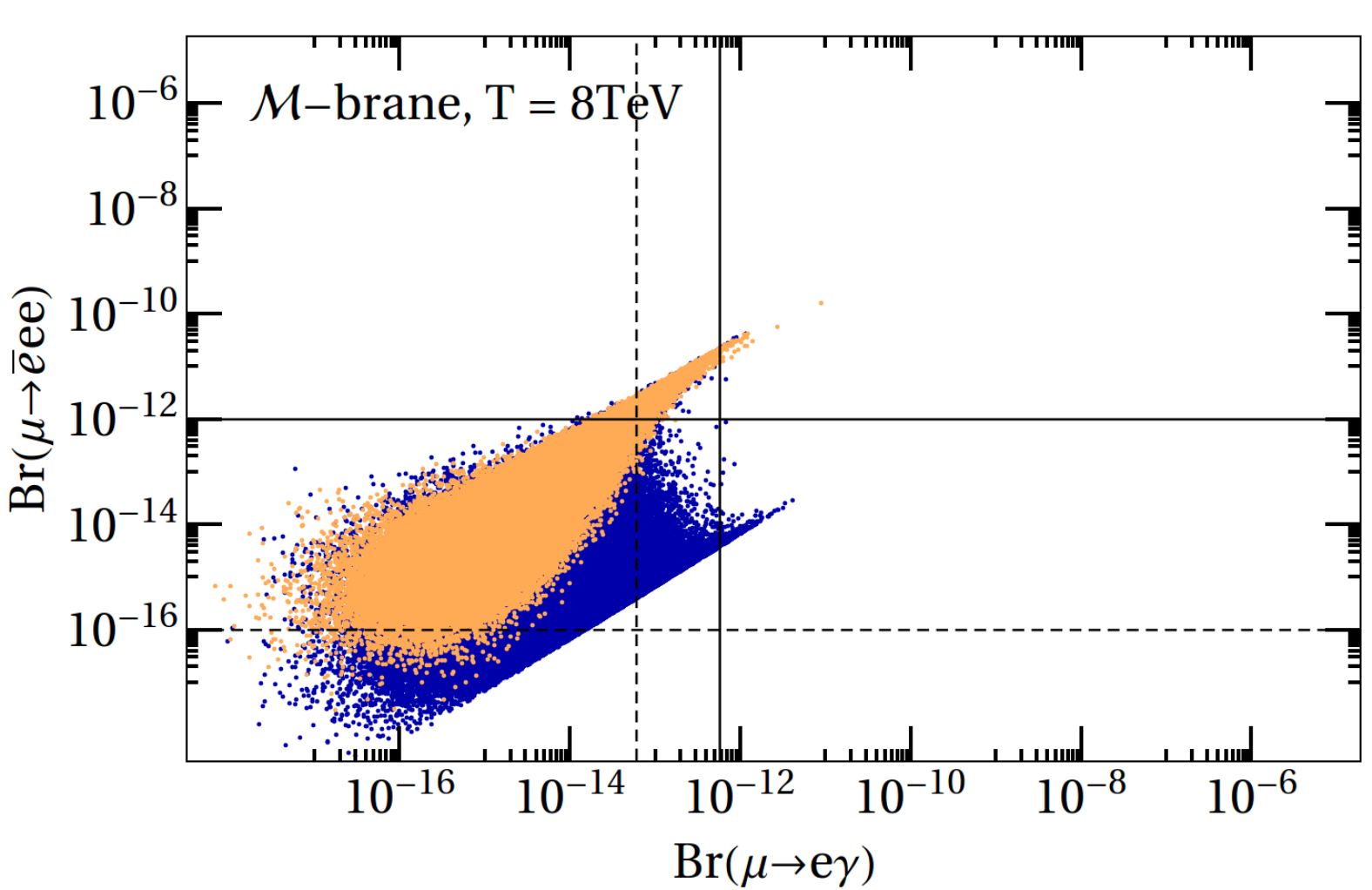}
\end{minipage}
\vskip0.1cm
\caption{\label{fig:MinimalMEGandM3e}  
2D scatter plots of the branching ratios $\mu\to e \gamma$ and 
$\mu \to 3 e$ for fixed $T$ ($4\,{\rm TeV}$ upper row, $8\,{\rm TeV}$ lower 
row) with $Y_{\rm max} = 0.5$ in orange (light grey) and $Y_{\rm max} = 3$ 
in blue (dark grey), respectively.  The {\it left panels} show the results 
for the \mbu models, the {\it right panels} for the \mbr\ ones. The current 
experimental bounds on the branching fractions are given by solid lines. 
The region above/to the left is excluded. The 
sensitivity of future experiments is shown by the dashed lines.}
\end{figure}

Figure \ref{fig:MinimalMEGandM3e} shows the values and correlation 
of the $\mu\to e \gamma$ and $\mu \to 3 e$ branching fractions in the 
minimal RS model for two different values of $T$,
$T=4\,{\rm TeV}$ (top) and  $T=8\,{\rm TeV}$ (bottom). $T=4\,{\rm TeV}$
is also roughly the lower bound on the KK scale from electroweak 
precision observables \cite{Casagrande:2008hr}. The left panels correspond 
to the \mbu\ scenario, the right panels to the \mbr\ case. Each plot 
shows the results for  $Y_{\rm max}=3$ in blue (dark grey)
and  for $Y_{\rm max}=1/2$ in orange (light grey). The current and expected 
future experimental upper bounds are shown by solid and dashed lines, 
respectively.

All four plots feature a sharp lower bound for $\mu \to 3e$ given the 
$\mu\to e\gamma$ rate, which is precisely given by the relation 
(\ref{eq:dipoleDomRelation}). $\mu \to 3e$ branching fraction values 
in the vicinity of this bound are dominated by the contributions from 
the dipole operator. For very large dipole coefficients or equivalently 
very large $\mu\to e \gamma$ branching fraction, the tri-lepton decay 
is always dominated by the dipole, and the two observables are strongly 
correlated. This generates the prominent thin 
line directed to the upper-right in the  \mbu\ model with large Yukawa 
couplings.

In the bulk Higgs case  $\mu\to e \gamma$  is, as expected, quite sensitive 
to the upper bound $Y_{\rm max}$.  This is a consequence of the 
$Y Y^\dagger Y$ terms in the dipole coefficient. They are naturally 
flavour-violating and scale with $Y_{\rm max}^2$. 
Consequently, the scan with larger Yukawa entries includes points with
substantially larger $\mu\to e \gamma$ branching fraction than the 
small Yukawa coupling scan. However, the dipole coefficient has two 
components. While the Higgs and the small Barr-Zee contributions scale 
as $Y_{\rm max}^2$ and vanish when the 5D Yukawa couplings go to zero, 
the gauge boson exchange contribution is not very sensitive to the Yukawa 
coupling size. In fact, for a generic anarchic Yukawa it grows
mildly with decreasing Yukawa size, see Figure~\ref{fig:GaugeDIpoleEffectSym}. 
Thus there has to be a smooth transition from the ''Higgs-dominated'' to the 
''gauge-dominated'' regime when the Yukawa coupling decreases.

To illustrate this point we included three curves 
in the plots for the bulk Higgs case defined as follows.
We chose three (random) Yukawa matrices with $Y_{\rm max}=3$ and scale 
the matrices down to $Y_{\rm max}=0.25$, keeping the relative size 
of the matrix entries fixed. The curves show the resulting trajectories.
For large Yukawa couplings the curves all run close to the 
dipole dominance bound.  With decreasing Yukawa couplings 
${\text{Br}}(\mu \to e \gamma)$ and ${\text{Br}}(\mu \to 3 e)$ first also 
decrease following the change in the dipole coefficient.
Then the growing effects of the tree-level operators begin to dominate
$\mu \to 3 e$ and the corresponding branching fraction begins to increase, 
while ${\text{Br}}(\mu \to e \gamma)$ continues to decrease. 
For even smaller Yukawa coupling the gauge-boson exchange contribution to  
${\text{Br}}(\mu \to e \gamma)$ exceeds the rapidly decreasing Higgs 
contribution and  ${\text{Br}}(\mu \to e \gamma)$ reaches a hard lower 
limit. 

The exactly brane-localized Higgs case displayed in the right panels of 
Figure~\ref{fig:MinimalMEGandM3e} behaves the same in this respect. 
However, since the leading Higgs contribution is suppressed for the exactly 
brane-localized Higgs, the range of values for ${\rm Br}(\mu\to e \gamma)$ is 
almost independent of $Y_{\rm max}$. A change in $Y_{\rm max}$ 
predominantly affects ${\rm Br}(\mu\to 3e)$, which increases for smaller 
$Y_{\rm max}$ due to the 
larger coefficients of four-fermion and fermion-Higgs operators. This also 
explains the drop shape of the scatter plot for $Y_{\rm max}=1/2$. Points 
with large $\mu\to 3e$ branching fraction arise from 
large tree-level Wilson coefficients either due to the structure of the 
Yukawa matrix or due to accidentally small couplings.
In both cases the process $\mu\to e \gamma$ also receives sizeable 
contributions from the tree operators
leading to the roughly linear correlation in the upper-right corner 
of the scatter points.

 \begin{figure}[t]
\begin{minipage}{0.47\textwidth}
   \includegraphics[width=1\textwidth]{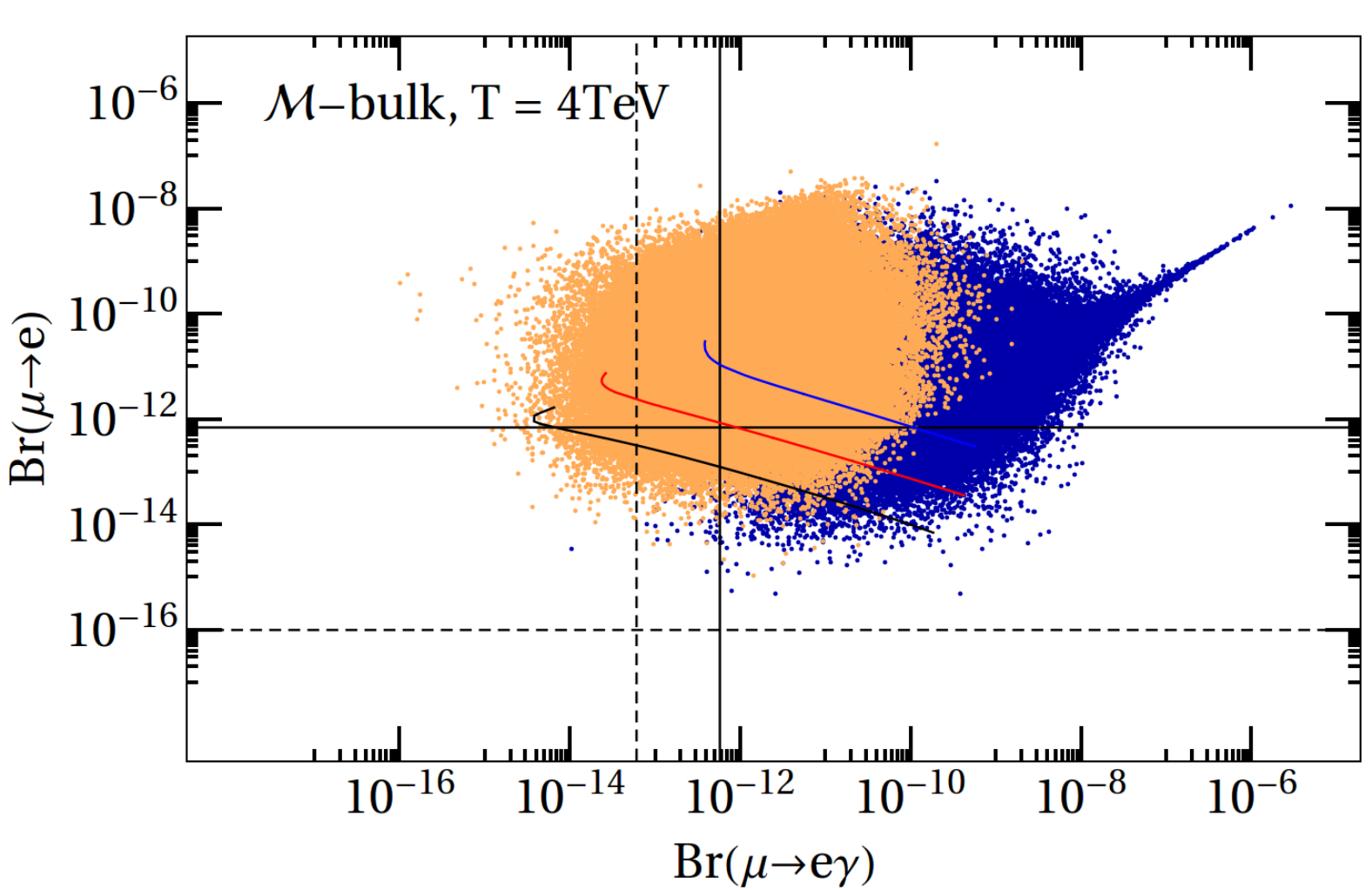}
\end{minipage}
\begin{minipage}{0.47\textwidth}
   \includegraphics[width=1\textwidth]{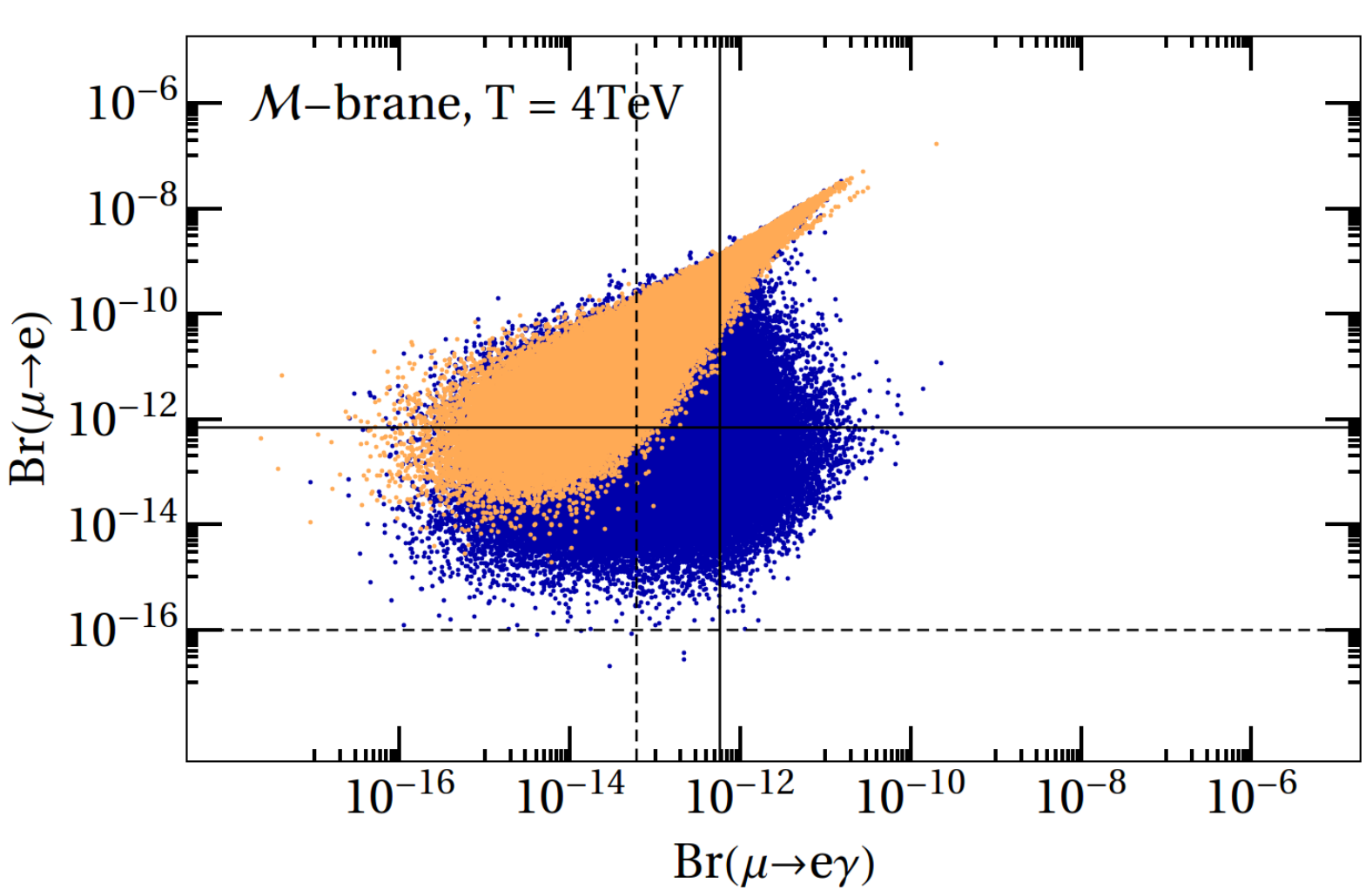}
\end{minipage}
\vskip0.3cm
\begin{minipage}{0.47\textwidth}
   \includegraphics[width=1\textwidth]{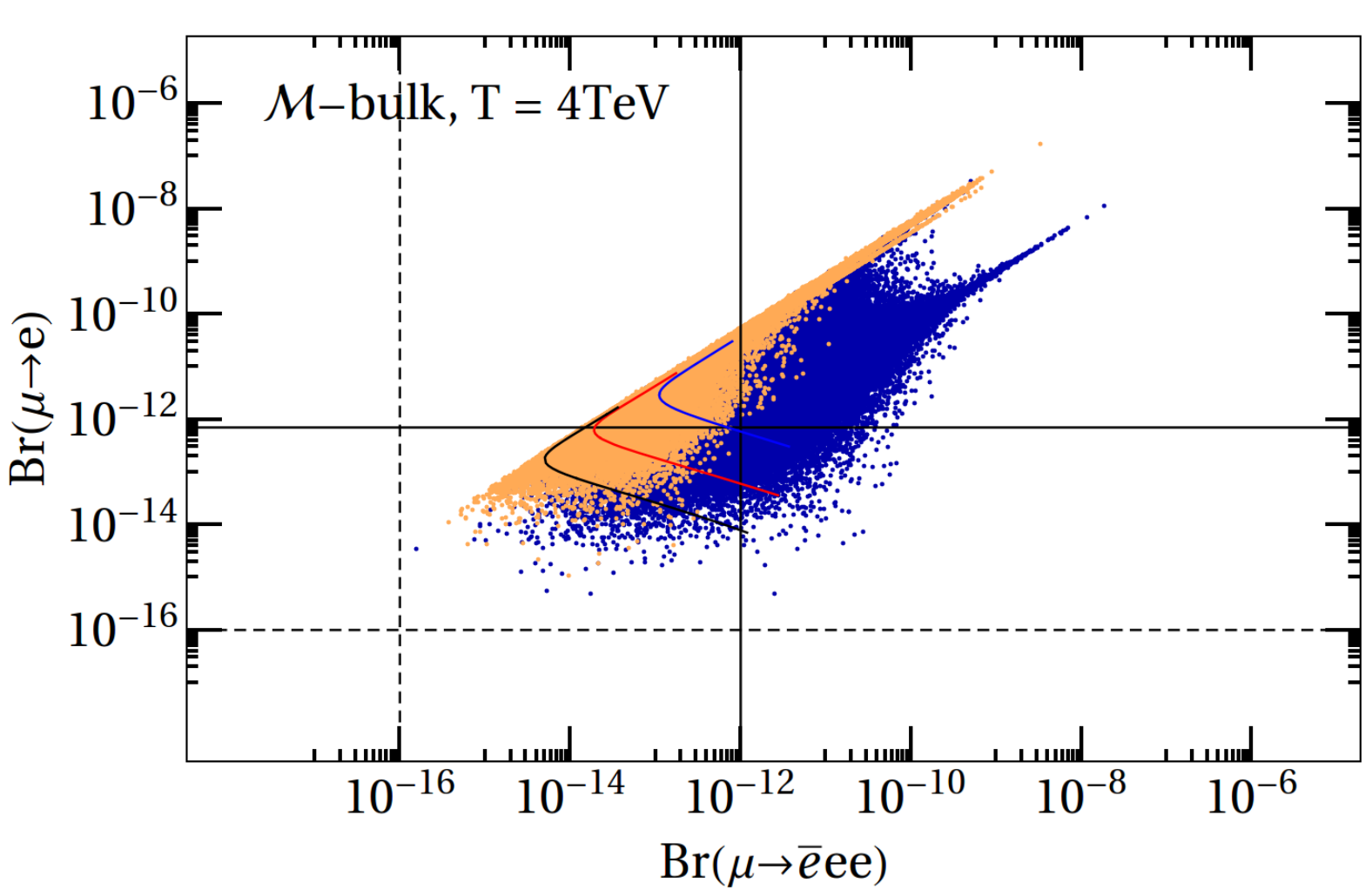}
\end{minipage}
\begin{minipage}{0.47\textwidth}
   \includegraphics[width=1\textwidth]{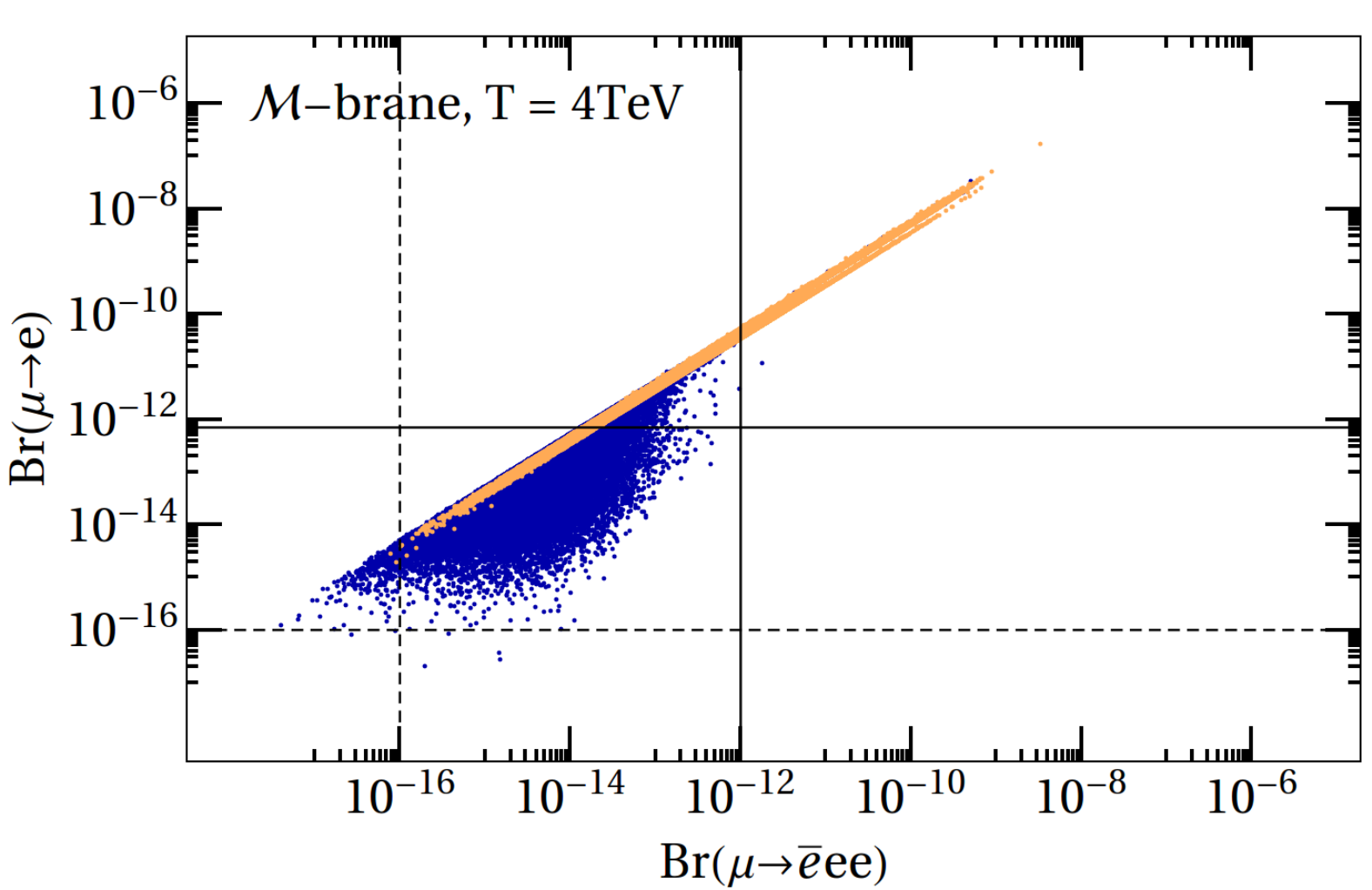}
\end{minipage}
\vskip0.1cm
\caption{\label{fig:MinimalMToECorrelation}  
Correlation of the branching ratios for $\mu\to e \gamma$ and $\mu \to e$ (upper row)
and for $\mu\to 3e $ and $\mu \to e$ (lower row), both for $T=4\;{\rm TeV}$.
The {\it left panels} show the results for the \mbu, the {\it right panels} 
for the \mbr\ model.}
\end{figure}

Values and correlations of $\mu \to 3e$ and $\mu\to e \gamma$ 
with muon conversion in gold are shown in 
Figure~\ref{fig:MinimalMToECorrelation} (colour coding as in the previous 
figure). The top row shows $\mu\to e \gamma$ against $\mu \to e$ 
(\mbu\ left, \mbr\ right). The two observables 
are essentially uncorrelated in the bulk Higgs 
scenario. This agrees with our previous observation that muon conversion 
is mostly insensitive to the dipole coefficient $a^A_{ij}$ which governs the 
$\mu\to e \gamma$ branching fraction.
Only in rare cases is the dipole operator large enough to dominate also 
muon conversion leading to the noticeable spike towards the right in the 
upper-left plot. In the exactly brane-localized Higgs case (upper-right 
panel), correlations are  absent only for $Y_\star=3$.
As mentioned before, for $Y_\star=1/2$ $\mu\to e \gamma$ 
receives non-negligible contributions from tree-level operators, which 
manifests itself in a weak correlation.

For the same reasons $\mu \to 3e$ and $\mu \to e$ (bottom row) are strongly 
correlated for small Yukawa couplings in \mbr\ model, but only 
feature a lower bound on the branching 
fraction of $\mu \to 3e$ for a given ${\text{Br}}(\mu \to e)$ 
in the other scenarios.
As noted in the previous subsection, the branching 
fraction of $\mu \to e$ 
decreases with increasing values for $Y_\star$. 
This effect can best be seen in 
the upper left panel of Figure~\ref{fig:MinimalMToECorrelation}.
The slopes of the three sample trajectories also verify this effect.

\subsubsection{Custodially protected model} 

\begin{figure}[p]
\begin{minipage}{0.47\textwidth}
   \includegraphics[width=1\textwidth]{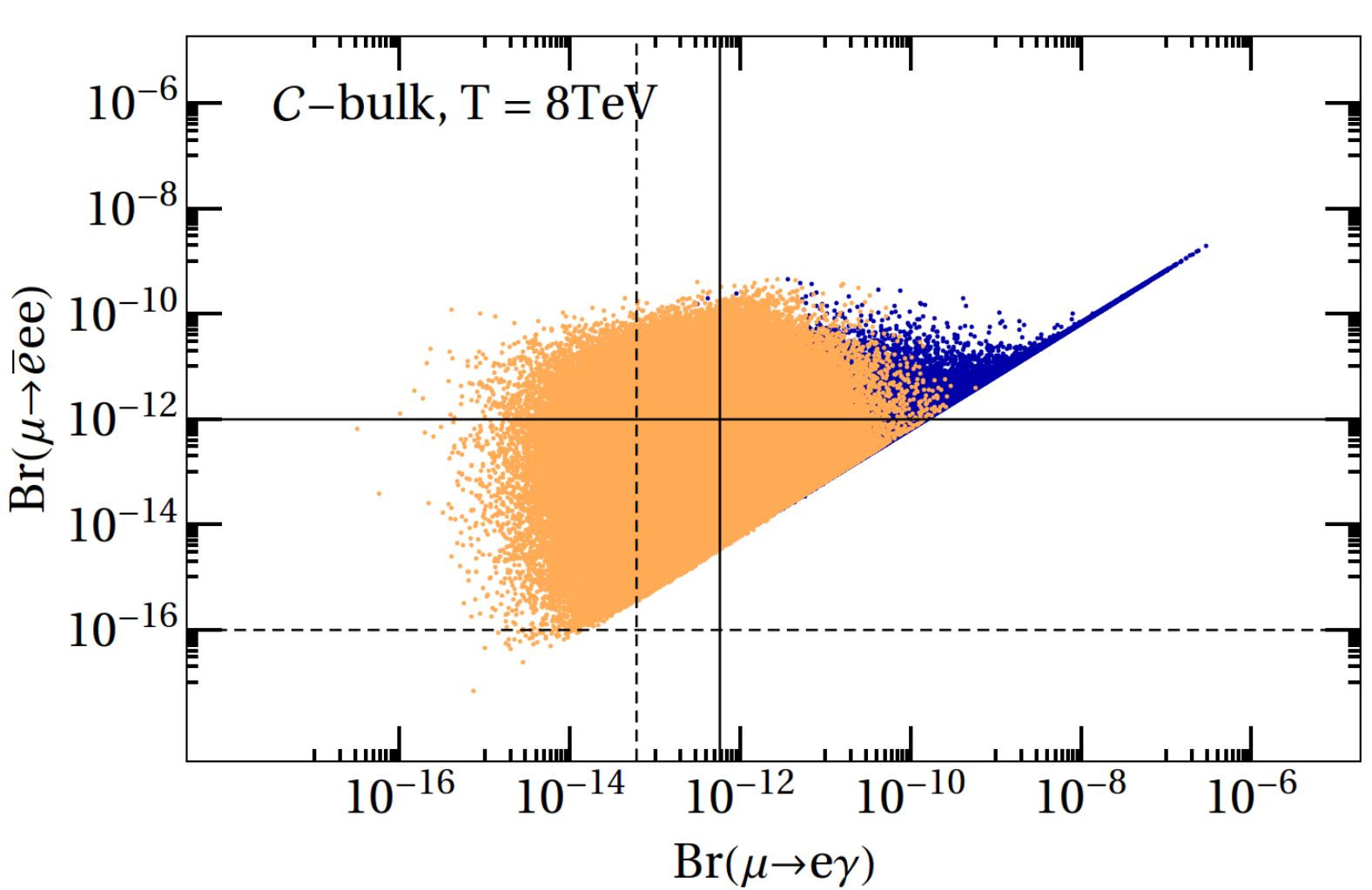}
\end{minipage}
\begin{minipage}{0.47\textwidth}
   \includegraphics[width=1\textwidth]{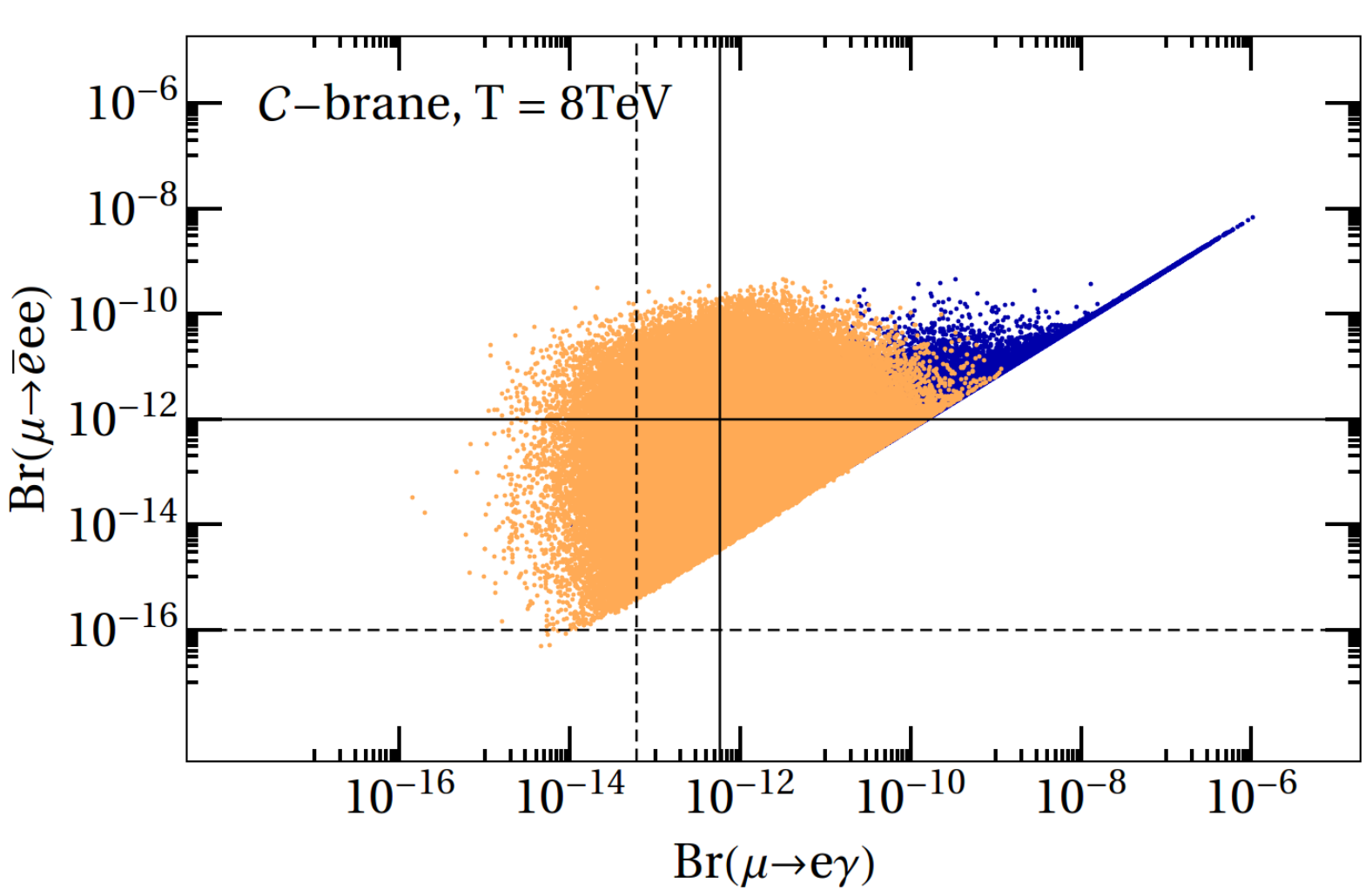}
\end{minipage}
\vskip0.3cm
\begin{minipage}{0.47\textwidth}
   \includegraphics[width=1\textwidth]{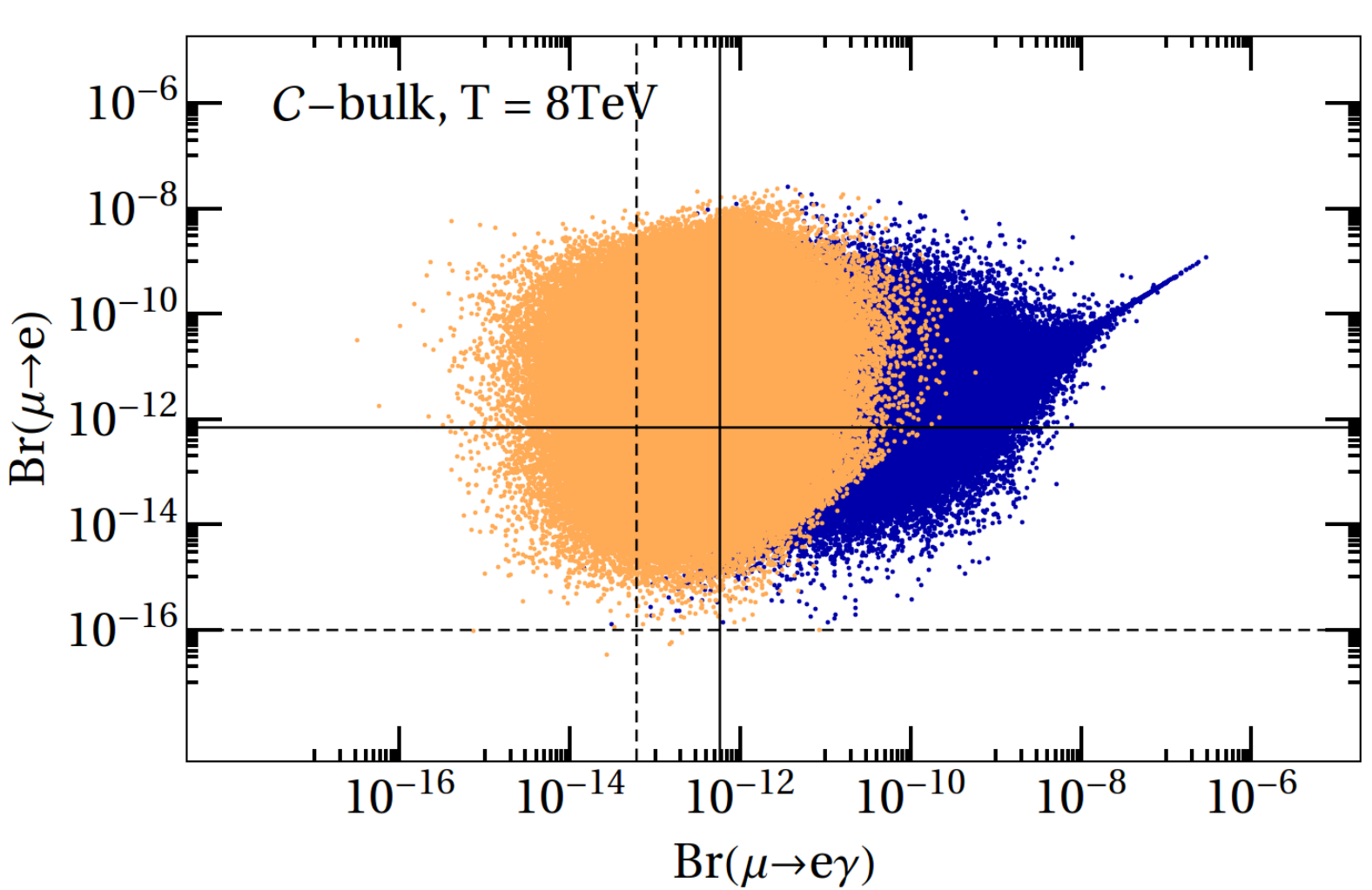}
\end{minipage}
\begin{minipage}{0.47\textwidth}
   \includegraphics[width=1\textwidth]{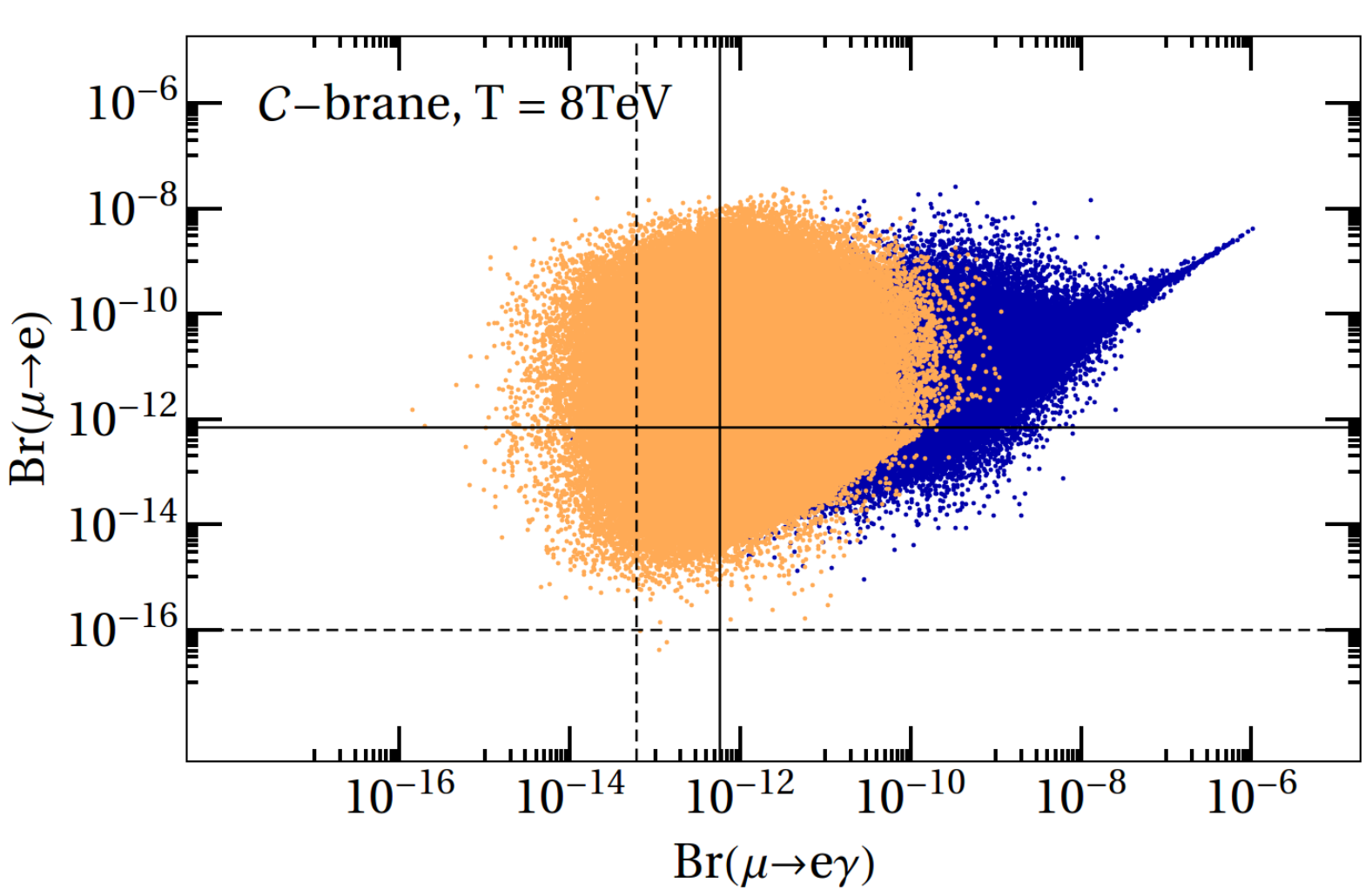}
\end{minipage}
\vskip0.3cm
\begin{minipage}{0.47\textwidth}
   \includegraphics[width=1\textwidth]{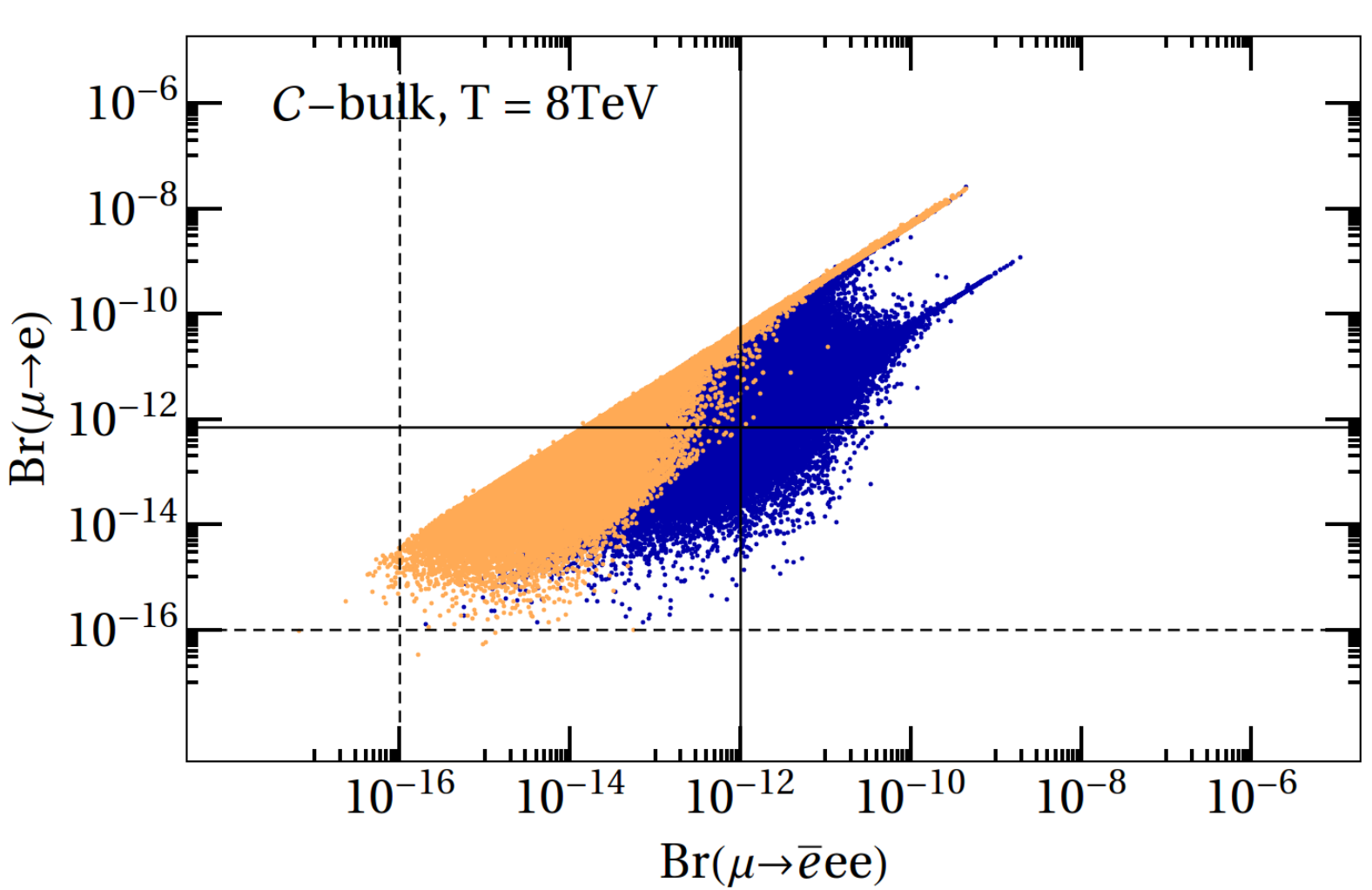}                              
\end{minipage}
\begin{minipage}{0.47\textwidth}
   \includegraphics[width=1\textwidth]{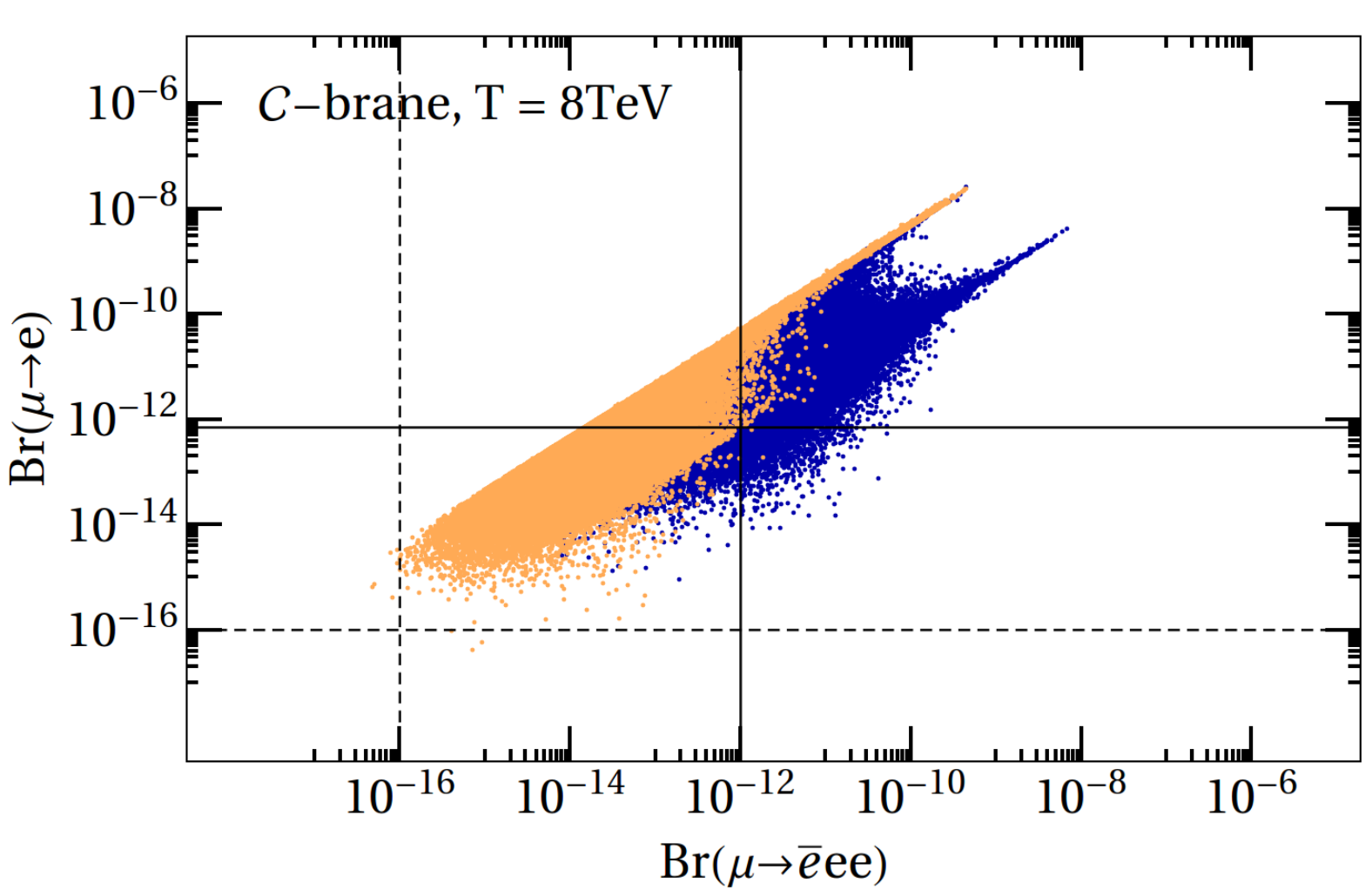}
\end{minipage}
\vskip0.1cm
\caption{\label{fig:CSCorrelation}  
Correlation of the branching ratios for $\mu\to e \gamma$ and $\mu \to 3e$ 
(top row), $\mu\to e \gamma$ and $\mu \to e$ (middle row), and $\mu\to 3e $ 
and $\mu \to e$ (bottom row) for $T=8\;{\rm TeV}$ in the custodially protected 
model. Model points with $Y_{\rm max}=1/2$ 
are indicated by orange (light grey) points. For $Y_{\rm max}=3$ we use blue 
(dark grey) points. The {\it left panels} show the results for the \cbu, 
the {\it right panels} for the \cbr\ model.}
\end{figure}

Figure~\ref{fig:CSCorrelation} shows the combined results for 
the custodially protected model. The left panels correspond to 
the bulk Higgs model \cbu\ and the right panels to the exactly 
brane-localized model \cbr.
The colour coding is the same as above.
Here the KK scale $T$ was fixed to $8\,\rm TeV$, since 
for $T$ around $4\, {\rm TeV}$ it is already non-trivial
to find points, which are not in conflict with the muon conversion 
bound. 

The broad picture for model \cbu\ is almost the same as for \mbu.
The shape of the distributions does not change.
Quantitatively, the custodially protected model generates significantly larger 
branching fractions. In particular, the  $\mu \to e \gamma$ branching 
fraction, which is most sensitive to the magnitude of the dipole 
operator coefficient, is typically enhanced by a factor of about five.
This was to be expected as the main difference in the 
custodially protected model is a larger gauge- and Higgs-contribution 
to the dipole coefficient $a^A_{ij}$.  Again the dipole operator creates 
a correlation of $\mu \to e\gamma$ and
$\mu \to 3 e$ especially for the larger value of $Y_{\rm max}$.

For the \cbr\ model the distributions are very different from the 
result in the \mbr\ model. This is a consequence of the 
new additional terms~\eqref{eq:HiggsresultsCustodialbrane} in $a^H_{ij}$.
This additional contribution to the dipole coefficient is only slightly 
larger than the corresponding contribution for a bulk Higgs.
The phenomenology for bulk and brane Higgs 
case is therefore quite similar in the custodial RS model. The bounds 
imposed by the non-observation of LFV are comparable, although 
more restrictive for the exactly brane-localized Higgs.
 
The fact that the sign of $a^H_{ij}$ depends on the 
Higgs localization does not lead to a noticeable effect. If the 
dipole operator is dominated by the Higgs contribution, a sign flip of 
the coefficients $a_{ij}$ only affects terms in \eqref{eq:BrMEG}, 
\eqref{eq:Brmu3e} and \eqref{eq:BRmToe} that come from an interference 
of the dipole with a four-fermion operator. In general, these terms do 
not provide the dominant contribution to the branching 
fractions. The situation would be different if the RS contribution could 
interfere with a sizeable SM contribution to LFV observables. An observed
enhanced or reduced rate could then be used to discriminate the brane from  
the bulk Higgs model.\footnote{This is precisely what is observed in Higgs 
production, see e.g.~\cite{Malm:2013jia}.}
 
\subsubsection{EDM constraint}

The randomly sampled Yukawa matrices also generate electric
dipole moments (EDMs) of the leptons through the non-hermitian
part of $\alpha_{ij}^A$, see (\ref{LFVLagrangianInBrokenPhase}).
In case of the electron the relation is given by
\begin{equation}
d_e =  \frac{Q_e e}{2 m_e} \,F_3(0) = 
m_e \, i \,(A_R-A_L)\,,
\end{equation}
where the form factor $F_3$ now refers to the flavour diagonal 
electron-photon vertex, and $A_{L,R}$ are defined as in (\ref{ARtotal}), 
(\ref{ALtotal}) with subscripts 12 replaced by 11, and $m_\mu$ by $m_e$.

\begin{figure}[p]
\begin{minipage}{0.47\textwidth}
   \includegraphics[width=1\textwidth]{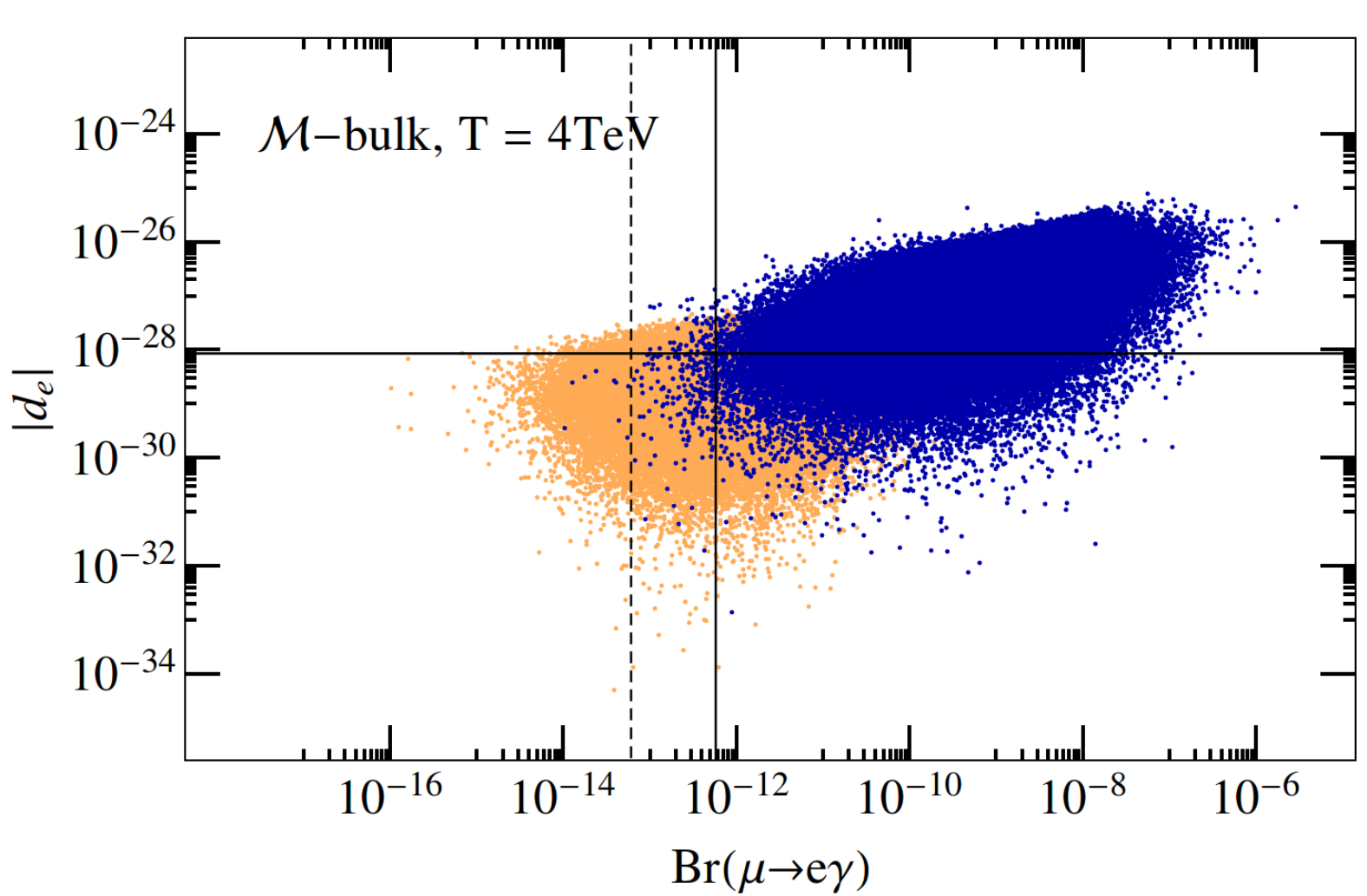}
\end{minipage}
\begin{minipage}{0.47\textwidth}
   \includegraphics[width=1\textwidth]{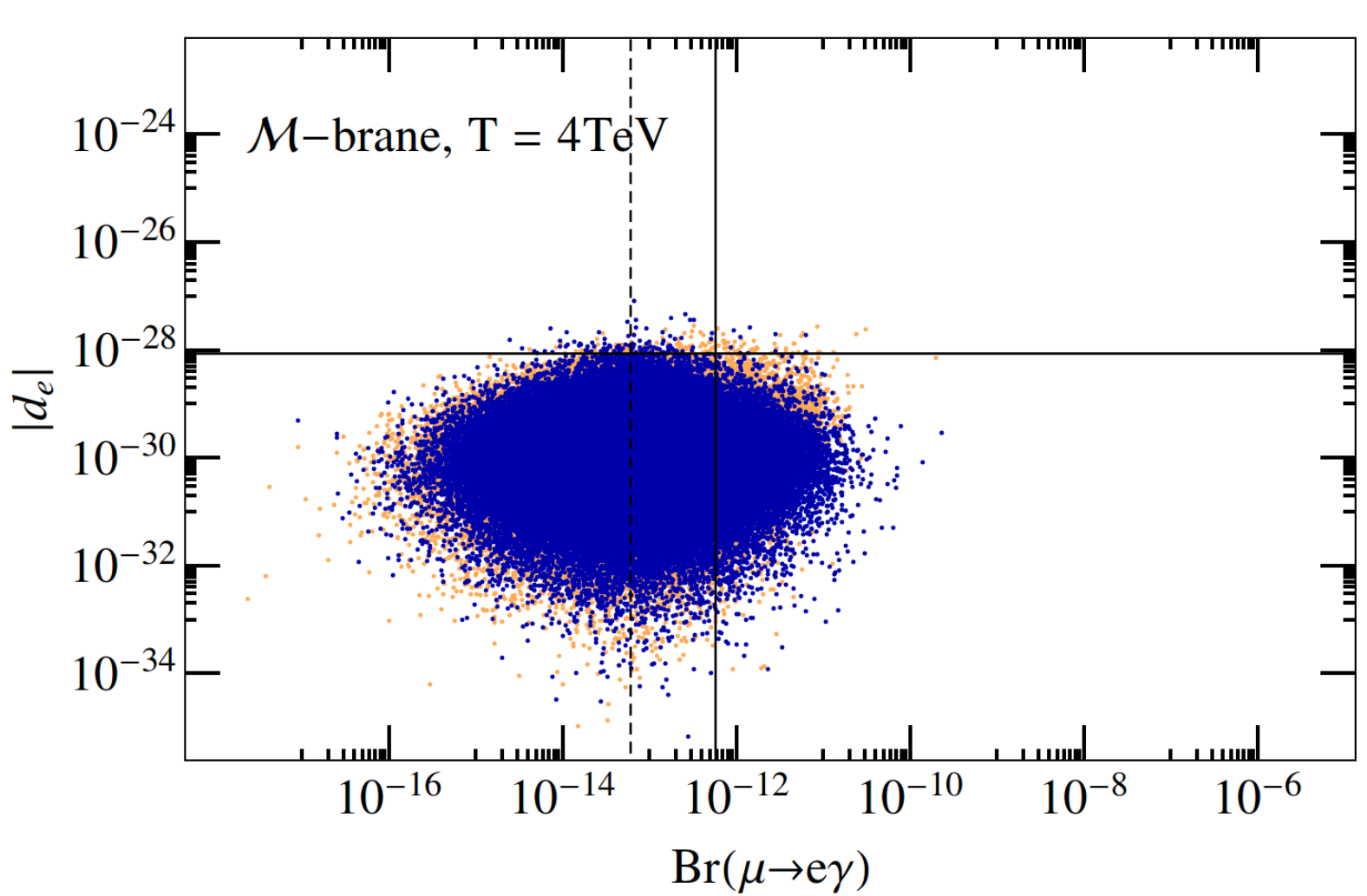}
\end{minipage}
\vskip0.3cm
\begin{minipage}{0.47\textwidth}
   \includegraphics[width=1\textwidth]{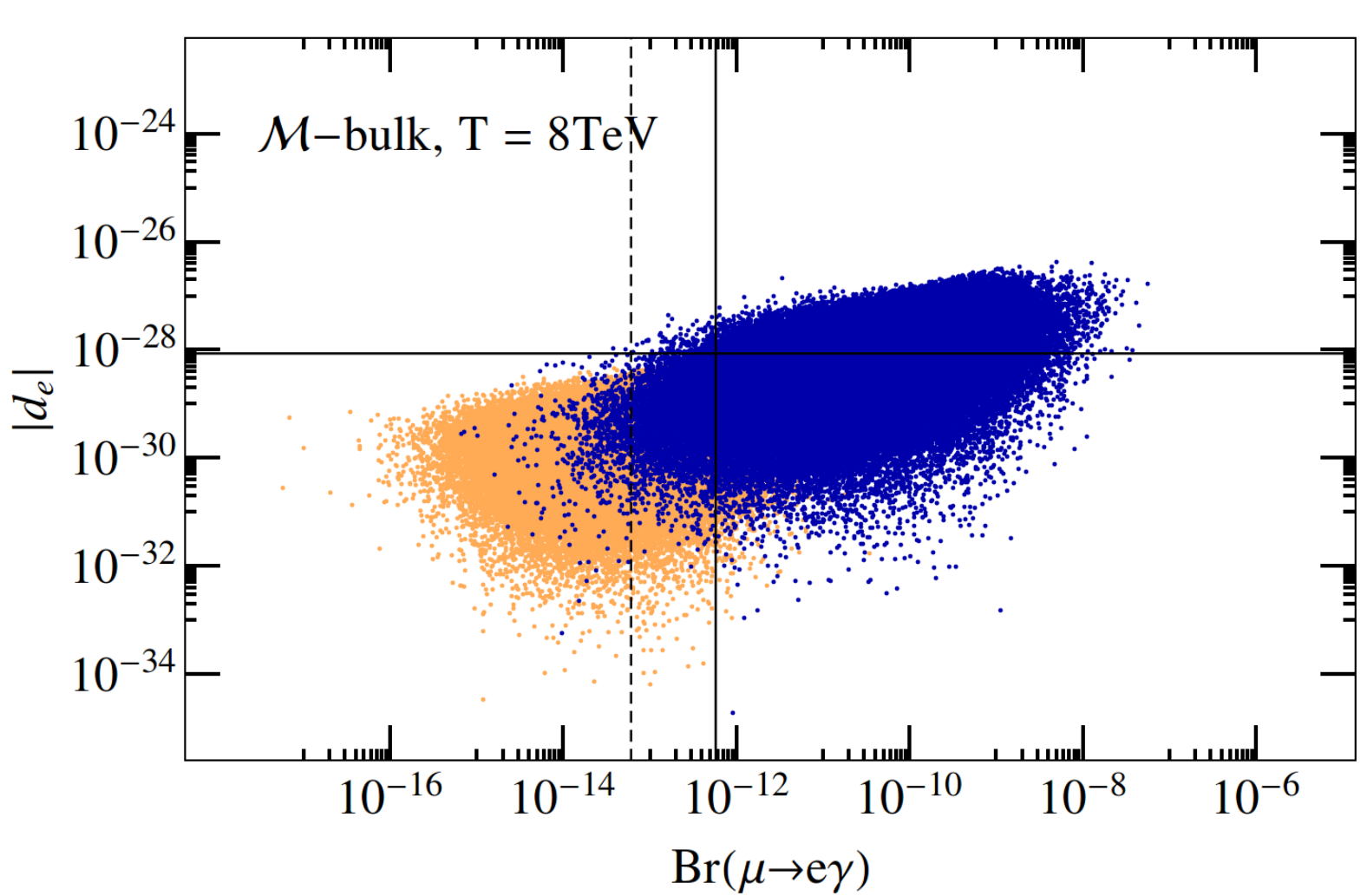}
\end{minipage}
\begin{minipage}{0.47\textwidth}
   \includegraphics[width=1\textwidth]{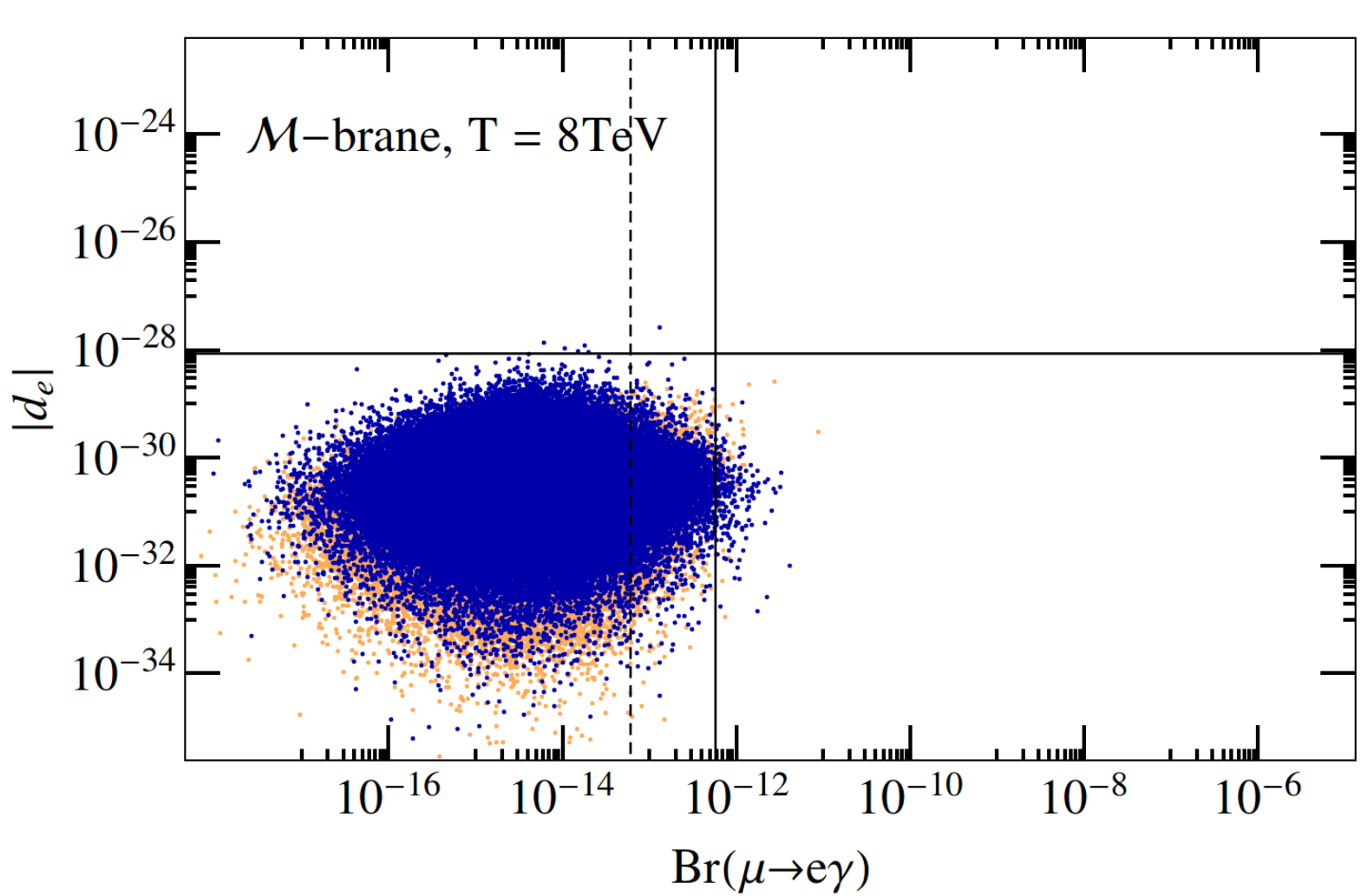}
\end{minipage}
\vskip0.3cm
\begin{minipage}{0.47\textwidth}
   \includegraphics[width=1\textwidth]{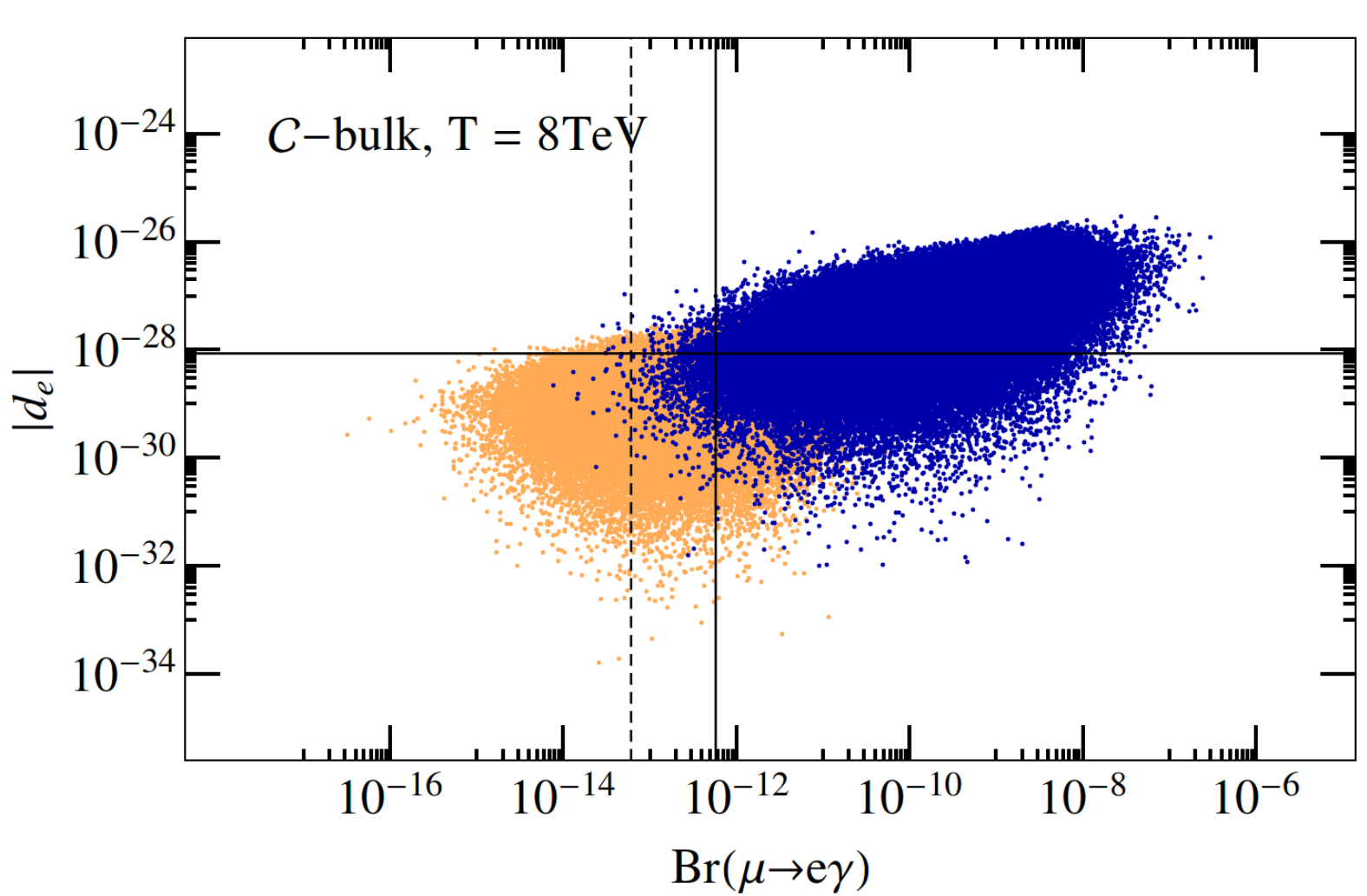}                              
\end{minipage}
\begin{minipage}{0.47\textwidth}
   \includegraphics[width=1\textwidth]{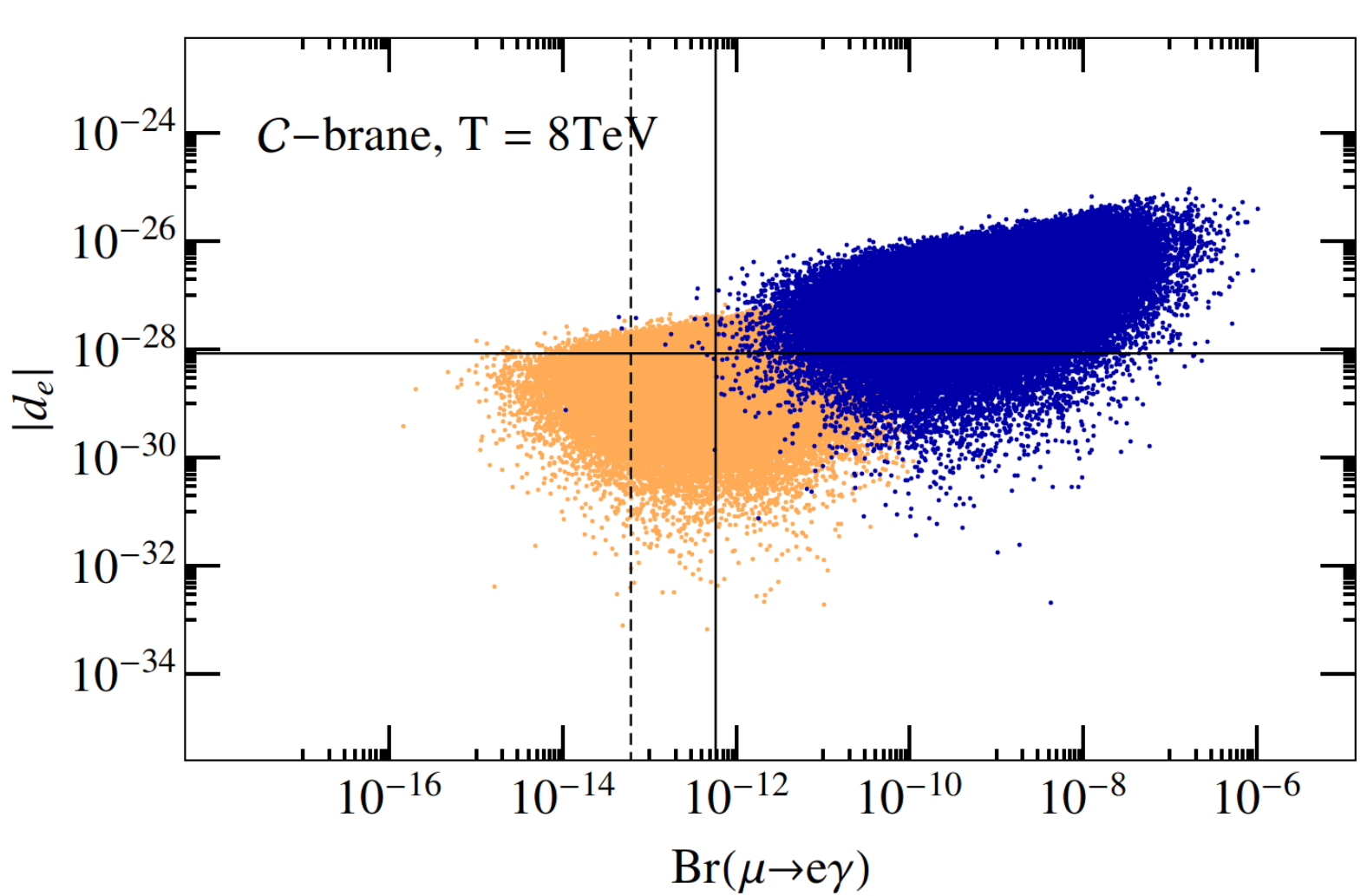}
\end{minipage}
\vskip0.1cm
\caption{\label{fig:EDMCorrelation1}  
Correlation of the electron EDM with the $\mu\to e \gamma$ 
branching ratio for all six scenarios. Data sets with $Y_\star=1/2$ 
are indicated by orange (light grey) points. For $Y_\star=3$ we use blue 
(dark grey) points. The {\it left panels} show the results for the bulk Higgs, 
the {\it right panels} for the exactly brane-localized Higgs.}
\end{figure}

We checked that the present experimental limit on the
electron electric dipole moment~\cite{Baron:2013eja},
\begin{equation}
\label{eldipoleconstraint}
|d_e| < 8.7 \cdot 10^{-29} \,e \;\mbox{cm} \quad \mbox{(at 90\% CL)}\,,
\end{equation}
does not affect our conclusions. That is, while up to 
90\% of the randomly scanned model points in the sample with 
$Y_{\rm max}=3$ (and up to 25\% in the $Y_{\rm max}=0.5$ sample) 
fail the EDM constraint for the bulk Higgs models, the EDM 
and the LFV observables are uncorrelated, so that the ranges 
covered by the scatter plots look almost exactly the same, when the points 
failing the EDM constraint are excluded. Only the 
tips of some of the spikes pointing to the upper-right in the scatter 
plots are cut-off in the $Y_{\rm max}=3$ sample, but these points 
are also excluded by LFV constraints (see solid lines in the previous 
figures). The predicted values of $|d_e|$ and the $\mu\to e\gamma$ 
branching fraction are shown in Figure~\ref{fig:EDMCorrelation1} for 
the different RS models and values of $T$ considered in this analysis. 
The featureless shape of the area filled by the sampled points illustrates the 
lack of specific correlations. The scaling of both observables with 
$Y_\star$ is visible (orange/light grey vs. blue/dark grey sample) 
whenever there is an unsuppressed contribution from Higgs exchange.

\begin{figure}[t]
\begin{minipage}{0.48\textwidth}
   \hskip-0.2cm\includegraphics[width=1\textwidth]{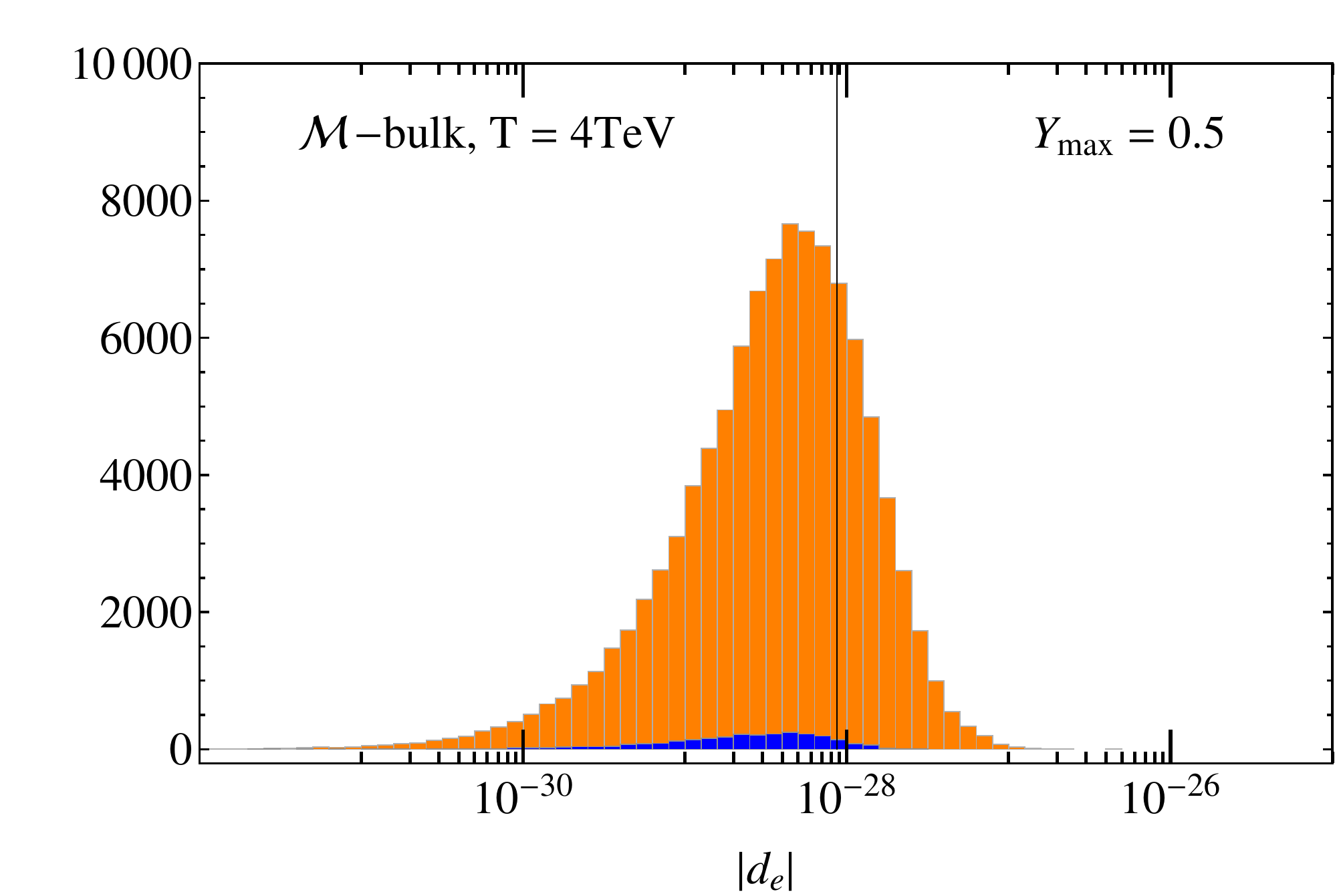}
\end{minipage}
\begin{minipage}{0.48\textwidth}
   \hskip-0.35cm\includegraphics[width=1\textwidth]{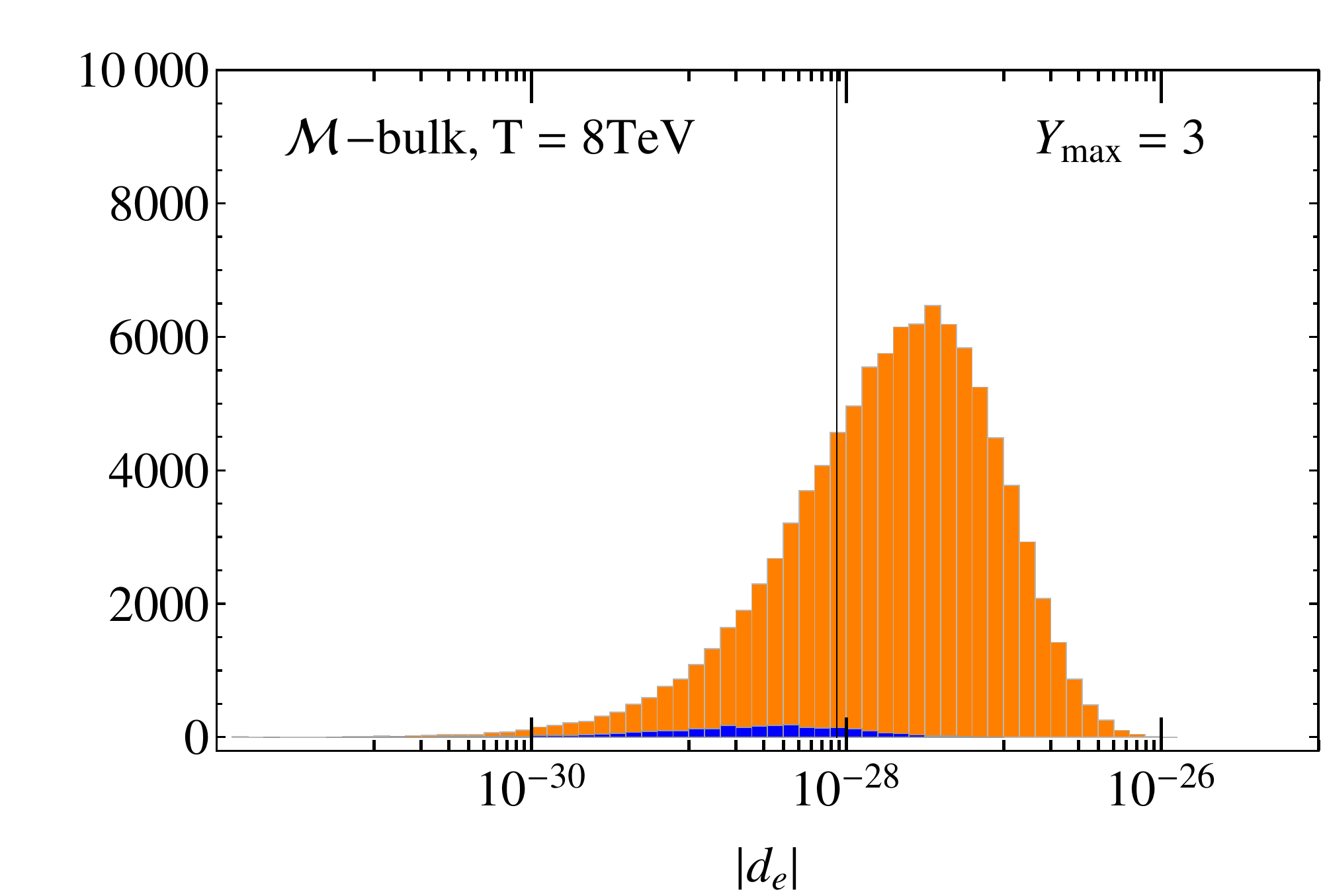}
\end{minipage}
\vskip0.3cm
\begin{minipage}{0.47\textwidth}
   \includegraphics[width=1\textwidth]{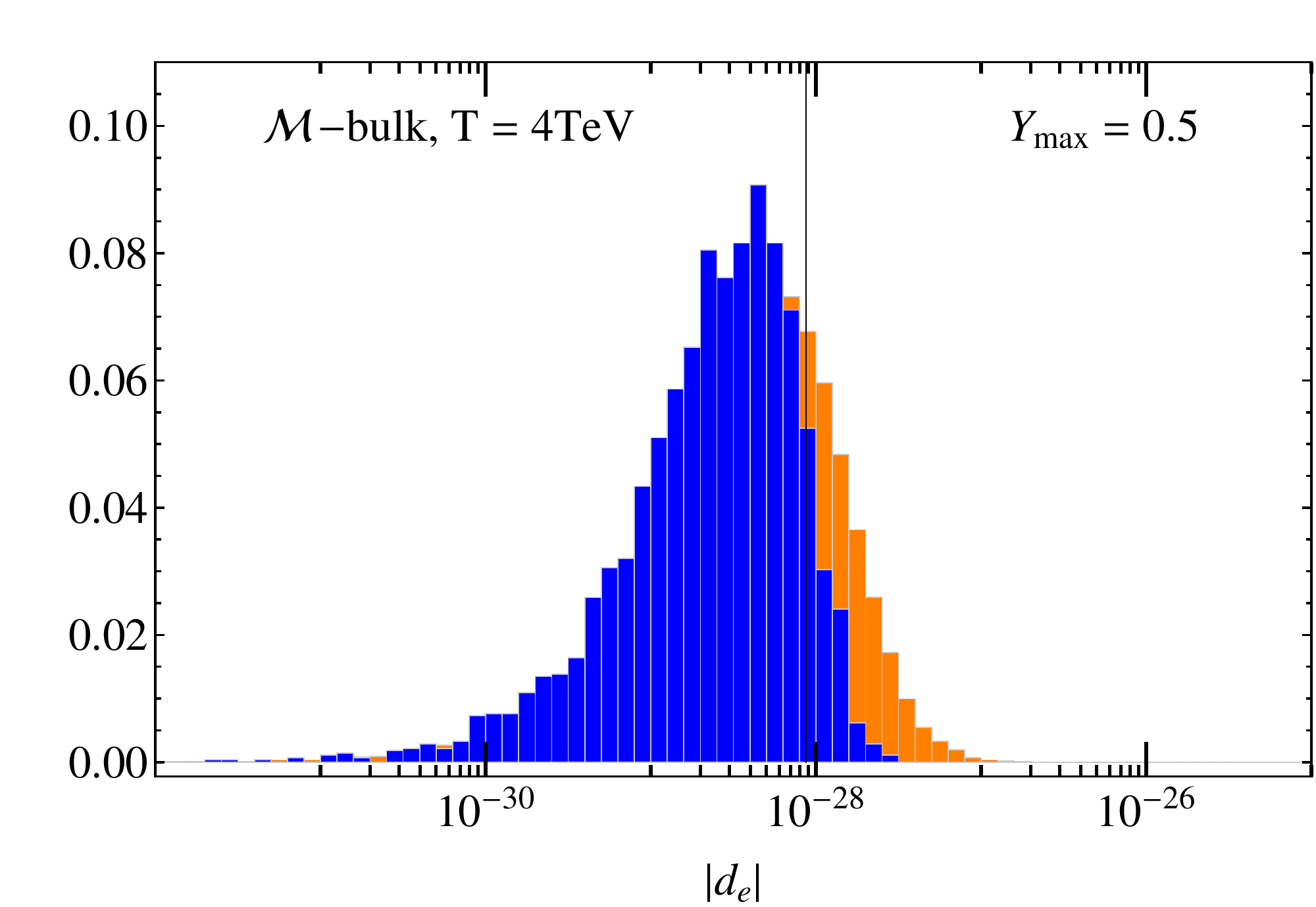}
\end{minipage}
\begin{minipage}{0.47\textwidth}
   \includegraphics[width=1\textwidth]{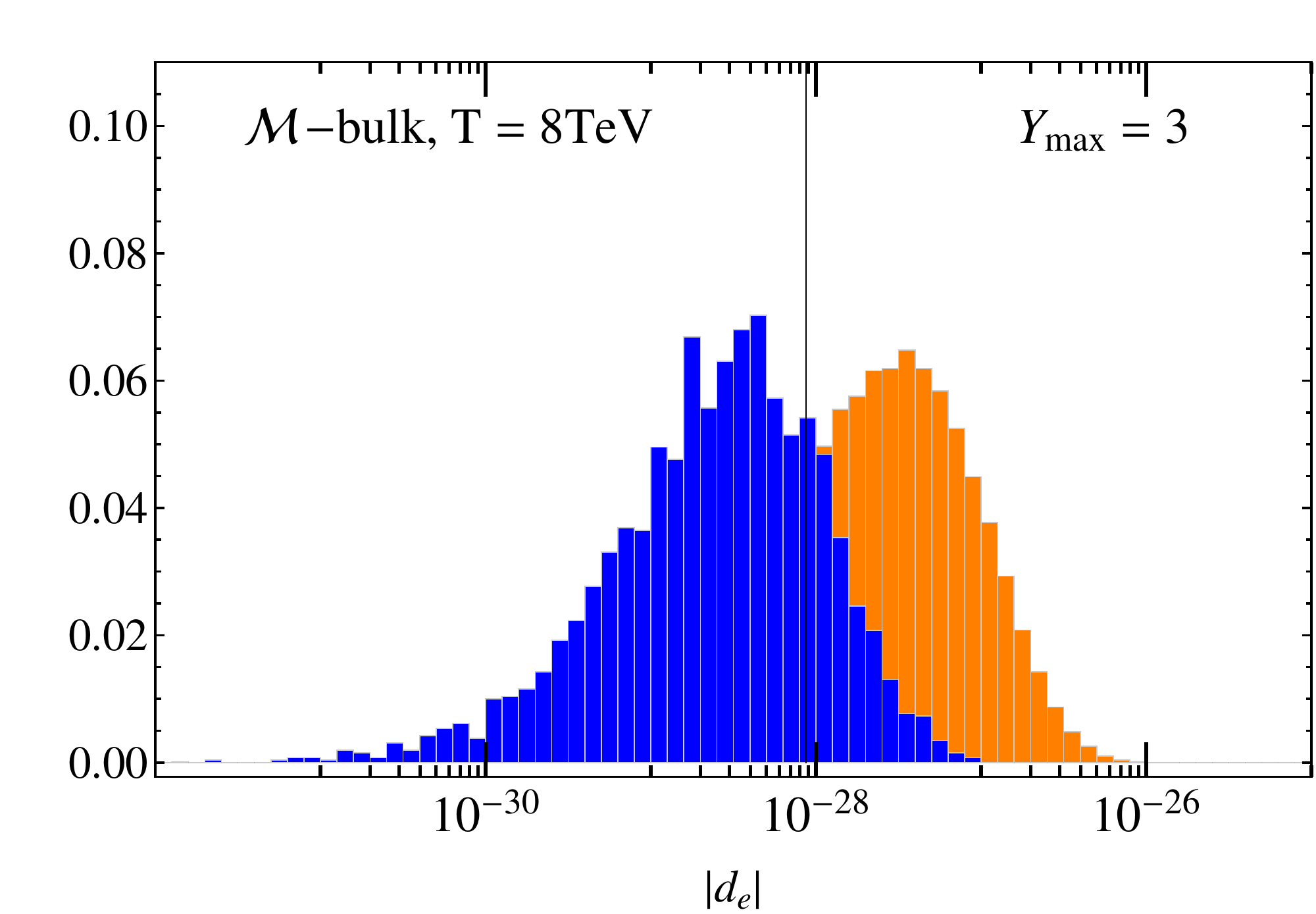}
\end{minipage}
\vskip0.1cm
\caption{\label{fig:EDM-LFV-interplay} 
Histograms of the electron EDM distribution of the \mbu\ sample 
before (orange/light grey) and after (darkgrey/blue) 
the LFV constraints are applied for $T=4\,$TeV, $Y_{\rm max}=0.5$ (left) 
and $T=8\,$TeV, $Y_{\rm max}=3$ (right). The first row shows the absolute 
number of representatives of the sample, the second the normalized 
distribution. The vertical line indicates the current bound 
(\ref{eldipoleconstraint}) on the EDM.}
\end{figure}

We can turn this question around and ask what are the expectations for 
value of the electron EDM after the LFV constraints on the RS models are taken 
into account. To this end we create histograms of $|d_e|$ for our samples 
before and after applying the {\em present$\,$} LFV constraints.  The \mbr\  
models are special, since the Higgs contribution is strongly suppressed. 
In these models 
the EDM distribution peaks at a few times $10^{-30} \,e \;\mbox{cm}$ 
for $T=4\,$TeV, well below the present limit (\ref{eldipoleconstraint}) 
and is hardly altered by the LFV constraints. The tail of the distribution 
above about a few times $10^{-29} \,e \;\mbox{cm}$ is, however, cut away 
independent of the value of the Yukawa coupling. 
This is different for the other models where the distribution and impact 
of constraints is sensitive to the maximally allowed Yukawa coupling 
$Y_{\rm max}$. This is illustrated on the example of the \mbu\ model 
in Figure~\ref{fig:EDM-LFV-interplay}, where the left two panels refer 
to $T=4\,$TeV, $Y_{\rm max}=0.5$ and the right ones to 
$T=8\,$TeV, $Y_{\rm max}=3$. The respective upper plot shows the 
distribution of $d_e$ before (light grey/orange) and after (dark grey/blue) 
the LFV constraints are applied. The second distribution is hardly visible, 
as most of the models in these two samples are excluded by the LFV 
constraints. In the second row we therefore show the normalized distribution 
for the two cases. In the first case, $T=4\,$TeV, $Y_{\rm max}=0.5$ the 
tail of larger values of $d_e$ is cut away, while the peak of the 
distribution remains at the same value as without the LFV constraints. 
In the second case, $T=8\,$TeV, $Y_{\rm max}=3$, most of the distribution 
without LFV constraints lies above the EDM bound (\ref{eldipoleconstraint}) 
indicated by the vertical line. When the LFV constraints are taken into 
account the expected value of the electron EDM reduces by more than an 
order of magnitude. We note that with LFV constraints applied the 
current EDM bound is always close to the upper end of the distribution. 
In other words, quite generally, the lepton-flavour violating processes put
similar generic bounds on the electron EDM as the current bound from 
the direct EDM measurement. Of course, these conclusions hold in the 
parameter space of RS models with anarchic Yukawa couplings in a statistical 
sense, and not for any particular model.


\subsection{A note on LFV $\tau$ decays}
\label{sec:tau}

Tau decays offer another opportunity to study LFV. However, the short 
lifetime of the $\tau$ and its high mass make it unsuited for 
studies in low-energy facilities. The best bounds on processes
like $\tau \to e \gamma$ or $\tau \to 3 \mu$  come from
Babar \cite{Aubert:2009ag}, Belle \cite{Hayasaka:2007vc} 
as well as LHCb \cite{Aaij:2013fia}.

The RS model naturally generates higher rates for 
$\tau \to \mu,e$ transitions than for $\mu \to e$ transitions, since there 
is a close relation of lepton masses with the corresponding zero-mode 
profiles, which also control the size of LFV. However, the fantastic 
sensitivity of past and future experiments searching for muon flavour 
violation still makes searches in the muon sector the most promising 
avenue, unless an additional flavour structure suppresses muon flavour 
violation.

 \begin{figure}
\begin{minipage}{0.47\textwidth}
   \includegraphics[width=1\textwidth]{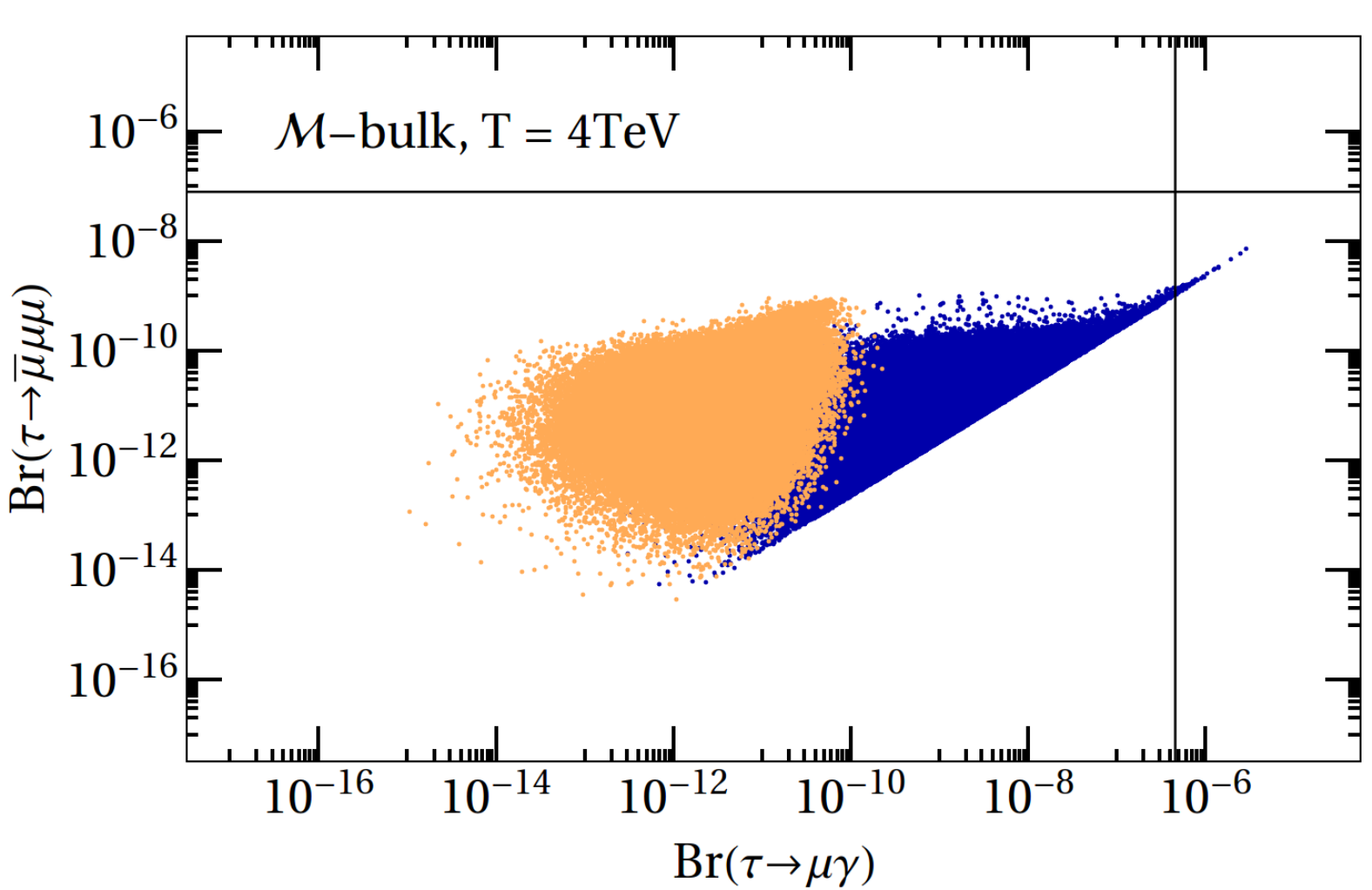}

    $\mbox{ }$
\end{minipage}
\begin{minipage}{0.47\textwidth}
  \includegraphics[width=1\textwidth]{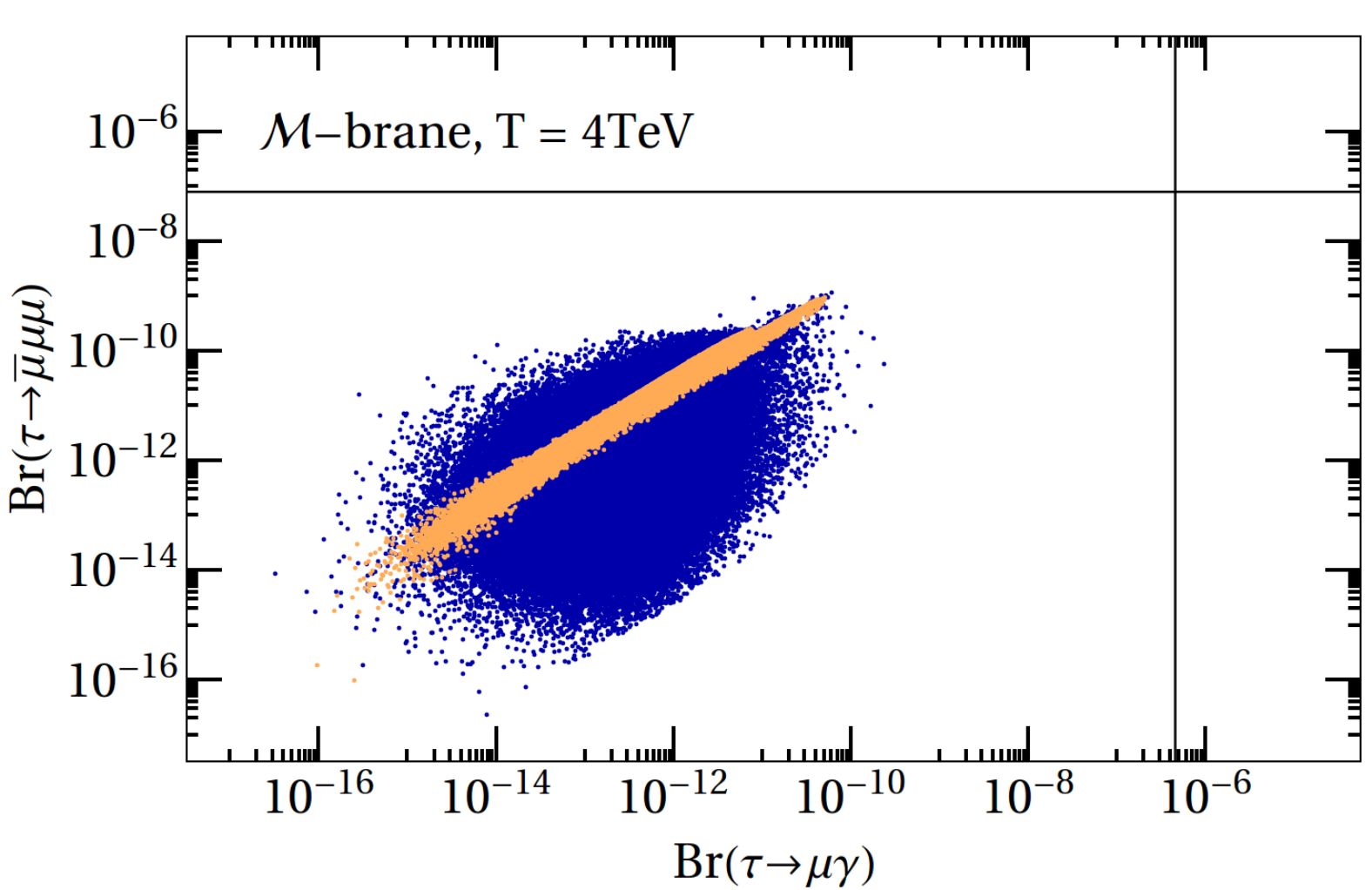}
    $\mbox{ }$
\end{minipage}
\vskip-0.1cm
\caption{\label{fig:tauCorrelation}  
Correlation of the branching ratios of $\tau\to \mu \gamma$ and 
$\tau \to 3\mu$ for $T=4\,\rm{TeV}$ in the minimal model. The left panel 
shows the bulk Higgs,
the right the exactly brane-localized Higgs case.}
\end{figure}

Nevertheless, it is instructive to provide the expectations for 
tau LFV in the RS model. In Figure~\ref{fig:tauCorrelation} we
show the values and correlation of the $\tau \to \mu\gamma$ and 
$\tau \to 3 \mu$ branching fractions.
The colour coding is the same as in the previous subsection.
The solid lines correspond to the current best upper bounds 
on the branching fractions. Compared to the bounds in the 
muon sector the current limits from tau decays are not restrictive even  
for $T=4\,{\rm TeV}$. An improvement of more than five orders of magnitude 
would be required for constraints as severe as those from muon decays.

Qualitatively, the $\tau \to \mu \gamma$ vs. $\tau \to 3\mu$ plot 
is similar to the corresponding ``muonic'' plots (first row in Figure
\ref{fig:MinimalMEGandM3e}). The main difference is the 
large effect of the four-fermion and fermion-Higgs operators. 
For the exactly brane-localized Higgs this generates the strong
correlation for small Yukawa couplings $Y_{\rm max}=0.5$. In the 
bulk Higgs case this effect 
prevents scatter points close to the dipole-dominance line.

\subsection{Discussion}
\label{sec:discussion}

In the following we summarize and emphasize the main conclusions from 
the phenomenological study. Based on the topologies of the decays 
we (naively) expected that $\mu \to e$ conversion in nuclei  
and $\mu \to 3e$ are driven by the current-current operators 
$(\bar{\ell}_e \Gamma \ell_\mu)\,(\bar \psi \Gamma \psi)$, 
whose Wilson coefficients are determined by the tree-level four-fermion
 and Higgs-fermion coefficients, $c_{ij}^a$ and $b_{ij}$, in the dimension-six 
SM effective Lagrangian~\eqref{EffectiveSMOperatorsDim6}. 
Whereas the rate for radiative muon decay should primarily be 
determined by the loop-induced dipole operator.
This situation would be ideal, as the two groups of processes
would then provide complementary information on the 
underlying RS model through the different parameter dependencies of 
the relevant Wilson coefficients.

However, and this is our main message \#1, we find that this is 
not generally the case. In particular, the process $\mu \to 3e$ can receive 
sizeable contributions from dipole operators. This can even be the 
dominant effect as indicated by the prominent dipole dominance 
line in the plots. In the minimal RS model with an exactly 
brane-localized Higgs and small Yukawa couplings the situation can 
also be reversed. Then $\mu \to e\gamma$ receives relevant contributions 
from tree-level coefficients and is no longer governed by the dipole 
operator alone. This makes it clear that it is necessary to 
consider all dimension-six operators for every observable to arrive at 
a reliable picture of LFV in the RS model. 

From our numerical analysis we also see that it is important to include 
both, the Higgs and the gauge-boson exchange contribution to the penguin 
diagrams that generate the dipole Wilson coefficient $\alpha^A_{ij}$, 
because they exhibit a fundamentally different 
dependence on the parameters---most notably the Yukawa coupling.
Only the full dipole coefficient including all contributions
gives an accurate description of the $\mu \to e \gamma$ decay
over a large range of Yukawa sizes. Without the Higgs contribution the 
dominant effect for large Yukawa couplings is completely missed. Whereas 
the presence of the gauge-boson exchange contribution  prevents the 
dipole from becoming irrelevant for small Yukawa couplings. This manifests 
itself in a lower bound 
on ${\rm Br}(\mu \to e \gamma)$, which depends only on the structure of 
the Yukawa matrix but {\it not} on the overall size of the couplings. 
This is our main message \#2: in RS models Higgs and gauge-boson induced 
flavour violation can both be important and only a full calculation 
of the dipole operator coefficient gives a reliable picture of LFV over 
all of the parameter space.

The scans suggest that anarchic RS models with minimal particle content
typically need KK scales $T$ larger than $4\,{\rm TeV}$ to be compatible 
with the current data on charged LFV. For models with a bulk Higgs the 
combination of $\mu \to e \gamma$ and muon conversion makes the bound on $T$ 
almost insensitive to the size of the Yukawa matrix. For an exactly 
brane-localized Higgs the situation is more complicated.
$T\gtrsim 4\,\rm TeV$ is still valid for small Yukawa couplings.
However, in this model the limit on $T$ comes mainly from muon conversion
data. As discussed above, muon conversion is dominated by tree-level operators 
whose Wilson coefficients decrease for increasing $Y_\star$. Thus, the 
bound on $T$ gets weaker the larger the Yukawa couplings. For 
$Y_{\rm max}=3$ we can still find numerous data point that satisfy all 
constraints for $T$ as low as $2\,\rm TeV$. In the custodially protected 
model on the other hand, the larger particle content generally leads to 
larger Wilson coefficients. Consequently, the lower bound on admissible KK 
scales $T$ is higher. It is noteworthy that in the custodially protected 
model the bound on $T$ is essentially independent of the Higgs localization 
for all three observables.

The limits on the KK scale $T$ will significantly improve in future 
experiments. Especially the next generation muon conversion 
searches will provide strong constraints on RS models. Our analysis was 
performed for a gold target nucleus---DeeMe \cite{Aoki:2012zza}, 
Mu2E \cite {Carey:2008zz} and COMET \cite{Cui:2009zz}  
use silicon and  aluminium target nuclei, respectively.
Aluminium and silicon have an approximately $20$ times smaller 
muon capture rate compared to gold, which enhances the branching fraction, 
but the wave-function overlap integrals 
($\mathcal{D}$,$\mathcal{V}$,$\mathcal{S}$) relevant to muon conversion 
are also smaller, see Table \ref{tab:InputParameters}.
Thus an expected lower bound on the branching fraction of 
$7\times 10^{-17}$ for aluminium is roughly equal to a bound around 
$10^{-16}$ in gold \cite{Kitano:2002mt}
(indicated by the dashed lines in the plots). This
combined with the expected improvement on ${\rm Br}(\mu \to e\gamma)$ 
during the next run of the MEG experiment 
\cite{Baldini:2013ke} could exclude the parameter space of anarchic 
RS models (custodially protected or minimal) up to a lowest KK resonance 
mass of 20~TeV, which corresponds to $T\gtrsim8 \,\rm TeV$.
The current $\mu\to 3e $ constraint is less constraining 
than the one from muon conversion. However, the proposed 
Mu3e experiment \cite{Mu3E} aims for sensitivity to a branching fraction  
of about $10^{-16}$. At this level $\mu\to 3e $ alone will be able to 
exclude anarchic models with $T\lesssim 5\,{\rm TeV}$.
Naturally, should LFV be observed in any of the 
new experiments, the model-dependent correlations 
among different processes can be used to further 
constrain the RS parameter space. Hence, our main message \#3: LFV 
violation provides very strong constraints on RS models, and future 
experiments will further strengthen them.

It is interesting to compare the charged LFV constraints on the RS model 
and its KK scale to those derived from other 
processes. The non-observation of direct KK gluon 
production forces $T$ to be larger than only about $1\,\rm TeV$. 
This cannot compete with the bounds from 
electroweak precision observables, notably the $S$ and $T$ parameter,
which are essentially model-independent. They only 
depend on the particle content of the model and to a lesser
degree on the 5D Higgs profile \cite{Cabrer:2011vu}. For the two 
models discussed in this work the electroweak precision observable 
bounds are $T>2.3\,\rm TeV$ (custodially protected) and $T>4\,\rm TeV$ 
(minimal) \cite{Casagrande:2008hr}. For not too large Yukawa couplings the 
RS contribution to $(g-2)_\mu$ is model-independent in the same 
sense~\cite{Beneke:2012ie,Moch:2014ofa}. However, in this case 
the SM contribution is non-zero and given the present situation 
of experiment and SM theory, the bounds on the KK scale 
are not competitive.    

If one allows for a somewhat stronger dependence of the bounds on the  
model parameters, Higgs production (and subsequent decay) is also an 
interesting observable. It depends more strongly on the (mainly quark) 
Yukawa matrices $Y_u,Y_d$ than the oblique parameters, 
but is still far less sensitive to its detailed structure than processes 
like $\mu\to 3e$, because Higgs production depends to leading order only 
on the traces of $Y_u^\dagger Y_u$, $Y_d^\dagger Y_d$. The trace 
of a product of anarchic matrices follows a narrower distribution than an 
individual matrix element. One finds that $T$ has to be larger than 
$2 \,(4) \,{\rm TeV}\;  \text{@}\, 95\%\,\rm CL$
for $Y_\star\approx 3 $ in the minimal (custodially protected) model  with a 
narrow bulk Higgs \cite{Malm:2013jia}. For smaller Yukawa couplings the bound 
becomes weaker as the effect on the production cross section  
decreases with $Y_\star$. For the exactly brane-localized Higgs the 
constraints are stronger,  
and one finds the same bounds on $T$ as above already for $Y_\star\approx 1$. 
As we have seen the situation is different for LFV observables.
In the minimal model an exactly  brane-localized Higgs leads to weaker 
bounds than a bulk Higgs, and in the custodially protected model
the bound from the bulk Higgs (with KK modes) is comparable 
to the one in the exactly brane-localized scenario. Thus for large quark 
Yukawa couplings and the exactly brane-localized Higgs, Higgs 
production provides at least equally strong bounds on the KK scale than 
the non-observation of charged LFV. The LHC will be able to
improve on this further in the future. In all other cases the next 
generation LFV experiments will be able to set the most stringent limits 
on the KK scale. Of course, this comparison assumes that the magnitude of 
anarchic Yukawa couplings is roughly the same in the quark and lepton 
sectors.

The quark-sector Yukawa couplings also enter the RS modification 
of meson oscillations, where in particular $\epsilon_K$, which measures 
CP violation in kaon mixing, is very sensitive to coloured KK states 
\cite{Csaki:2008zd,Blanke:2008zb}. If we follow \cite{Csaki:2008zd} 
and estimate this effect from the dominant left-right four-quark operator
$(Q_i \gamma_\mu T^A Q_i)(D_j \gamma^\mu T^A D_j)$ along the lines 
of Section~\ref{sec:simpleestimates}, we find that $T$ in excess of  
$8\,\rm TeV$ is needed to avoid conflict with experimental data.  
We have seen in the previous section that for any given data point 
$\mu\to e\gamma$  can differ by orders of magnitude from the simple 
estimate, since cancellations may or may not be present for a given set 
of 5D Yukawa and mass parameters. The same is true in the quark sector and 
much smaller modifications of $\epsilon_K$ than indicated by the naive estimate
are possible, see \cite{Blanke:2008zb}. We confirmed this independently 
using the results of \cite{Moch:2015oka}. Thus, from an analysis similar 
to the one performed in the previous section, as well as from 
\cite{Blanke:2008zb}, the limit of the TeV brane scale from $\epsilon_K$ 
is around $T=7\,\rm TeV$ --- quite close to the one expected from 
the next generation LFV experiments. However, contrary to $\epsilon_K$, 
whose ultimate reach is limited by the theoretical precision of the 
SM prediction, the potential of LFV observables, in particular of 
$\mu\to e$ conversion, is not yet exhausted. In any case, given that 
the quark and lepton Yukawa couplings may exhibit different patterns, 
independent tests of the RS model in both the quark and lepton flavour 
sectors in a wide range of observables are not redundant.

Our findings can be compared to the results of \cite{Agashe:2006iy}, which 
provided the first detailed analysis of lepton flavour violation
in the minimal RS model in the KK mode picture. The branching ratio of 
$\mu \to e \gamma$ is determined from the Higgs-exchange contribution 
to the dipole Wilson coefficient alone, 
which is computed via one-loop diagrams involving the Higgs zero-mode and 
first fermion KK excitation. 
Muon conversion and $\mu \to 3e$ are computed from the tree-level Wilson 
coefficients, while the dipole contribution is neglected. 
Both exactly brane-localized and bulk Higgs scenarios are investigated. 
We can compare with our full results only for the bulk Higgs case 
as their exactly brane-localized Higgs result for the dipole operator is 
cut-off dependent. Despite these caveats the overall size of the bound 
on the KK scale $T$ for a bulk Higgs is compatible with
the one found above. The main difference is the dependence of the 
branching fractions on the model parameters. As~\cite{Agashe:2006iy} 
only includes the Higgs contribution,\footnote{The absence of the KK modes 
does affect the qualitative characteristics of the Higgs contribution.} 
the dipole coefficient has a straightforward dependence on the Yukawa 
coupling size $Y_\star$. The identification of the 
dependence on the Yukawa coupling size as a distinguishing feature of 
tree-level and loop-induced observables, i.e. muon conversion and 
tri-lepton decay as opposed to $\ell'\to \ell \gamma$, is however valid only 
for medium-size Yukawa couplings, since otherwise the neglected 
gauge-boson contribution with its different dependence on $Y_\star$ 
becomes relevant for small $Y_\star$, and for large $Y_\star$ 
the tri-lepton decay is dominated by the dipole operator and therefore 
effectively also loop-induced.

We can also compare our results with \cite{Csaki:2010aj}. Here the 
Higgs-exchange contribution to the dipole operator is not considered. 
The dipole coefficient is computed in the 5D framework from a subset of 
gauge-boson exchange diagrams including a dimension-eight effect  
with three Yukawa matrices. Comparing orders of magnitudes 
their results for $\mu \to e\gamma$ and $\mu \to 3e$ are similar to our 
exactly brane-localized Higgs case in the minimal model. In particular, 
the lower bound on the $\mu \to e\gamma$ branching fraction for small 
Yukawa couplings is also present in their estimates.


\section{Summary}
\label{sec:conclusion}

In this paper we have undertaken a comprehensive study of 
charged lepton flavour violation in Randall-Sundrum models. The approach 
follows the strategy developed in \cite{Beneke:2012ie}, that is we 
assume that the KK scale is significantly larger than the electroweak 
scale and match the RS theory to the SM effective 
theory including dimension-six operators by integrating out the 
extra dimension in a fully 5D quantum-field-theoretical framework. 
Some of the Wilson coefficients of the dimension-six operators 
could be taken from \cite{Beneke:2012ie,Moch:2014ofa} and the remaining 
ones were computed here. We considered the RS model with minimal 
field content and an extended model with additional fields to protect 
low-energy precision measurements from custodial symmetry violating 
contributions. We further considered three implementations of the 
IR-brane localized Higgs field. The exactly localized case and 
the limiting cases of a bulk Higgs with and without KK Higgs excitations. 
The third scenario was motivated by the recent work \cite{Agashe:2014jca}, 
which demonstrated the non-decoupling of Higgs KK modes in the KK mode 
picture. We confirm this finding within the 5D framework.

Our calculation is the first complete one of dimension-six effects 
(more precisely those not suppressed by powers of small lepton 
masses), and considerably sharpens previous results 
from~\cite{Agashe:2006iy,Csaki:2010aj,Beneke:2014sta}. This 
concerns in particular the Wilson coefficient of the electromagnetic 
dipole operator $\bar L_i \sigma^{\mu\nu}E_j F_{\mu\nu}$. Not only 
does it depend on the way the Higgs is localized 
near the IR brane, which has phenomenological consequences. It also 
receives three contributions with different dependence on the 
magnitude of the anarchic 5D Yukawa matrices, which can all be 
important in certain parameter regions. Amongst these the gauge-boson 
exchange contribution is computationally the most demanding.  
As already emphasized in the 
discussion of our results in the previous section the interplay of 
the three contributions leads to distinctive features in the 
scan of the parameter space and it is important to include them all. 
We also find  that $\mu \to 3e$ can receive 
sizeable contributions from dipole operator, while in some 
cases $\mu \to e\gamma$ receives relevant contributions 
from tree-level operators and is no longer governed by the dipole 
operator alone.

Assuming generic anarchic Yukawa matrices we 
studied the typical range for the branching fractions 
of $\mu\to e\gamma$, $\mu \to 3e$, $\mu N  \to e N$ as well 
as $\tau\to \mu\gamma$, $\tau \to 3\mu$
in both the minimal and the custodially protected RS model. The 
combination of $\mu\to e\gamma$ and $\mu N  \to e N$ 
currently provides the most stringent constraints on the parameter 
space of the models. A typical lower limit on the KK scale $T$ is 
around $2 \,{\rm TeV}$ in the minimal model (up to 4~TeV in the 
bulk Higgs case with large Yukawa couplings), and around $4 \,{\rm TeV}$ 
in the custodially protected model, which corresponds to a mass of up 
to 10~TeV for the first 
KK excitations, far beyond the lower limit from the non-observation 
of direct production at the LHC. 
The next-generation LFV experiments will push the
lower limit on the KK scale $T$ further up. Given their projected 
sensitivity each of $\mu \to e \gamma$, $\mu N \to e N$ and 
$\mu \to 3e$ will contribute to improving the current bound.
When combining all searches, the non-observation of lepton-flavour 
violation will exclude anarchic RS models without 
additional flavour symmetries up to a KK scale of $T \sim 8~{\rm TeV}$,  
which corresponds to KK gluon masses of about $20 \rm \,TeV$. 

We also correlated the electric dipole moment predicted by 
the RS models with lepton flavour violating observables in the presently 
considered scenario of random anarchic Yukawa matrices. We find that 
the non-observation of charged LFV in current experiments imposes a similar 
bound on the electron EDM as the one set by the direct EDM measurement.

\subsubsection*{Acknowledgements}
We are grateful to A.~Crivellin for suggesting to include muon conversion 
in the analysis and for correspondence regarding \cite{Crivellin:2013hpa} and 
the Barr-Zee terms. We are further grateful to K.~Agashe for correspondence 
on KK Higgs contributions. 
The work of M.B.~and P.M.~is supported in part by the Gottfried Wilhelm
Leibniz programme of the Deutsche Forschungsgemeinschaft (DFG). 
The work of J.R.~is supported by STFC UK. We thank the Munich Institute 
for Astro- and Particle Physics (MIAPP) of the DFG cluster of excellence 
``Origin and Structure of the Universe'' for hospitality during part 
of the work. Feynman diagrams were drawn with the help of
Axodraw \cite{Vermaseren:1994je} and JaxoDraw \cite{Binosi:2003yf}. 


\appendix


\section{The bulk Higgs}
\label{sec:Higgs}

For the bulk Higgs we follow~\cite{Cacciapaglia:2006mz,Davoudiasl:2005uu}. 
The 5D Higgs action reads 
\begin{align}
\label{HiggsLagrangian}
S_{\Phi} = \int d^4x \int_{\frac{1}{k}}^{\frac{1}{T}} dz\,
\frac{1}{(kz)^5} & 
\bigg[\,g^{MN} \!\left(D_M\Phi \right)^\dagger \left(D_{N} \Phi\right)  
- \frac{\mu^2}{z^2} \Phi^\dagger \Phi \nn \\ 
&\quad - \delta(\sqrt{g_{55}}(z-1/T))V_{1/T}  
- \delta(\sqrt{g_{55}}(z-1/k))V_{1/k}\bigg]\,.
\end{align}
The brane potentials are
\begin{align}
\label{branepotentials}
 V_{1/k}=m_{1/k}\,\Phi^\dagger \Phi\,, 
&& V_{1/T} = - m_{1/T}\,\Phi^\dagger \Phi + \lambda  
\left( \Phi^\dagger \Phi\right)^2
\end{align}
with $m_{1/k}=(2+\beta)k$. We define $\beta =\sqrt{4+\mu^2}$.
Note that in \cite{Cacciapaglia:2006mz} the IR brane potential is, 
up to normalization, written as 
\begin{align}
 V_{1/T} = \frac{\tilde\lambda}{2 k^2} \left[\Phi^\dagger \Phi 
-\frac{v_{\rm TeV}^2}{2}\right]^2
\end{align}
where the coupling constant $\tilde\lambda$ is dimensionless.
For our purposes the form of \eqref{branepotentials}
is more convenient.
The choice for the UV potential parameter $m_{1/k}$ leads to a Higgs 
vacuum expectation value (vev) that rises towards 
the IR brane for positive $\beta$.  

\subsection{Zero mode and vacuum expectation value}

The zero-mode equations of motion are given by 
\begin{eqnarray}
&& \left( z^3 \partial_z z^{-3}\partial_z 
-\frac{\mu^2}{z^2} \right)\Phi^{(0)}(z)=-m_0^2\Phi^{(0)}(z)
\\
&& \left.\frac{\partial_z \Phi^{(0)}}{\Phi^{(0)}}\right|_{z=1/k}=m_{1/k}\,, 
\qquad
\left.\frac{\partial_z \Phi^{(0)}}{\Phi^{(0)}}\right|_{z=1/T}=
m_{1/T}\,\frac{T}{k}
\label{eq:bc1}
\end{eqnarray}
The boundary conditions ensure that the boundary terms arising from 
integration by parts vanish. $m_0^2$ is the squared mass of 
the zero mode in the {\it unbroken} phase. As 
we will see below it is tachyonic and of the order of the 
physical Higgs mass, i.e.~much smaller than the KK scale $T$.

The general solution of the differential equation is
\begin{align}
\label{modeequation}
\Phi^{(0)}(z)=\mathcal{N}_0 z^2 \left( \mathcal{J}_{\beta}(m_0 z)
+ C \mathcal{Y}_{\beta}(m_0 z)\right)\,.
\end{align}
The UV-brane boundary condition can be used to determine   
\begin{align}
C=&-\frac{\mathcal{J}_{\beta+1}(m_0/k)}{\mathcal{Y}_{\beta+1}(m_0/k)}
\approx \frac{2^{-2(1+\beta)}\pi}{\Gamma(1+\beta)\Gamma(2+\beta)}
\left(\frac{m_0}{k}\right)^{2+2\beta} 
\qquad \text{for}\; m_0\ll k\,.
\end{align}
Observing that $\mathcal{J}_{\beta}(x)/\mathcal{Y}_{\beta}(x) 
\propto x^{2\beta}$ for small arguments, we find that
\begin{align}
\mathcal{J}_{\beta}(m_0 z) \gg C{\mathcal{Y}_{\beta}(m_0 z)} 
\end{align}
for $m_0 \ll T$.
We can use this approximation for the zero mode to obtain
\begin{align}
\Phi^{(0)}(z)= \mathcal{N}_0 z^2 \mathcal{J}_{\beta}(m_0 z)\approx 
\mathcal{N}_0 z^{2+\beta} +\mathcal{O}(m_0 z)\,.
\end{align}
The overall  normalization is given by
\begin{equation}
\int_{\frac{1}{k}}^{\frac{1}{T}} \!\frac{dz}{(kz)^3} \,{\Phi^{(0)}(z)}^2=1 
\quad\Rightarrow\quad
\mathcal{N}_0\approx \sqrt{
\frac{2 (1 + \beta)}{1 - \epsilon^{2 + 2 \beta}}}\,k^{3/2} T^{1+\beta}\,.
\end{equation}
Up to higher terms in $m_0/T$ the zero-mode mass is determined by the equation
\begin{align}
\left.\frac{\partial_z \Phi}{\Phi}\right|_{z=1/T}=\,
T(2+\beta)-m_0 \frac{\mathcal{J}_{\beta+1}(m_0 /T)}
{\mathcal{J}_{\beta}(m_0 /T)} =m_{1/T}\frac{T}{k}\,.
\end{align}
Expanding the Bessel function for small argument, we find
\begin{equation}
m_{1/T}\frac{T}{k} -T(2+\beta)=- \frac{m_0^2}{2 (1+\beta)T} \,+ 
\;{\text{higher-order terms}}\,,
\end{equation}
which implies 
\begin{equation}
m_0^2\approx 2(1+\beta)\left( 2+\beta -\frac{m_{1/T}}{k}\right) T^2\,.
\label{eq:m0}
\end{equation}
Note that $m_0$ must be small compared to $T$, otherwise 
the expansions above would not have been allowed. We 
return to this point below. 

The 5D profile of the vev is not needed in 
our computation, since it is done in the unbroken electroweak phase. 
The Higgs vev only enters 
at the 4D level in the effective Lagrangian---as a low-energy parameter 
determined from experiment.
Still it is instructive to see how the vev profile arises.
To this end we substitute $\Phi\to \frac{1}{\sqrt{2}}(v+h)$
into the Lagrangian \eqref{HiggsLagrangian} and expand all terms 
(see \cite{Malm:2013jia} for a more detailed derivation). 
We can use that 4D derivatives on $v$ vanish.
This leads to the equations
\begin{align}
(-\partial_z +\frac{T}{k} m_{1/T} - 3 \frac{T}{k} \lambda v^2)h|_{z\to 1/T} 
=0\\
  (\partial_z -m_{1/k})h|_{z\to 1/k}=0\\
(-\partial_z +\frac{T}{k} m_{1/T} -  \frac{T}{k} \lambda v^2)v|_{z\to 1/T} =0\\
  (\partial_z -m_{1/k})v|_{z\to 1/k}=0
\end{align}
along with the standard equation for Higgs bulk profiles.
This gives the solution
\begin{align}
v(z)=\mathcal{N}_v z^{2+\beta}\,,
\end{align}
which is strongly IR localized already for moderately large,  
positive values of $\beta$.
The IR boundary condition determines
\begin{align}
\mathcal{N}_v^{\,2}= \left( m_{1/T} -(2+\beta)k \right)\frac
1\lambda T^{4+2\beta}\,.
\end{align}
Equivalently, by requiring that $W$ boson acquires the correct mass
\begin{align}
\mathcal{N}_v=\sqrt{\frac{2(1+\beta)}{1-\epsilon^{2+2\beta}} }\,
\, T^{\beta+1} k^{3/2} v_{\rm SM}.
\end{align}

With this input we can compute the physical Higgs mass
\begin{eqnarray}
m_H^2&=&m_0^2+6(1+\beta)\frac{T^2}{k} \lambda \,v(1/T)^2
\nn\\[-0.2cm]
&=& 2(1+\beta)
\left((2+\beta)k-m_{1/T} +3 \lambda \,{v(1/T)^2} \right) \frac{T^2}{k}\,.
\end{eqnarray}
Using
\begin{align}
&\lambda\frac{v(1/T)^2}{k}=\frac{m_{1/T}}{k}-(2+\beta)
\end{align}
this result can be rewritten into
\begin{align}
m_H^2=4 (1+\beta)\lambda\frac{v(1/T)^2}{k} T^2
\stackrel{!}{\approx} (125 {\rm GeV})^2
\end{align}
and
\begin{align}
&m_0^2=-\frac{m_H^2}{2}<0\;.
\end{align}  
Thus we find $|m_0^2|\ll T^2$, which was necessary to justify the 
expansion in the broken phase. We note that the requirement that 
$m_H\approx 125\,$GeV implies a fine-tuning between the parameters 
$m_{1/T}$ and $(2+\beta) k$, see (\ref{eq:m0}). We further note the relations 
\begin{align}
\label{eq:lambdaHiggs}
\lambda \, v(1/T)^2&=\frac{m_H^2}{4 (1+\beta)T^2} k\,,\\
m_{1/T}&= (2+\beta) k 
 + \underbrace{\frac{m_H^2}{4 (1+\beta)T^2}}_{\ll 1} k\,.
\end{align}   

\subsection{Higgs Propagator}

The 5D Higgs propagator is determined by the equations 
 \begin{eqnarray}
\label{propagatorequations}
 \left[ p^2 - \frac{\mu^2}{z^2} +z^3 \partial_z z^{-3} \partial_z \right] 
\Delta_\Phi(p,z,z^\prime)
& =& i (kz^\prime)^3 \delta(z-z^\prime)\,,\\
\left.\partial_z \Delta_\Phi(p,z,z^\prime)\right|_{z=1/k}&=& m_{1/k} 
\Delta_\Phi(p,1/k,z^\prime)\,,\\
\left.\partial_z \Delta_\Phi(p,z,z^\prime)\right|_{z=1/T}&=&m_{1/T} 
\frac{T}{k} \Delta_\Phi(p,1/T,z^\prime)\,,
\end{eqnarray}
which can be solved in the standard way, see e.g.~\cite{Beneke:2012ie}. 
After Wick rotation to euclidean space, the full Higgs propagator is 
given by
\begin{eqnarray}
\label{higgspropagator}
\Delta_{\phi}(p,z,z')&=&
\Theta(z-z')\,i k^3 z^2 z'^2 
\nn\\
&& \hspace*{-2cm}
\times\,\frac{\left(I_{\beta +1}\left(\frac{p}{k}\right) 
K_{\beta }(p z')+K_{\beta
   +1}\left(\frac{p}{k}\right) I_{\beta }(p z')\right) 
\left(I_{\beta +1}\left(\frac{p}{T}\right) K_{\beta }(p
   z)+K_{\beta +1}\left(\frac{p}{T}\right) I_{\beta }(p z)\right)}
{I_{\beta +1}\left(\frac{p}{k}\right) K_{\beta
   +1}\left(\frac{p}{T}\right)-K_{\beta +1}\left(\frac{p}{k}\right) 
I_{\beta +1}\left(\frac{p}{T}\right)}\nn \\
&&\hspace*{-2cm}+\,\{z\leftrightarrow z'\}\,,
\end{eqnarray}
where $K$ and $I$ are modified Bessel functions.
It is useful to not only have the full propagator, but also the zero-mode 
subtracted propagator.  We only work to leading accuracy in $v/T$, that is 
we approximate
\begin{align}
m_{1/T}\frac{T}{k} = (2+\beta) T\,.
\end{align}
The Higgs zero mode is then massless, and its profile is proportional 
to the vev profile derived previously.   
The zero mode can readily be removed from Euclidean propagator via
\begin{align}\label{higgspropagatorZMS}
&\Delta_{\phi}^{\rm ZMS}(p,z,z')=\Delta_{\phi}(p,z,z')-\frac{i}{(-p^2)} 
\,\Phi^{(0)}(z)\Phi^{(0)}(z')\,,
\end{align}
since removing the zero mode  
corresponds to removing the pole at $p^2=0$ from the full propagator.

\subsection{Yukawa matrix scaling}
\label{app:yukawascaling}

For the bulk Higgs field the Yukawa coupling 
develops a dependence on the Higgs 5D mass $\mu$ or, equivalently, $\beta$. 
To see how this dependence arises let us compare the situation 
with the delta-regularized narrow bulk Higgs~\eqref{eq:HiggsRegulator}.
In the latter case, we find for the 
4D SM lepton Yukawa matrix the standard expression
\begin{align}
\label{deltaYukawa}
y_{ij}& = f_{L_i}^{(0)}(1/T) g_{E_i}^{(0)}(1/T) \frac{T^3}{k^4} Y_{ij} 
+{\text{higher terms}}\nn \\
&=  \sqrt{\frac{1-2c_{L_i}}{1-\epsilon^{1-2c_{L_i}}}}
    \sqrt{\frac{1+2c_{E_j}}{1-\epsilon^{1+2c_{E_j}}}} \,Y_{ij}\,.
\end{align}
For the bulk Higgs the bulk action contains the interaction term
\begin{align}
S \supset  - \int_{\frac{1}{k}}^{\frac{1}{T}} \frac{dz}{(kz)^5} 
\int d^4x \,{Y}^{\beta}_{ij} \,\bar L_i(x,z) \Phi(x,z) E_j(x,z) + 
\text{h.c.}\,,
\end{align}
where $L,E,\Phi$ are 5D fields, and ${Y}^{\beta}$ is the dimensionful 
bulk Higgs Yukawa coupling. Inserting zero modes and integrating 
over $z$, we obtain (up to terms suppressed by powers of $\epsilon$) 
\begin{align}\label{betaYukawa}
y_{ij} &= Y^{\beta}\sqrt{\frac{1-2c_{L_i}}{1-\epsilon^{1-2c_{L_i}}}}
             \sqrt{\frac{1+2c_{E_j}}{1-\epsilon^{1+2c_{E_j}}}}
             \frac{\sqrt{2(1+\beta)}\,k^{1/2}}{2-c_{L_i}+c_{E_j}+\beta}\;.
\end{align}
Since the SM Yukawa coupling should remain finite for large $\beta$, 
the bulk-Higgs Yukawa coupling scales as
\begin{align}
Y^{\beta}\propto  \frac{2-c_{L_i}+c_{E_j}+\beta}{\sqrt{2(1+\beta)}} 
\stackrel{\beta \to \infty}{\rightarrow} \frac{\sqrt{\beta}}{\sqrt{2}}\,.
\end{align} 
Comparing the expressions \eqref{deltaYukawa} and \eqref{betaYukawa}
we identify 
\begin{align}
\label{YukawaRelation}
Y^{\beta} = \frac{Y}{\sqrt{k}} \frac{2-c_{L_i}+c_{E_j}+\beta}
{\sqrt{2(1+\beta)}}\,.
\end{align}

\subsection{KK Higgs example: Four-fermion operators}
\label{sec:example}

\begin{figure}
\begin{center}\begin{minipage}{0.4\textwidth}
\includegraphics[width=0.65\textwidth]{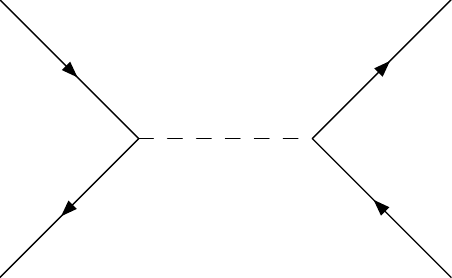}
\end{minipage}
\begin{minipage}{0.49\textwidth}
\includegraphics[width=\textwidth]{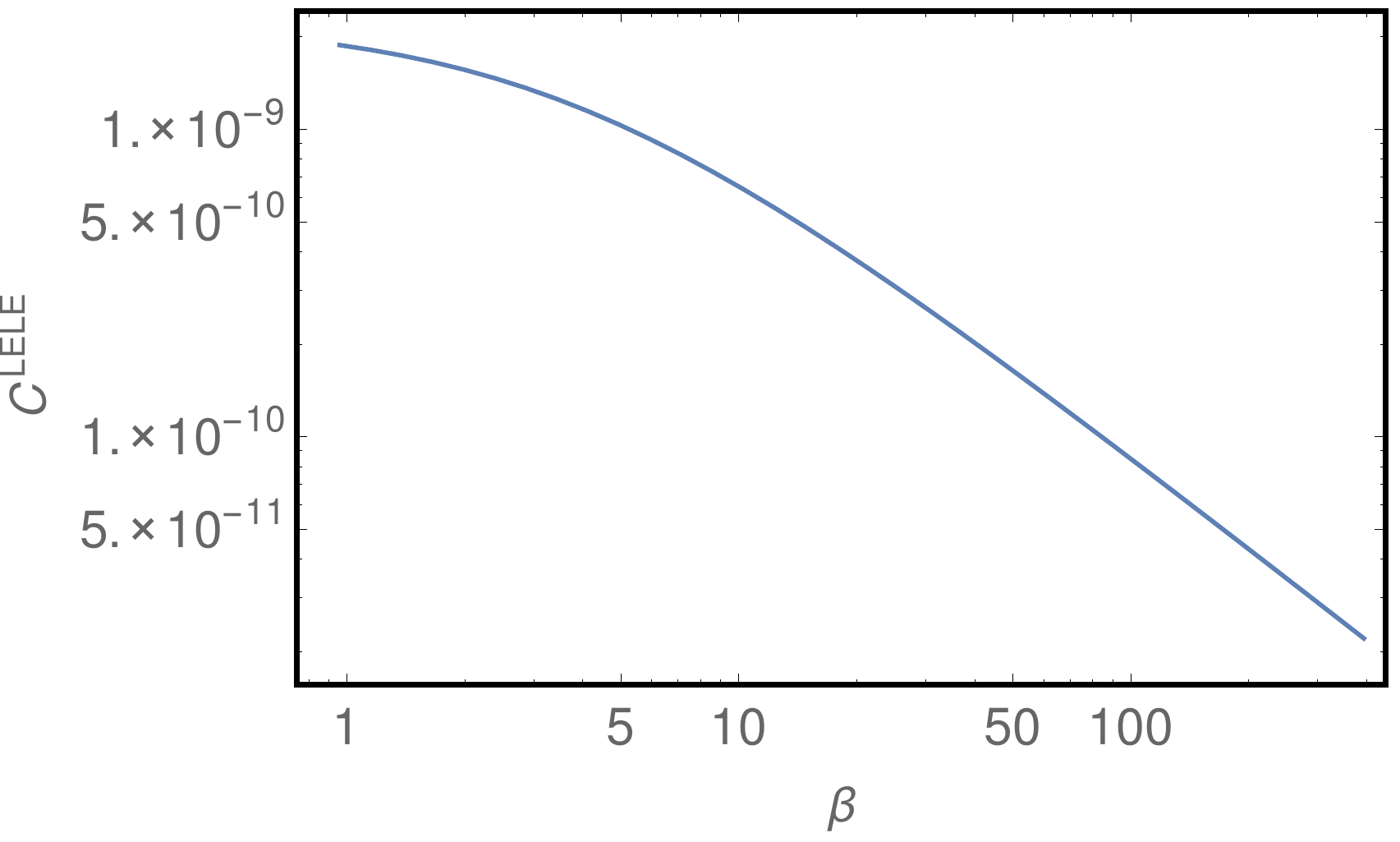} 
\end{minipage}
\end{center}
\vskip-0.4cm
\caption{{\it Left panel:} Diagram contributing to the matching onto 
$\bar L_i E_j\,\bar L_k E_l$. The intermediate Higgs propagator is 
zero-mode subtracted to remove long-distance 
contributions. {\it Right panel:} Wilson coefficient $C^{LELE}_{1111}$ as a 
function of $\beta$ for $T=1\,{\rm TeV}$,
$c_{L_1}=-c_{E_1}=0.6$ and $Y=\mathds{1}$. 
\label{Higgs4FermionDiag}}
\end{figure}

To gain some intuition for the properties of the Higgs KK modes 
we consider the example of the Feynman diagram in 
Figure~\ref{Higgs4FermionDiag}, which might contribute to the matching 
of four-fermion operators of the form $\bar L_i E_j\;\bar L_k E_l$. 
The corresponding Wilson coefficient is given by
\begin{align}
\label{vanishing4fermionCoeff}
\frac{C^{LELE}_{ijkl}}{T^2}= i Y^{\beta}_{ij} Y^{\beta}_{kl} 
\int_{\frac{1}{k}}^{\frac{1}{T}}\frac{dx}{(kx)^5}
\int_{\frac{1}{k}}^{\frac{1}{T}}\frac{dy}{(ky)^5} 
\,f^{(0)}_{L_i}(x) g^{(0)}_{L_j}(x) \Delta_{\phi}^{\rm ZMS}(p=0,x,y)  
f^{(0)}_{L_k}(y) g^{(0)}_{L_l}(y)\,.
\end{align}
For vanishing four-momentum exchange the zero-mode subtracted Higgs 
propagator has the particularly simple form
\begin{eqnarray}
\Delta^{\rm ZMS}_{\phi}(0,x,y)&=&
-\frac{i k^3 x^{2-\beta } y^{2-\beta }}{2 \beta}
\bigg[T^{2 \beta } x^{2 \beta }  y^{2 \beta } 
\left( \beta\left( T^2 \left(x^2+y^2\right)-2\right)-\frac{2}{2+\beta}\right) 
\nn \\
&& +\,x^{2 \beta }  \theta (y-x) +y^{2 \beta} \theta(x-y)\bigg],
 \label{eq:HiggsPropExp}
\end{eqnarray}
where we dropped terms suppressed by powers of $T/k$.
With this expression the integrals over $x$ and $y$ in 
\eqref{vanishing4fermionCoeff} are straightforward, and $C^{LELE}_{ijkl}$ can  
be determined analytically for all values of $\beta$. 
$C^{LELE}_{ijkl}$ vanishes as $1/\beta$ in the limit $\beta \to \infty$, 
as illustrated in the right panel of Figure~\ref{Higgs4FermionDiag}. 

This result can be understood by looking at the defining expression.
The scaling with $\beta$ is determined by three 
factors: the Yukawa matrix scaling, the scaling of the Higgs propagator, 
and the scaling of the integration variables $x$, $y$ in the relevant 
integration regions. The two Yukawa couplings each contribute a factor 
of $\sqrt{\beta}$. The Higgs propagator is slightly more complicated. 
Let us examine the three terms square brackets in \eqref{eq:HiggsPropExp} 
separately. The first term (without step-functions) does not feature an 
immediate suppression for large $\beta$, since the $1/\beta$ in the 
prefactor of the square bracket is cancelled. The suppression arises only 
after integration over the bulk coordinates. To see this, we write 
$x$ and $y$ in the overall factor $(T^2 x y)^{\beta}$ as 
$1/T(1-x_0/\beta)$ and $1/T(1-y_0/\beta)$, respectively, such that 
$x_0$ and $y_0$ measure the distance of $x$, $y$ from the IR brane in 
units of $1/(\beta T)$, the typical scale for Higgs KK excitations. 
We then find factors of the form $(1-x_0/\beta)^\beta$ and 
$(1-y_0/\beta)^\beta$, which behave as $e^{-x_0 }$ and $e^{-y_0 }$ for 
large $\beta$, respectively. Hence the first term counts as 
$\mathcal{O}(1)$ only for $x$ and $y$ within $1/(\beta T)$ of the 
IR brane. The 5D coordinate integrals then count as $1/(\beta T)$
each, and the contribution of the first term in \eqref{eq:HiggsPropExp} to 
the Wilson coefficient is of order $(\sqrt{\beta})^2 \times 1 
\times 1/\beta^2$, which vanishes for large $\beta$.
The remaining two terms in \eqref{eq:HiggsPropExp} have different properties.
There is a global factor of $1/\beta$, but there is no requirement that  
$x$, $y$ are close to the IR brane. Let us focus on the second term, which 
is non-zero only for~$y>x$. It contains the factor $(x/y)^{\beta}$, which 
ensures that the contribution to the Wilson coefficient is exponentially 
suppressed if $x \ll  y\,(1-1/\beta)$.  Changing integration variables from 
$x,y$ to $y,x-y$ shows that the integral over $x-y$ counts as $1/(\beta T)$, 
while the integral over $y$ is effectively unconstrained. Thus the 
overall total scaling is  $(\sqrt{\beta})^2 \times 1/\beta \times 1/\beta$, 
which also vanishes for large $\beta$. 
The same argument with $x \leftrightarrow y$ ensures that the third term 
in \eqref{eq:HiggsPropExp} does not contribute to $C^{LELE}_{ijkl}$ for 
$\beta \to \infty$.


\section{KK Higgs contributions}
\label{sec:KKHiggs}

\begin{figure}[t]
 \centering
 \begin{minipage}{0.48\textwidth}
 \includegraphics[height=0.6\textwidth]{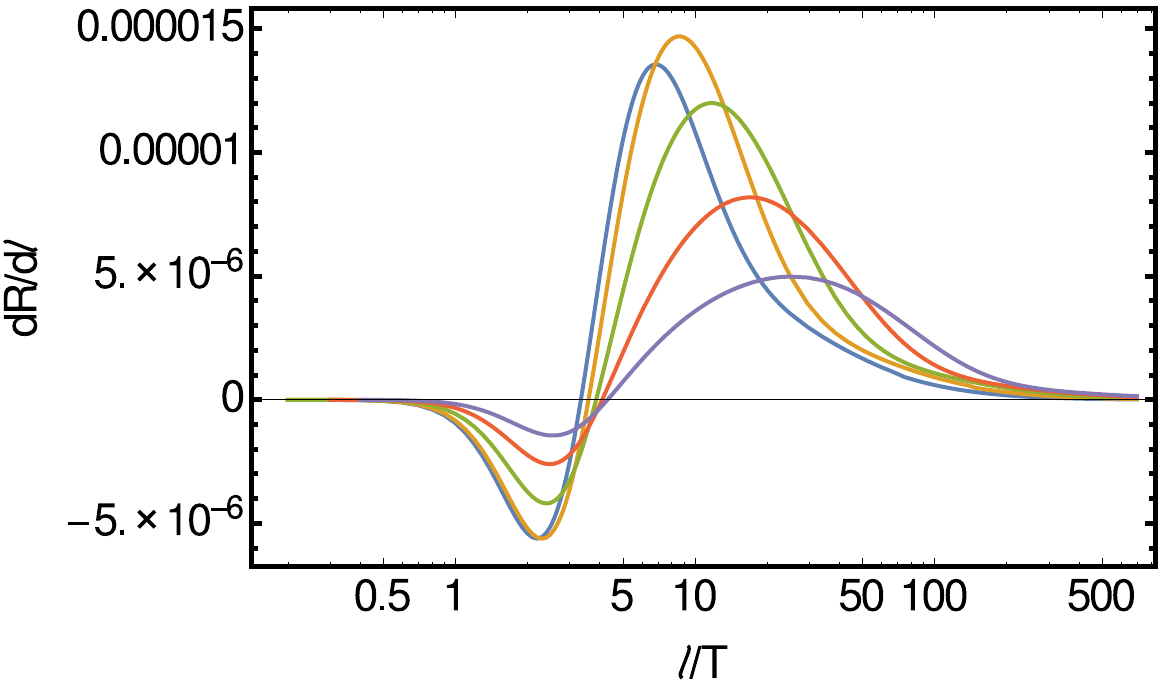}
 \end{minipage}
 \hspace*{0.2cm}
 \begin{minipage}{0.49\textwidth}
 \includegraphics[height=0.6\textwidth]{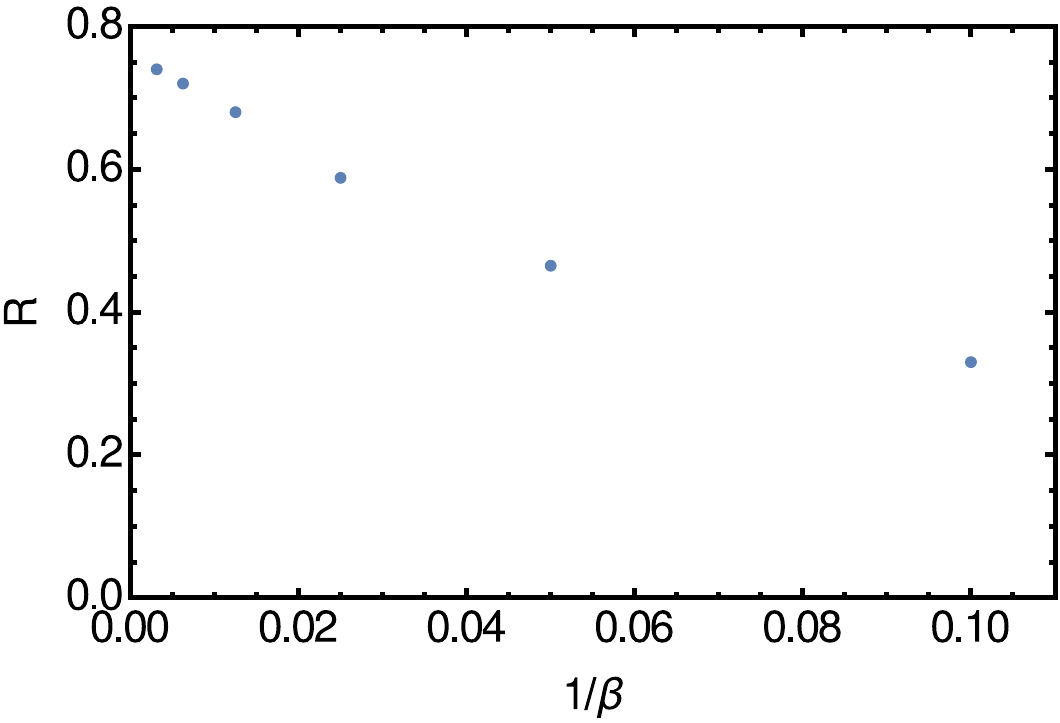}
\end{minipage}
\caption{{\it Left panel:} $dR/dl$ as a function of the loop momentum 
for $\beta=$~10, 20, 40, 80, 160 (curves from left to right)  
for the  diagrams in the last line of \eqref{eq:RRatio}.
{\it Right panel:} Corresponding ratio $R$ as a function of $1/\beta$ 
(no uncertainties shown).}
 \label{fig:KKHiggsInternalPlot}
\end{figure}

The contribution of KK Higgs modes to $a^H_{ij}$ is proportional to the 
corresponding contribution for the Higgs zero-mode for each diagram 
topology. It is therefore convenient to study the ratio of the two 
contributions, 
\begin{align}
 R=\frac{a^{H}_{\rm KK}}{a^H_{\text{ZM}}}\,.
\end{align}
Up to small corrections this ratio is also independent of 
the flavour of the propagating states. The 
Higgs KK contribution can then be obtained by multiplying
the zero-mode result by the corresponding $R$.  
It should be noted that not all topologies shown below 
actually contribute to $a^H_{ij}$ in a specific RS model, 
either because the combination of SU(2) and U(1) group factors vanishes 
or because the model does not have Feynman rules that allow 
for the particular diagram to exist. 
Note that we do not separate contributions from 
wrong- and correct-chirality Higgs couplings. The numbers refer to the 
sum of both type of contributions, and are given by:

\begin{align}
\label{eq:RRatio}
& \includegraphics[width=2.5cm]{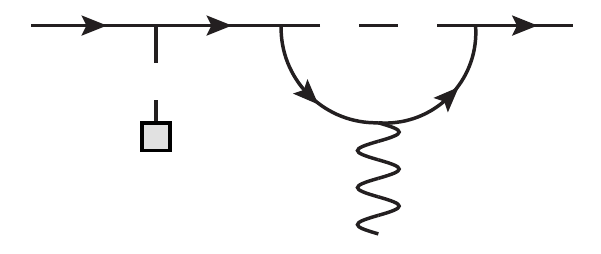}  R\approx 0.27(0.01)
&&   \includegraphics[width=2.5cm]{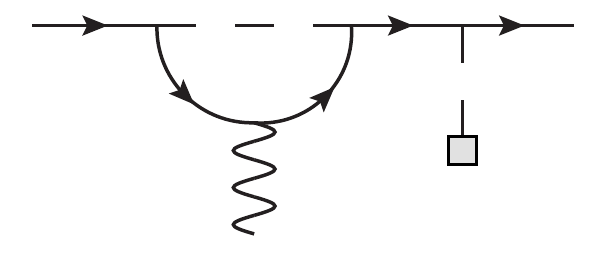}  R\approx 0.27(0.01)          \nn  \\ \; \nn  \\
& \includegraphics[width=2.5cm]{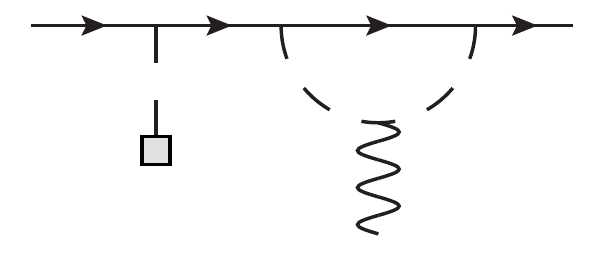}  R\approx 0.08(0.01)
&&   \includegraphics[width=2.5cm]{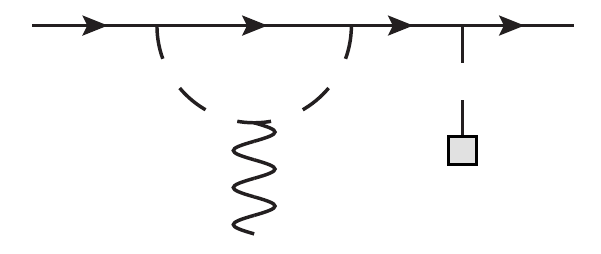}  R\approx 0.08(0.01)          \nn  \\ \; \nn  \\
 &  \includegraphics[width=2.5cm]{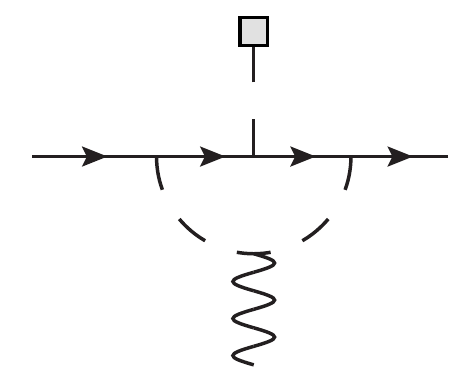}  R\approx 0.15(0.05)
&& \nn  \\ \; \nn  \\
 &      \includegraphics[width=5.5cm]{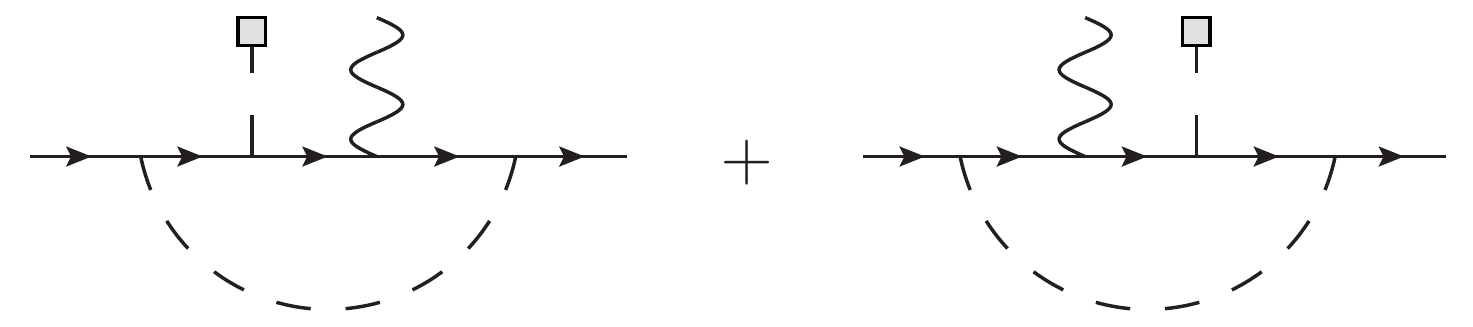}  R\approx 0.77(0.08) 
&&\\[-0.4cm]
 \nn
\end{align}
Our error estimates are shown in parenthesis. It arises from the numerical 
integration error and an estimate for the extrapolation error to 
$\beta=\infty$, since a numerically stable evaluation is possible only 
up to $\beta \approx 200-300$. 
The numerically most challenging diagrams are the ones where the
photon is emitted from the Higgs, since they contain products of KK Higgs 
propagators. We also note that the KK Higgs contribution converges 
relatively slowly as $\beta \to \infty$, if the diagrams 
involve an external Higgs attachment to a fermion line 
in the loop, as illustrated in  
Figure~\ref{fig:KKHiggsInternalPlot}.
The typical scaling with powers of $\beta$ in the different momentum 
regions  discussed in the main text 
does not set in until $\beta\sim 40$. This behaviour agrees with 
observations made in \cite{Agashe:2014jca}.
 
\end{document}